\documentclass[12pt, draftclsnofoot, onecolumn]{IEEEtran}
\usepackage[compress]{cite}
\usepackage{graphicx}
\usepackage{color}
\usepackage{amsmath}
\usepackage{caption}
\usepackage{capt-of}
\usepackage{hyperref}
\usepackage{amssymb}
\usepackage{amsfonts}
\usepackage{psfrag}
\usepackage{booktabs}
\usepackage{listofitems}
\usepackage{array}
\usepackage{subfig}
\usepackage[named]{algo}
\usepackage{makecell}
\usepackage{multirow}
\usepackage{algorithmic}
\usepackage{algorithm}
\usepackage{algorithm,setspace}
\usepackage{framed}
\usepackage[table,xcdraw]{xcolor}
\usepackage{comment}
\usepackage{stackengine}
\usepackage{hhline}
\usepackage[table,xcdraw]{xcolor}
\usepackage{tablefootnote}
\usepackage{tikz}
\usepackage{longtable}

\newcommand\xrowht[2][0]{\addstackgap[.5\dimexpr#2\relax]{\vphantom{#1}}}

\IEEEoverridecommandlockouts

\DeclareMathOperator*{\argmaxA}{arg\,max}

\begin{document}
\title{Safety Analysis for Laser-based Optical Wireless Communications: A Tutorial}

\author{\IEEEauthorblockN{Mohammad Dehghani Soltani,  Elham Sarbazi, Nikolaos Bamiedakis, Priyanka \\ de Souza, Hossein Kazemi, Jaafar M. H. Elmirghani, Ian H. White, Richard V. Penty, Harald Haas and Majid Safari}
\thanks{M. D. Soltani and M. Safari are with the Institute for Digital Communications (IDCOM), School of Engineering, the University of Edinburgh, Edinburgh, UK. e-mails: \{msoltan, majid.safari\}@ed.ac.uk.}
\thanks{E. Sarbazi, H. Kazemi and H. Haas are with the LiFi Research and Development Centre, Department of  Electronic \& Electrical Engineering, University of Strathclyde, Glasgow, UK. e-mails: \{e.sarbazi, h.kazemi, harald.haas\}@strath.ac.uk.}
\thanks{Jaafar M. H. Elmirghani is with the School of Electronic and Electrical Engineering, University of Leeds, Leeds, UK. e-mails: J.M.H.Elmirghani@leeds.ac.uk.}
\thanks{Ian H. White is with the University of Bath, Bath, UK. e-mail: I.H.White@bath.ac.uk}
\thanks{N. Bamiedakis and R. Penty are with the Centre for Photonic Systems, Electrical Engineering
Division, Department of Engineering, University of Cambridge, Cambridge CB3 0FA, U.K. e-mails:\{nb301,rvp11\}@cam.ac.uk.}
\thanks{$^*$\textit{Corresponding author: M. D. Soltani, msoltan@ed.ac.uk}}
} 

\maketitle

\vspace{-1cm}

\begin{abstract}
Light amplification by stimulated emission of radiation (laser) sources have many advantages for use in high data rate optical wireless communications. In particular, the low cost and high-bandwidth properties of laser sources such as vertical-cavity surface-emitting lasers (VCSELs) make them attractive for future indoor optical wireless communications. In order to be integrated into future indoor networks, such lasers should conform to eye safety regulations determined by the international electrotechnical commission (IEC) standards for laser safety.   
In this paper, we provide a detailed study of beam propagation to evaluate the received power of various laser sources, based on which as well as the maximum permissible exposure (MPE) defined by the IEC 60825-1:2014 standard, we establish a comprehensive framework for eye safety analyses. This framework allows us to calculate the maximum allowable transmit power, which is crucial in the design of a reliable and safe laser-based wireless communication system.  
Initially, we consider a single-mode Gaussian beam and calculate the maximum permissible transmit power. Subsequently, we generalize this approach for higher-mode beams. It is shown that the $M$-squared-based approach for analysis of multimode lasers ensures the IEC eye safety limits, however, in some scenarios, it can be too conservative compared to the precise beam decomposition method.
Laser safety analyses with consideration of optical elements such as lens and diffuser, as well as for VCSEL array have been also presented. Skin safety, as another significant factor of laser safety, has also been investigated in this paper. We have studied the impacts of various parameters such as wavelength, exposure duration and the divergence angle of laser sources on the safety analysis by presenting insightful results.   
\end{abstract}


\section{Introduction}
\label{Section1}
The number of mobile users is increasing rapidly, with expected numbers to exceed $5.7$ billion, that is anticipated to generate $79$\% of the global data traffic \cite{VNI_2022}. It is forecast that smartphones, laptops, tablets and wireless sensors will struggle to get their required share of the radio frequency (RF) spectrum without a significant capacity increase \cite{cave2002review}. In the 5th generation ($5$G) and beyond of cellular networks, $1000$-fold capacity growth is expected with respect to the 4th generation ($4$G) and long-term evolution (LTE) wireless networks \cite{5GNet2014}. Therefore, the RF spectrum is becoming congested and wireless devices have to cope with increased co-channel interference. As a result, the data throughput of the devices is severely affected, and the established connection may be poor.

Optical wireless communication (OWC) enables wireless connectivity harnessing a huge spectrum, which is available in infrared, visible or ultraviolet bands \cite{KhalighiFSO2014}. Moreover, OWC is license free thus leading to a cost-effective service \cite{OWCMurat2014}. It is a solution to alleviate RF spectrum congestion for both indoor and outdoor scenarios \cite{OWC2020,Koonen2018JLT}. OWC enables the creation of smaller communication cells in the network, known as attocells \cite{LiFi2016} which is a vital key to unlock the path to exponential capacity growth for indoor scenarios \cite{Elgala2011Mag}.
Indoor OWC can offload heavy traffic loads from congested RF wireless networks, therefore making the room available for extreme low-capacity streams such as in the Internet of Things (IoT). It can ensure higher security compared to the RF counterpart in the physical layer because optical signals do not penetrate walls or opaque objects \cite{1455777}. Outdoor applications of OWC, which are widely referred to as free-space optical (FSO) communications, include satellite-to-satellite communications, inter-building connections as well as satellite-to-plane communications \cite{Kaushal2017OWC}. Underwater OWC is another application which has recently gained much attention since it is able to provide a much higher transmission bandwidth, and, as a result, much higher data rate in comparison to acoustic and RF counterparts \cite{7450595,7593257}. Vehicle-to-vehicle (V2V) communication via the optical spectrum is another recent area of focus, where experimental results have confirmed its reliability even under heavy-fog conditions \cite{7323787}.
All the aforementioned applications of OWC and its capability to offer low-cost, high-speed, secure and reliable communication links have made OWC an indispensable piece of future generations of communications. 


Compared to light emitting diodes (LEDs), light amplification by stimulated emission of radiation (laser) diodes provide a larger modulation bandwidth that can achieve multi-Gb/s data rates, making them appealing to be used for Tb/s indoor networks \cite{Wang2019Laser,9217158,9217368} required for the next generations of wireless networks. A data rate of $4~\times~12.5$~Gbps has been successfully demonstrated by experiments in \cite{5770159} via optical fiber over the entire beam footprint of about 1 m.
Among different types of laser diodes, vertical cavity surface emitting lasers (VCSELs) are one of the strongest candidates to fulfil this role due to several outstanding features such as \cite{Iga2000JSTQE}: high-speed modulation (bandwidths over $28$ GHz) \cite{10.1117/12.2001497,6800720,haglund201530}, high power conversion efficiency and low cost.  
Furthermore, high aggregate bit-rates exceeding $100$ Gbit/s have been confirmed by means of VCSEL arrays through experiments in many studies \cite{8084331,6924758,6995226,9217158,9217368}. 
These attributes make VCSELs noteworthy for many applications, particularly for high-speed indoor networks \cite{Liu19VCSEL}. 
However, one of the major challenges of utilizing laser sources as transmitters is to meet eye safety regulations \cite{IEC_Std608251}. In the following section, we review the relevant standards, research papers and handbooks on laser safety.

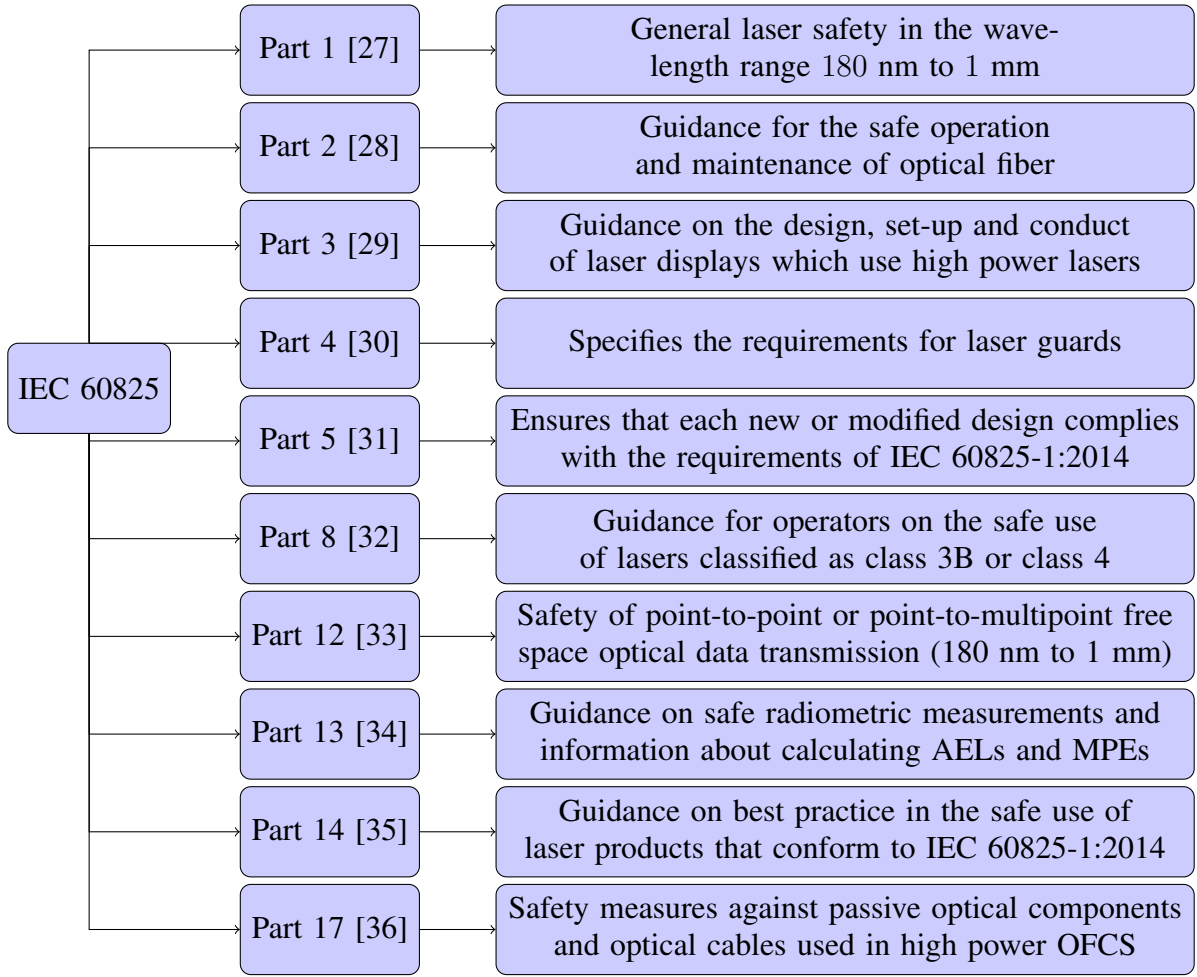
\begin{figure}[!th]
\begin{center}
\begin{tikzpicture}
\tikzstyle{Box} = [rectangle, rounded corners, minimum height=1.2cm,text centered, draw=black, fill=white]

\node (B1) [Box,draw=black, fill=blue!20]{IEC 60825};
\node (B2) [Box,draw=black, fill=blue!20,right of=B1, yshift=4.5cm, xshift=2.2cm,text width=2.1cm]{Part 1 \cite{IEC_Std608251}};
\node (B3) [Box,draw=black, fill=blue!20, below of=B2, yshift=-0.3cm,text width=2.1cm]{Part 2 \cite{IEC60825-2}};
\node (B4) [Box,draw=black, fill=blue!20, below of=B3, yshift=-0.3cm,text width=2.1cm]{Part 3 \cite{IEC60825-3}};
\node (B5) [Box,draw=black, fill=blue!20, below of=B4, yshift=-0.3cm,text width=2.1cm]{Part 4 \cite{IEC60825-4}};
\node (B6) [Box,draw=black, fill=blue!20, below of=B5, yshift=-0.3cm,text width=2.1cm]{Part 5 \cite{IEC60825-5}};
\node (B7) [Box,draw=black, fill=blue!20, below of=B6, yshift=-0.3cm,text width=2.1cm]{Part 8 \cite{IEC60825-6}};
\node (B8) [Box,draw=black, fill=blue!20, below of=B7, yshift=-0.3cm,text width=2.1cm]{Part 12 \cite{IEC60825-7}};
\node (B9) [Box,draw=black, fill=blue!20, below of=B8, yshift=-0.3cm,text width=2.1cm]{Part 13 \cite{IEC60825-8}};
\node (B10) [Box,draw=black, fill=blue!20, below of=B9, yshift=-0.3cm,text width=2.1cm]{Part 14 \cite{IEC60825-9}};
\node (B10-1) [Box,draw=black, fill=blue!20, below of=B10, yshift=-0.3cm,text width=2.1cm]{Part 17 \cite{IEC60825-10}};

\draw [->] (B1) |- (B2);
\draw [->] (B1) |- (B3);
\draw [->] (B1) |- (B4);
\draw [->] (B1) |- (B5);
\draw [->] (B1) |- (B6);
\draw [->] (B1) |- (B7);
\draw [->] (B1) |- (B8);
\draw [->] (B1) |- (B9);
\draw [->] (B1) |- (B10);
\draw [->] (B1) |- (B10-1);

\node (B11) [Box,draw=black, fill=blue!20,right of=B2, xshift=5.85cm, text width=9cm]{General laser safety in the wavelength range $180$ nm to $1$ mm};
\node (B12) [Box,draw=black, fill=blue!20,right of=B3, xshift=5.85cm, text width=9cm ]{Guidance for the safe operation and maintenance of optical fiber};
\node (B13) [Box,draw=black, fill=blue!20,right of=B4, xshift=5.85cm, text width=9cm ]{Guidance on the design, set-up and conduct of laser displays which use high power lasers};
\node (B14) [Box,draw=black, fill=blue!20,,right of=B5, xshift=5.85cm, text width=9cm ]{Specifies the requirements for laser guards};
\node (B15) [Box,draw=black, fill=blue!20,,right of=B6, xshift=5.85cm, text width=9cm]{Ensures that each new or modified design complies with the requirements of IEC 60825-1:2014};
\node (B16) [Box,draw=black, fill=blue!20,,right of=B7, xshift=5.85cm, text width=9cm]{Guidance for operators on the safe use of lasers classified as class 3B or class 4};
\node (B17) [Box,draw=black, fill=blue!20,,right of=B8, xshift=5.85cm, text width=9cm]{Safety of point-to-point or point-to-multipoint free space optical data transmission (180 nm to 1 mm)};
\node (B18) [Box,draw=black, fill=blue!20,,right of=B9, xshift=5.85cm, text width=9cm]{Guidance on safe radiometric measurements and information about calculating AELs and MPEs};
\node (B19) [Box,draw=black, fill=blue!20,,right of=B10, xshift=5.85cm, text width=9cm]{Guidance on best practice in the safe use of laser products that conform to IEC 60825-1:2014};
\node (B20) [Box,draw=black, fill=blue!20,,right of=B10-1, xshift=5.85cm, text width=9cm]{Safety measures against passive optical components and optical cables used in high power OFCS};

\draw [->] (B2) -- (B11);
\draw [->] (B3) -- (B12);
\draw [->] (B4) -- (B13);
\draw [->] (B5) -- (B14);
\draw [->] (B6) -- (B15);
\draw [->] (B7) -- (B16);
\draw [->] (B8) -- (B17);
\draw [->] (B9) -- (B18);
\draw [->] (B10) -- (B19);
\draw [->] (B10-1) -- (B20);

\end{tikzpicture}
\end{center}
\caption{IEC 60825 standard series on the safety of laser products along with their main focus.} 
\label{Fig-IEC}
\end{figure}

\begin{figure}[t!]
\begin{center}
\begin{tikzpicture}
\tikzstyle{Box} = [rectangle, rounded corners, minimum height=1.2cm,text centered, draw=black, fill=white]

\node (B1) [Box,draw=black, fill=red!20]{ANSI};
\node (B2) [Box,draw=black, fill=red!20,right of=B1, yshift=4.5cm, xshift=1.6cm]{ANSI Z136.1 \cite{ANSI2014}};
\node (B3) [Box,draw=black, fill=red!20, below of=B2, yshift=-0.3cm]{ANSI Z136.2 \cite{ANSI-Z136.2}};
\node (B4) [Box,draw=black, fill=red!20, below of=B3, yshift=-0.3cm]{ANSI Z136.3 \cite{ANSI-Z136.3}};
\node (B5) [Box,draw=black, fill=red!20, below of=B4, yshift=-0.3cm]{ANSI Z136.4 \cite{ANSI-Z136.4}};
\node (B6) [Box,draw=black, fill=red!20, below of=B5, yshift=-0.3cm]{ANSI Z136.5 \cite{ANSI-Z136.5}};
\node (B7) [Box,draw=black, fill=red!20, below of=B6, yshift=-0.3cm]{ANSI Z136.6 \cite{ANSI-Z136.6}};
\node (B8) [Box,draw=black, fill=red!20, below of=B7, yshift=-0.3cm]{ANSI Z136.7 \cite{ANSI-Z136.7}};
\node (B9) [Box,draw=black, fill=red!20, below of=B8, yshift=-0.3cm]{ANSI Z136.8 \cite{ANSI-Z136.8}};
\node (B10) [Box,draw=black, fill=red!20, below of=B9, yshift=-0.3cm]{ANSI Z136.9 \cite{ANSI-Z136.9}};

\draw [->] (B1) |- (B2);
\draw [->] (B1) |- (B3);
\draw [->] (B1) |- (B4);
\draw [->] (B1) |- (B5);
\draw [->] (B1) |- (B6);
\draw [->] (B1) |- (B7);
\draw [->] (B1) |- (B8);
\draw [->] (B1) |- (B9);
\draw [->] (B1) |- (B10);

\node (B11) [Box,draw=black, fill=red!20,right of=B2, xshift=5.85cm, text width=9cm]{General safe use of lasers for industry, military, research and development};
\node (B12) [Box,draw=black, fill=red!20,right of=B3, xshift=5.85cm, text width=9cm ]{Safe use of optical communication systems utilizing laser diode and LED sources};
\node (B13) [Box,draw=black, fill=red!20,right of=B4, xshift=5.85cm, text width=9cm ]{Safe use of high power lasers (class 3B and class 4) in health care};
\node (B14) [Box,draw=black, fill=red!20,,right of=B5, xshift=5.85cm, text width=9cm ]{Guidance on safe measurements of the classification and hazard evaluation};
\node (B15) [Box,draw=black, fill=red!20,,right of=B6, xshift=5.85cm, text width=9cm]{Safe use of lasers in educational institutions};
\node (B16) [Box,draw=black, fill=red!20,,right of=B7, xshift=5.85cm, text width=9cm]{Safe use of lasers  outdoors, e.g., construction, laser lightshows and military};
\node (B17) [Box,draw=black, fill=red!20,,right of=B8, xshift=5.85cm, text width=9cm]{Guidance on the test methods and protocols used to provide eye protective equipment};
\node (B18) [Box,draw=black, fill=red!20,,right of=B9, xshift=5.85cm, text width=9cm]{Guidance on the safe use of lasers in research, development, or testing};
\node (B19) [Box,draw=black, fill=red!20,,right of=B10, xshift=5.85cm, text width=9cm]{Policies to ensure laser safety in both public and private manufacturing environments};

\draw [->] (B2) -- (B11);
\draw [->] (B3) -- (B12);
\draw [->] (B4) -- (B13);
\draw [->] (B5) -- (B14);
\draw [->] (B6) -- (B15);
\draw [->] (B7) -- (B16);
\draw [->] (B8) -- (B17);
\draw [->] (B9) -- (B18);
\draw [->] (B10) -- (B19);

\end{tikzpicture}
\end{center}
\caption{Classification of ANSI standards on laser safety including their main focus.} 
\label{Fig-ANSI}
\end{figure}
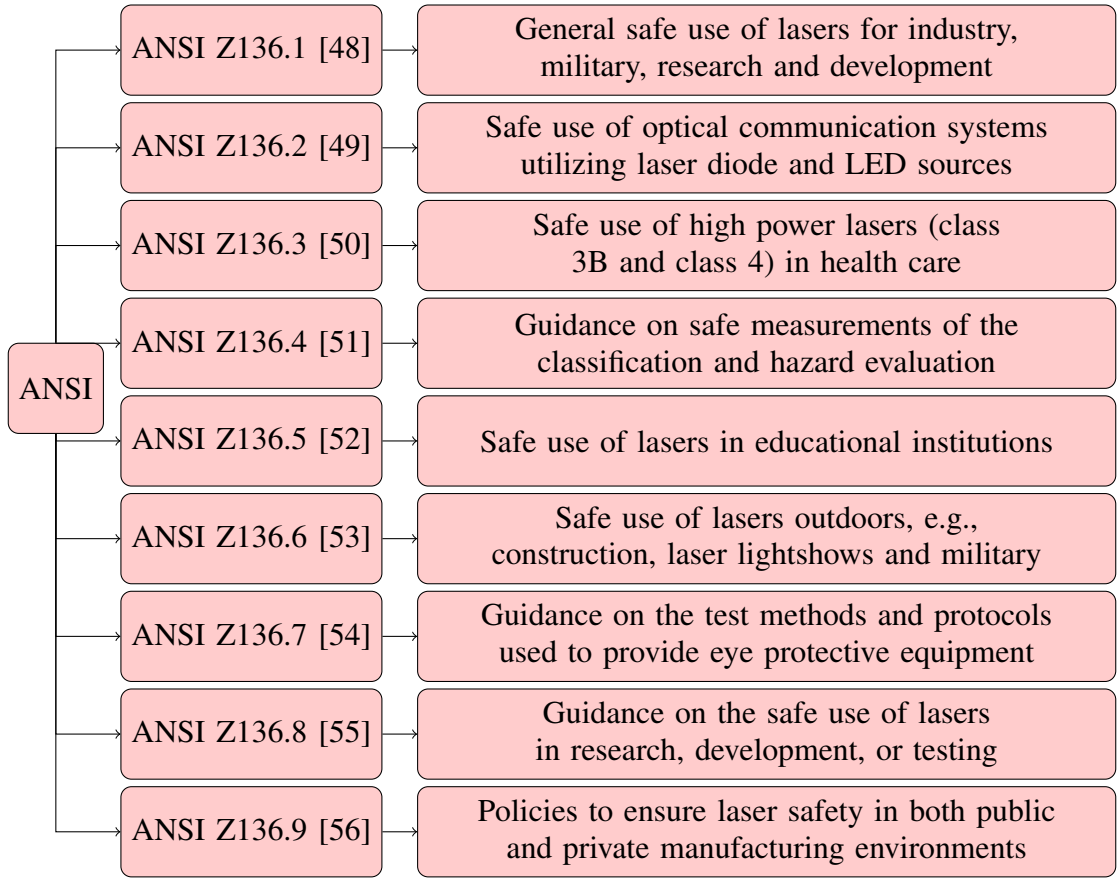

\subsection{Literature Review of Laser Safety}
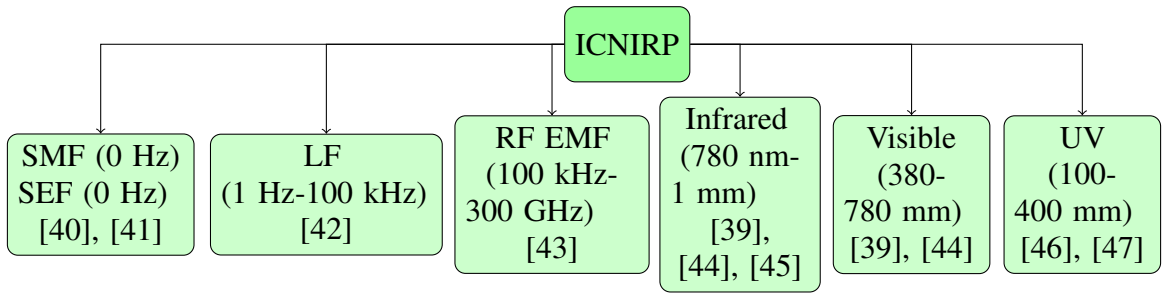
\begin{figure}[!th]
\begin{center}
\begin{tikzpicture}
\tikzstyle{startstop} = [rectangle, rounded corners, minimum height=1cm,text centered, draw=black, fill=white]

\node (ICNIRP) [startstop,draw=white, fill=white]{};
\node (B0) [startstop,xshift=-3 cm,fill=green!40] {ICNIRP};
\node (B1) [startstop, below of=ICNIRP, yshift=-1cm, xshift=-10cm, text width=2.2cm, fill=green!20] {SMF (0 Hz)  SEF (0 Hz) \newline \cite{ICNIRP_Magnetic,bernhardt2003exposure}};
\node (B2) [startstop, below of=ICNIRP, yshift=-1cm, xshift=-7 cm, text width=2.8cm, fill=green!20] {\hspace{1.1cm}{LF} \newline (1 Hz-100 kHz) \newline \cite{ICNIRP_LF}};
\node (B3) [startstop, below of=ICNIRP, yshift=-1cm, xshift=-4 cm, text width=2.3cm, fill=green!20] {\hspace{0.4cm}{RF EMF} \newline (100 kHz- 300 GHz) \newline \cite{ICNIRP_EMF}};
\node (B4) [startstop, below of=ICNIRP, yshift=-1cm, xshift=-1.5 cm, text width=1.85cm, fill=green!20] {\hspace{0.2cm}{Infrared} \newline (780 nm- 1 mm)\newline \cite{international2013visibleinfrared,international2013guidelines,international2013farinfrared} };
\node (B5) [startstop, below of=ICNIRP, yshift=-1cm, xshift=0.78 cm, text width=1.8cm, fill=green!20] {\hspace{0.3cm}{Visible} \newline (380- \\780 mm) \newline \cite{international2013visibleinfrared,international2013guidelines}};
\node (B6) [startstop, below of=ICNIRP, yshift=-1cm, xshift=3.05 cm, text width=1.8cm, fill=green!20] {\hspace{0.6cm}{UV} \newline (100-\\ 400 mm)\newline \cite{ICNIRP_UV,ICNIRP_UV_workers}};
%
\draw [->] (B0) -| (B1);
\draw [->] (B0) -| (B2);
\draw [->] (B0) -| (B3);
\draw [->] (B0) -| (B4);
\draw [->] (B0) -| (B5);
\draw [->] (B0) -| (B6);

\end{tikzpicture}
\end{center}
\caption{Classification of ICNIRP guidelines on limits of exposure.} 
\label{Fig-ICNIRP}
\end{figure}

With the increased use of lasers in OWC for high-speed links, the safety of lasers\footnote{We note that safety of optical fiber is out of the focus of this paper. The IEC has released a standard for safety of optical fiber communication systems (OFCS) (part 2 of IEC 60825) \cite{IEC60825-2}.} has emerged as a topic of interest. 
There are several standards on the safety of lasers.
IEC 60825 consists of 10 parts\footnote{EN 60825 is European requirements documents which follow the IEC versions and BS EN 60825 is the UK national requirements documents \cite{StdEye2014}. The USA has always had its own regulation on lasers which is known as 21 CFR 1040.10 \cite{21CFR1040.10:web}. This is a USA government regulation rather than a standard, which has not been updated for about 30 years and is consequently very outdated compared to the IEC / EN standard.}, under the general title safety of laser products. Fig.~\ref{Fig-IEC} illustrates the principle target of each part. The most relevant parts to OWC design with lasers are part 1 \cite{IEC_Std608251} and 12 \cite{IEC60825-7}, where general laser safety and the safety of free space optical communications are specified respectively.
In addition, the American national standards institute (ANSI) has published a 9 part ANSI Z-136 series of standards. This series, along with an outline of the primary focus of their work, are presented in Fig.~\ref{Fig-ANSI}. Parts 1 and 2 are the most relevant with regards to the design of safe optical wireless links, where the safe use of lasers and the safety of fixed terrestrial point-to-point free-space links are discussed, respectively. The international commission on non-ionizing radiation protection (ICNIRP) has specified the maximum level of exposure to coherent laser sources for wavelengths between $180$ nm to $1$ mm \cite{international2013guidelines}\footnote{This includes the revision of ICNIRP on maximum levels of exposure laser radiation for wavelengths between 400 nm and 1.4 $\mu$m, which was released in October 2000.}. Table~\ref{table:standards} summarizes the most relevant standards on eye safety for laser products. It is worth mentioning that ICNIRP have determined exposure limits for the whole electromagnetic spectrum including static magnetic fields (SMF) \cite{ICNIRP_Magnetic}, static electric fields (SEF) \cite{bernhardt2003exposure}, low frequencies (LF) from $1$ Hz to $100$ kHz \cite{ICNIRP_LF}, RF electromagnetic fields (EMF) from $100$ kHz to $300$ GHz \cite{ICNIRP_EMF}, Infrared from $780$ nm to $1$ mm \cite{international2013visibleinfrared,international2013guidelines,international2013farinfrared}, visible spectrum from $380$ nm to 
$780$ nm \cite{international2013visibleinfrared,international2013guidelines} and ultraviolet (UV) from $100$ nm to $400$ nm \cite{ICNIRP_UV,ICNIRP_UV_workers}. Fig.~\ref{Fig-ICNIRP} illustrates the ICNIRP classification on limits of exposure for the electromagnetic spectrum. 
\begin{table*}[t]
\caption{Related standards on eye safety for laser products.}
\label{table:standards}
\centering
\begin{tabular}{|l|l|l|}
\hline
\multicolumn{1}{|c|}{Year} & \multicolumn{1}{|c|}{Standard} & \multicolumn{1}{|c|}{Remark}  
\\
\hline
\hline
2014 & IEC 60825-1 \cite{IEC_Std608251} & General laser safety \\
\hline
2019 & IEC 60825-12 \cite{IEC60825-7} & Free space optical communications\\
\hline
2014 & ANSI Z-136.1 \cite{ANSI2014} &  Safe use of lasers \\
\hline
2012 & ANSI Z-136.2 \cite{ANSI-Z136.2} & \begin{tabular}[c]{@{}l@{}} Safe use of optical communications systems utilizing laser\\ diodes or LEDs, including end-to-end optical fiber based\\ links, fixed terrestrial point-to-point free-space links, or a\\ combination of both.\end{tabular} \\
\hline
2013 & ICNIRP \cite{international2013guidelines} & \begin{tabular}[c]{@{}l@{}}Maximum levels of exposure to laser radiation for wavelengths\\ between 180 nm and 1000 $\mu$m.\end{tabular}\\
\hline
\end{tabular}
\end{table*}

Laser safety has been discussed in plenty of research papers with different applications such as industry, medicine, military, education and so on since the late 1960s \cite{5391898,Ps1978ExperimentalDO,dmitriev1979nonlinear,Ps1985Birng,galbiati2001evaluation,  schulmeister2011review,10.1117/12.2029383,HEUSSNER201538,ramos2018evaluation,heussner2019eye,Jiao:19,Jiao:19-experiments,sliney2019revisiting,Jean:21}. The main contributions and achievements of these studies are outlined in Table~\ref{table:relevant_work}. There are many research papers on laser safety, however, we have included relatively more recent papers in Table~\ref{table:relevant_work}, in which important factors such as MPE, exposure duration and subtense angle have been evaluated.
There are also several handbooks on general techniques for the safe use of lasers in different environments such as medicine, industry, education, outdoors and so on. Some of the well-know books on laser safety are presented in Table~\ref{table:books}. They mainly focus on the safety analysis for ideal Gaussian beams, the radiation hazards, protective techniques and laser classifications.

\begin{center}
\renewcommand{\arraystretch}{0.7}
\begin{longtable}{|l|l|l|}
\caption{Selected papers on laser safety.}
\label{table:relevant_work}\\
\hline 
\multicolumn{1}{|c|}{Year} & \multicolumn{1}{c|}{Paper} & \multicolumn{1}{c|}{Main contribution/achievement} \\ 
\endfirsthead
\multicolumn{3}{c}%
{{ \tablename\ \thetable{} -- continued from previous page}} \\
\endhead
\hline \multicolumn{3}{|r|}{{Continued on next page}} \\ \hline
\endfoot
\endlastfoot
\hline
\hline
1968 & Makous and Gould \cite{5391898} & \begin{tabular}[c]{@{}l@{}} Development of a theoretical estimate of retinal irradiance\\ for threshold damage \end{tabular} \\
\hline
1978 & Avdeev \emph{et al.} \cite{Ps1978ExperimentalDO} & \begin{tabular}[c]{@{}l@{}}Determination of MPEs for the wavelength of 1.54 $\mu$m\\ and pulse durations of $40$ ns and $1$ ms to be $0.16$ J/cm$^2$\\ and $0.3$ J/cm$^2$, respectively.\end{tabular} \\
\hline
1979 & Dmitriev \emph{et al.} \cite{dmitriev1979nonlinear} & \begin{tabular}[c]{@{}l@{}} Possibility of nonlinear effect and damage on the retina\\ for ns pulsed lasers. \end{tabular} \\
\hline
1985 & Birngruber \emph{et al.} \cite{Ps1985Birng} & \begin{tabular}[c]{@{}l@{}}
A thermal damage model for retinal lesions in accordance\\ with experimental measurements in the regime of $1$ ms to\\ $300$ ms.\end{tabular}\\
\hline
2001 & Galbiati \emph{et al.} \cite{galbiati2001evaluation} &  \begin{tabular}[c]{@{}l@{}}Determination of the location and size of the apparent\\ source for laser beams.\end{tabular}\\
\hline
2011 & Schulmeister \emph{et al.} \cite{schulmeister2011review} & \begin{tabular}[c]{@{}l@{}}The dependency of the maximum angular subtense on\\ exposure durations. \end{tabular} \\
\hline
2013 & Gustafsson \cite{10.1117/12.2029383} & Atmospheric influence on the NOHD has been evaluated. \\
\hline
2015 & Heussner \emph{et al.} \cite{HEUSSNER201538} & \begin{tabular}[c]{@{}l@{}} A prediction model for ocular damage is demonstrated,\\ which is validated for long-term exposure.  \end{tabular} \\
\hline
2018 & Ramos \emph{et al.} \cite{ramos2018evaluation} & \begin{tabular}[c]{@{}l@{}} Experimental measurements of nano seconds laser pulses\\ to the mammalian eye are presented. \end{tabular} \\
\hline
2019 & Heussner \emph{et al.} \cite{heussner2019eye} & \begin{tabular}[c]{@{}l@{}}A software based on the photochemical, thermomechanical\\ and photomechanical damages is developed to evaluate the\\ eye safety of laser products.\end{tabular} \\
\hline
2019 & Jiao \emph{et al.} \cite{Jiao:19} & \begin{tabular}[c]{@{}l@{}}A series of experiments were conducted to determine the\\ retinal damage thresholds induced by a 420-750 nm\\ supercontinuum source and a CW 532 nm laser, where no\\ significant difference could be found between them.  \end{tabular} \\
\hline
2019 & Jiao \emph{et al.} \cite{Jiao:19-experiments} & \begin{tabular}[c]{@{}l@{}}The retinal damage thresholds are determined for two\\ groups of chinchilla grey rabbits with the ocular axial\\ lengths of 15.97 and 17.25 mm, respectively, using a\\ 1319 nm CW laser.   \end{tabular} \\
\hline
2019 & Sliney \cite{sliney2019revisiting} & \begin{tabular}[c]{@{}l@{}}Laser products that intentionally radiate towards the face \\necessitate higher reliability and protection to be assured\\ of always remaining below the Class 1, however, the\\ current limits are considered sound. \end{tabular} \\
\hline
2021 & Jean \emph{et al.} \cite{Jean:21} & \begin{tabular}[c]{@{}l@{}}Computer models are developed to predict injury thresholds\\ in the wavelength range from $1050$ nm to $10.6$ $\mu$m.\end{tabular} \\
\hline
\end{longtable}
\vspace{-1cm}
\end{center}

\begin{table*}[b]
\caption{Handbooks on laser safety.}
\label{table:books}
\centering
\begin{tabular}{|l|l|l|}
\hline
\multicolumn{1}{|c|}{Year} & \multicolumn{1}{|c|}{Book} & \multicolumn{1}{|c|}{Remark}  \\
\hline
\hline
1985 & Winburn \cite{winburn1985practical} &  \begin{tabular}[c]{@{}l@{}}This book includes characteristics of lasers; eye\\ components; skin damage thresholds; classification\\ of lasers by ANSI Z136.1; selecting laser-protective\\ eyewear; hazards associated with lasers.\end{tabular} \\
\hline
1997 & Henderson \cite{henderson1997guide} & \begin{tabular}[c]{@{}l@{}}A guide for the safe use of laser products, which embraces\\ hazards of laser radiation, laser classification, approaches\\ for hazard control and biological effect of laser radiation.\end{tabular} \\
\hline
2003 & Henderson \emph{et al.} \cite{henderson2003laser} & \begin{tabular}[c]{@{}l@{}}A handbook that covers laser radiation hazards,\\ laser classification and protective measures and safety.\end{tabular} \\
\hline
2013 & Sliney \emph{et al.} \cite{sliney2013safety} & \begin{tabular}[c]{@{}l@{}} Topics such as the effect of laser radiation on the eye and\\ skin, laser hazard classification and laser safety for outdoor\\ applications are covered. \end{tabular} \\
\hline
2014 & Barat \cite{barat2014laser} & \begin{tabular}[c]{@{}l@{}}A handbook that determines general protective approaches\\ for the safe use of lasers. \end{tabular} \\
\hline
\end{tabular}
\end{table*}

\subsection{Scope and structure of the Tutorial}
Most of the literature discussing laser eye safety has involved calculations using single-mode Gaussian beams, and the literature suggests that laser classification for non-single-mode-Gaussian beams should be evaluated experimentally. Furthermore, there is a lack of a solid analytical framework for eye safety where various beam propagation models, as well as optics such as lenses and diffusers, are considered. Furthermore, to the best of the authors' knowledge, there is no such a wide study with the main focus of utilizing laser sources as  transmitters for indoor OWC where long time exposure is a possibility. 

In this study, we provide in-depth assessments of beam propagation for ideal single-mode Gaussian and multimode beams. We also investigate the beam propagation for arrays and beam shaping optics including lenses and diffusers. Beam propagation investigations have been supported by demonstrating extensive results. 
We have also developed a generalized framework which recommends the maximum power operation of non-single-mode Gaussian beams, based upon known laser parameters, rather than purely experimental evaluation. This allows us to calculate the theoretical maximum power at which we can operate our lasers and feed into our link budgets. The calculated maximum permissible transmit power of the laser source ensures both eye and skin safety, which is derived for various cases including single mode Gaussian/non-Gaussian beams and multimode beams. The multimode results are based on experimental measurements reported in the literature. The eye safety of multimode lasers is investigated using different approaches based on analytical methods and the calculation of $M$-squared method. A comparison between these techniques is carried out in this paper. It is confirmed that the $M$-squared-based approach is a dependable method to determine the maximum transmit power, however, in some scenarios, it provides a significantly lower transmit power level compared to the precise beam decomposition method. For those scenarios, the power budget can be improved by evaluating the maximum transmit power based on the presented precise beam decomposition method.
We have also presented unique algorithms to obtain the maximum transmit power of a laser source with optical components such as lenses and diffusers. We further note that algorithms proposed in this study consider not only the parameters of a laser including the wavelength and beam waist but also the exposure duration and the most hazardous position. 
We have also presented perceptive results of eye safety by evaluating various parameters and delivering in-depth discussions. To the best of the authors' knowledge, there are no papers that provide such wide mathematical analyses for eye safety supported by a broad range of results on laser sources of interest in modern OWC systems. 

The remainder of this paper is organized as follows.
Laser safety terminology is described in Section~\ref{LaserSafetyTerm}. Laser beam propagation models, including single transverse mode, higher-mode, using an optical element such as a lens or a diffuser and for an array of VCSELs are detailed in Section~\ref{Sec:LaserBeamPropagation}. Eye safety analyses along with insightful results for each of the described beam propagation model are presented in Section~\ref{Sec: eye safety analysis}. This section has been wrapped up with skin safety. Finally, conclusions are drawn in Section~\ref{Sec: Conclusion}. 
The acronyms used in the paper are listed in Table~\ref{table:acronyms}.

\begin{table}[]
\caption{List of acronyms}
\label{table:acronyms}
\centering
\begin{tabular}{|l|l|}
\hline
\multicolumn{1}{|c|}{\textbf{Acronym}} & \multicolumn{1}{|c|}{\textbf{Description}}                      \\ \hline \hline
AEL              & Accessible emission limit                 \\ \hline
ANSI             & American national standards institute     \\ \hline
CW               & Continuous wave                           \\ \hline
EL               & Exposure limit                            \\ \hline 
EMF              & Electromagnetic field                     \\ \hline
FSO              & Free-space optical                        \\ \hline
FWHM             & Full width half maximum power             \\ \hline
ICNIRP           & International commission on non-ionizing radiation protection \\ \hline
IEC              & International electrotechnical commission \\ \hline
IoT              & Internet of things                        \\ \hline
ISO              & International organization for standardization \\ \hline
LASER            & Light amplification by stimulated emission of radiation \\ \hline
LF               & Low frequency                             \\ \hline
LED              & Light emitting diodes                     \\ \hline
LTE              & Long-term evolution                       \\ \hline
MHP              & Most hazardous position                   \\ \hline
MPE              & Maximum permissible exposure              \\ \hline
MRP              & Most restrictive position                 \\ \hline
NOHD             & Nominal ocular hazard distance            \\ \hline
OFCS              & Optical fibre communication system        \\ \hline
OWC              & optical wireless communication            \\ \hline
PMN              & Principal mode number                     \\ \hline
RF               & Radio frequency                           \\ \hline
SEF              & Static electric field                     \\ \hline
SMF              & Static magnetic field                     \\ \hline
VCSEL            & Vertical-cavity surface-emitting laser    \\ \hline
V2V              & Vehicle-to-vehicle                        \\ \hline
UV               & Ultraviolet                               \\ \hline
\end{tabular}
\end{table}

\section{Laser Safety Terminology}
\label{LaserSafetyTerm}
There are various standards specifying safe levels of laser radiation \cite{StdEye2014, IEC_Std608251, international2013guidelines, ANSI2014}. We will briefly review the measures and conditions specified in these standards for the safe use of lasers. In this section, we discuss various laser radiation hazards and in order to clarify our later discussions, we define the nomenclatures that will be used throughout this article.


\subsection{Spectral Regions and Laser Radiation Hazards}
\label{Termin-wavelength}


Laser radiation can cause biological damages through temperature effects due to the absorbed energy and also through some photochemical reactions. Damage is mostly due to temperature effects with the critical organs  considered being the eyes and skin. These biological effects vary depending upon the wavelength of the laser radiation, the exposure duration and the tissue being exposed. In this article, we denote the wavelength by $\lambda$ and mainly focus on visible and infrared wavelength regimes\footnote{It is worth mentioning that the algorithms presented in this paper can be readily extended to other wavelengths as well.}, since the commercially available laser sources operate in these spectral regions. In the following, we briefly discuss the primary hazards in different spectral regions.

\subsubsection{Effects of visible and near infrared radiation}
The primary effect on the eye in this spectral region is damage to the retina. The ability of the cornea and the eye lens to focus these wavelengths directly onto a small spot on the retina makes these wavelengths significantly more hazardous than the longer infrared wavelengths, despite the equivalent received optical irradiances on the surface of the eye. 
Injury to the skin in the visible and near infrared wavelength range ($400-1400$~nm) results from temperature rises above $45^\circ$C \cite{international2013guidelines}.



\subsubsection{Effects of mid and far infrared radiation}
For wavelengths greater than $1400$~nm, the laser radiation primarily causes damage to the cornea and the skin, as the radiation at these wavelengths is absorbed by water, and the vitreous humor of the eye protects the retina from damage \cite{international2013guidelines}.

\subsection{Exposure Duration}
The exposure duration is defined as the time span of a pulse, a pulse train or a continuous laser radiation on the human body\footnote{For a single pulse, this is the duration between the half-peak power point of the leading edge and the corresponding point on the trailing edge. For a train of pulses (or subsections of a train of pulses), this is the duration between the first half-peak power point of the leading pulse and the last half-peak power point of the trailing pulse.} \cite{StdEye2014}. Laser safety standards specify exposure durations between 100 fs and 30 ks (about 8 hours) \cite{StdEye2014, IEC_Std608251, international2013guidelines, ANSI2014}. The applicable exposure duration itself is dependent on the wavelength and the nature of the exposure, i.e. intended or accidental exposure, intra-beam exposure, or exposure to diffused radiation. As an example, for accidental intra-beam viewing of a continuous visible light laser radiation by a person with normal health condition, the exposure time is limited to $0.25$~s which is equal to the aversion response time \cite{schulmeister2011review}.

It should also be noted that the longer the exposure duration is, the more damage to the eye and skin it can cause. In this paper, the exposure duration is denoted by $t_{\rm{ex}}$ and we derive the maximum transmit power that ensures eye safety regulations for the highest exposure duration.



\begin{figure}[t!]
 \centering
\includegraphics[width=0.9\textwidth]{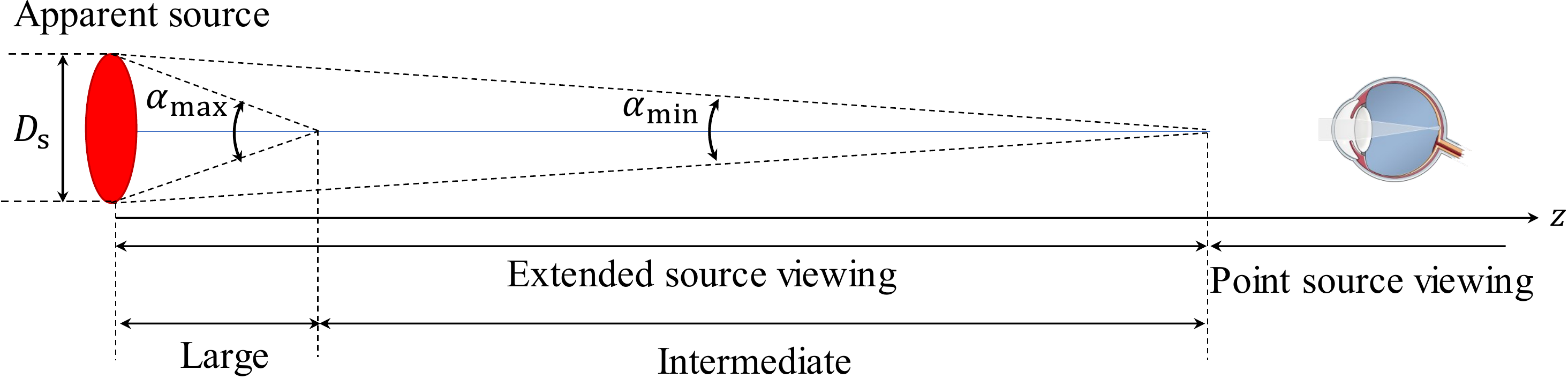} 
\caption{Geometric representation of the angular subtense, $\alpha$, of the apparent source.}
\label{Fig-SubtenseAngle}
\end{figure}

\subsection{Angular Subtense of the Apparent Source}
The apparent source is defined as the real or virtual object that forms the smallest retinal image by eye accommodation. Human eye accommodation refers to the act of physiologically adjusting the focal length of the eye lens to keep an object in focus on the retina as its distance from the eye varies.

The angular subtense of the apparent source is the angle subtended by the apparent source as viewed from a point in space\footnote{The angular subtense of the source should not be confused with the divergence angle of the beam.}. It relates to the ability of the eye to focus the laser’s power onto a small spot on the retina. (see Fig.~\ref{Fig-SubtenseAngle}). The angular subtense, denoted by $\alpha$, is calculated as:
\begin{equation}
\label{Eq:SubtenseAngle}
    \alpha = 2\tan^{-1}{\left(\dfrac{D_\mathrm{s}}{2z}\right)}
\end{equation}
where $D_\mathrm{s}$ is the source size (assuming that the source is circular) and $z$ is the distance of the apparent source from the considered position in space. For non-circular sources, the effective angular subtense is the arithmetic mean of two angular subtense values calculated for the maximum and minimum dimensions of the apparent source \cite{international2013guidelines}.

If $\alpha<\alpha_{\rm{min}}=1.5$~mrad, the source is considered as a \textit{point source}, otherwise, it is regarded as an \textit{extended source} \cite{StdEye2014,henderson2003laser}. According to the IEC $60825-1$ standard \cite{IEC_Std608251}, the extended sources are further classified into two types depending on their angular subtense; intermediate sources and large sources. For intermediate sources the condition $\alpha_{\rm{min}}<\alpha<\alpha_{\rm{max}}$ holds, while for large sources $\alpha\geq\alpha_{\rm{max}}$, where $\alpha_{\rm{max}}=100$~mrad. In practice, by using optical elements, a point source can be turned into an extended source.







\subsection{Maximum Permissible Exposure}
The maximum permissible exposure (MPE), also known as the exposure limit (EL), is the maximum irradiance level to which the eye or skin can be exposed without consequential injury immediately or after a long time. The MPE is reported in units of J$/$m$^2$ or W$/$m$^2$ and depends on the laser wavelength, the exposure time and the type of tissue at risk\footnote{In the context of eye safety, MPE also depends on the image size on the retina where the image size is specified in terms of the angular subtense of the apparent source.}. The MPE values are set by the ICNIRP \cite{international2013guidelines} and in collaboration with the IEC \cite{IEC_Std608251} and the ANSI \cite{ANSI2014} standards. Since the injury levels for eye and skin tissues are different, these standards have specified different sets of MPEs for ocular and skin exposure hazards.

The MPE not only depends on the wavelength and the exposure duration, but also on whether a source is viewed as a point source or an extended one, determined by the subtense angle. 
Table~\ref{table_eye_safety} summarizes the eye-safe MPE expressions for point and extended sources in different wavelength regimes and for their corresponding applicable exposure duration. To obtain these MPEs, a limiting aperture diameter of $7$ mm has been considered. It is defined in terms of the diameter of a circular area over which the average irradiance or radiant is considered \cite{IEC_Std608251}.
Note that for the range of wavelengths considered in this table, eye safety is more restricting than skin safety, hence, the provided MPEs suffice for the laser radiation safety analysis.



\setlength\doublerulesep{0.2cm}

\begin{table}[t!]
\renewcommand{\arraystretch}{1.5}
\centering
\caption[blah]{MPE calculation for eye safety analysis\footnotemark \cite{StdEye2014}.}
\label{table_eye_safety} 
\begin{tabular}{|c|c|c|c|c|}
\hline
\multicolumn{5}{|c|}{Spectral regions and exposure durations} \\
\hline    
$\lambda$ [nm] & \multicolumn{2}{c|}{$700\ \leq\lambda< 1050$ } & $1050\ \leq\lambda< 1150$  & $1150\ \leq\lambda< 1400$  \\
\hline
\makecell{$t_{\rm{ex}}$ [s]} & $10^{-3}$ to 10 & 10 to $3\times10^4$ & $10^{-3}$ to 10 & 10 to $3\times10^4$ \\
\hline
\hline
\multicolumn{5}{|c|}{Expressions for MPE [W$/$m$^2$]} \\
\hline
\makecell{Point source} & $18t_{\rm{ex}}^{0.75}{\rm{C}}_4/t_{\rm{ex}}$ &10${\rm{C}}_4{\rm{C}}_7$ & $90t_{\rm{ex}}^{0.75}/t_{\rm{ex}}$ & $10C_4C_7$ \\
\hline
\makecell{Extended source} & \multicolumn{2}{c|}{\xrowht{40pt} $\begin{aligned}
& 18t_{\rm{ex}}^{0.75}{\rm{C}}_4{\rm{C}}_6/t_{\rm{ex}}, \ {\rm if}\ t_{\rm{ex}}\leq T_2 \\
& 18{\rm{C}}_4{\rm{C}}_6T_2^{-0.25}, \quad\; {\rm if}\ t_{\rm{ex}}> T_2\end{aligned}$} & 
\multicolumn{2}{c|}{$\begin{aligned}
& 90t_{\rm{ex}}^{0.75}{\rm{C}}_6{\rm{C}}_7/t_{\rm{ex}}, \ {\rm if}\ t_{\rm{ex}}\leq T_2 \\
& 90{\rm{C}}_6{\rm{C}}_7T_2^{-0.25}, \quad\; {\rm if}\ t_{\rm{ex}}> T_2\end{aligned}$}   \\
\hline
\hline
\multicolumn{5}{|c|}{Required coefficients and parameters for calculating MPE} \\
\hline    
${\rm{C}}_4$ & \multicolumn{2}{c|}{$10^{0.002(\lambda-700)}$}  & \multicolumn{2}{c|}{5}\\
\hline
${\rm{C}}_6$ &  \multicolumn{4}{c|}{\xrowht{70pt} ${\rm C}_6=\begin{cases} 1, & \alpha\leq\alpha_{\rm min} \\[5pt]
\dfrac{\alpha}{\alpha_{\rm min}}, & \alpha_{\rm min}<\alpha\leq\alpha_{\rm max} \\[10pt]
\dfrac{\alpha_{\rm max}}{\alpha_{\rm min}}, & \alpha>\alpha_{\rm max}
\end{cases}$}  \\
\hline
${\rm{C}}_7$ &  \multicolumn{4}{c|}{\xrowht{50pt} $C_7=\begin{cases} 1, & \lambda< 1150\\
10^{0.018(\lambda-1150)}, & 1150\ \leq\lambda< 1200 \\
8, & 1200\ \leq\lambda< 1400
\end{cases}$} \\
\hline
$T_2$ [s] &  \multicolumn{4}{c|}{\xrowht{50pt} $T_2\!=\!\begin{cases} 10, & \alpha\leq\alpha_{\rm min}\\
10\times10^{\frac{(\alpha-\alpha_{\rm min})}{98.5}}, & \alpha_{\rm min}<\alpha\leq\alpha_{\rm max} \\
100, & \alpha>\alpha_{\rm max}
\end{cases}$} \\
\hline    
\end{tabular}
\end{table}

\footnotetext{{For $10^{-3}\leq t_{\rm{ex}}<0.25$, we have $\alpha_{\rm max}=200t_{\rm{ex}}^{0.5}$ mrad while for $t_{\rm{ex}}>0.25$, we have $\alpha_{\rm max}=100$ mrad.
}}

\subsection{Most Hazardous Position}
For a given laser beam, the most hazardous position, MHP, denoted by $z_{\rm haz}$, is defined as the location at which the ratio of the exposure level to the MPE value is at its maximum. It can be expressed mathematically as \cite{schulmeister2015classification}:
\begin{equation}
    z_{\rm haz}=\ \argmaxA_{z} \ \frac{I(z)}{\rm MPE},
    \label{z_haz}
\end{equation}
where $I(z)$ is the exposure level (i.e., irradiance in W$/$m$^2$). We note that MPE depends on the subtense angle which is a function of $z$ (see \eqref{Eq:SubtenseAngle}). 
It is clear from (\ref{z_haz}) that the MHP for different laser beam profiles can be different. Therefore, it is necessary to carefully specify MHP based on the characteristics of the laser source, e.g., whether the beam is single-mode or multimode, and whether the source is a single element or an array. 

The MHP also depends upon the type of laser hazard; i.e, the tissue being exposed (retinal, corneal or skin). For corneal and skin hazards, the MHP is where the laser beam has the smallest area and hence, the highest irradiance. 
For retinal hazards, the MHP is not necessarily where the irradiance is the highest, but where the cornea and the eye lens focus the highest optical power onto the smallest retinal spot. It is this irradiance that can be most damaging to the retina. Therefore, the MHP in this case depends on the angular subtense of the apparent source. This means that various points in space with different distances from the eye should be evaluated in order to specify the MHP. 

According to \cite{StdEye2014, IEC_Std608251, international2013guidelines, ANSI2014}, the closest evaluation position which needs to be considered in a retinal hazard MPE analysis is $10$ cm from the apparent source. 
This is based on the assumption of a near point of $10$ cm, i.e. $10$ cm is the closest position at which a healthy human eye is able to image the apparent source onto the retina. Closer distances result in higher optical powers entering the eye, but forming blurred, less hazardous images.
%

\subsection{Laser Classification}
Lasers are often categorized, based on their operating wavelength and transmit power, into four main classes, with a few subclasses \cite{henderson2003laser}.    
Old classification includes, Class 1, Class 2, Class 3A, Class 3B and Class 4. The revised version was demonstrated in 2001 under the IEC and EN classification \cite{IEC60825-1:2001}, which specifies the following classes:
\begin{itemize}
\item Class 1: This class is eye-safe under all operating conditions, even for long term intentional viewing.
\item Class 1M: This class is safe for an unaided eye, but might be hazardous when viewed with optical instruments such as binoculars.
\item Class 2: These are visible lasers. The class is safe for unintentional exposure durations less than $0.25$ s.
\item Class 2M: These are visible lasers. This class is safe for unintentional exposure durations less than $0.25$ s, but may be hazardous when viewed with collecting optics such as binoculars.
\item Class 3R: Unintentional or accidental exposure to direct, reflected beam or radiation in this class is considered as low risk, but potentially hazardous.
\item Class 3B\footnote{Terms `M', `R' are derived from ``magnifying optical viewing instruments", ``reduced or relaxed requirements", respectively; and `B' is based on the old classification.}: Radiation in this class is very likely to be hazardous. For a continuous wave laser the maximum output power into the eye must not exceed $500$~mW. The radiation can be hazardous to both the eye and skin. However, viewing of the diffuse reflection is safe.
\item Class 4: This is the highest class of laser radiation and it is severely dangerous. Radiation in this class is very hazardous, and viewing of the diffuse reflections may be dangerous as well.
\end{itemize}

These classes are summarized in Table~\ref{TableLaserClassifications}. It is noted that higher wavelengths can accommodate higher power and still be a class 1 product.

\setlength\doublerulesep{0.1cm}
\begin{table}[]
\renewcommand{\arraystretch}{1.5}
\centering
\caption[blah]{Laser classifications.}
\label{TableLaserClassifications}
\begin{tabular}{|l||l|l|l|}
\hline
\multicolumn{1}{|c|}{Class}    & \multicolumn{1}{|c|}{Risk or injury}  & \multicolumn{1}{|c|}{Exposure duration} & \multicolumn{1}{|c|}{Viewing condition} \\ \hline \hline
\rowcolor[HTML]{32CB00} 
Class 1  & eye-safe    & all durations     & also with optics \\ \hline
\rowcolor[HTML]{34FF34} 
Class 1M & eye-safe    & all duration      & naked eye only  \\ \hline
\rowcolor[HTML]{67FD9A} 
Class 2  & eye-safe    & less than $0.25$ s  & also with optics \\ \hline
\rowcolor[HTML]{9AFF99} 
Class 2M & eye-safe    & less than $0.25$ s  & naked eye only \\ \hline
\rowcolor[HTML]{FFCB2F} 
Class 3R & low risk        & less than $0.25$ s  & also with optics  \\ \hline
\rowcolor[HTML]{FD6864} 
Class 3B & \begin{tabular}[c]{@{}l@{}} diffuse reflections are safe\\ low risk for skin injury \end{tabular}   & all duration  & also with optics    \\ \hline
\rowcolor[HTML]{FE0000} 
Class 4  & \begin{tabular}[c]{@{}l@{}} severely eye-hazardous\\ diffuse reflections may be hazardous\end{tabular} & all duration    & also with optics \\ \hline
\end{tabular}
\end{table}

\section{Laser Beam Propagation}
\label{Sec:LaserBeamPropagation}
In this section, we review single mode and multimode laser beams, and beam transformation by optical elements such as lenses and diffusers. The understanding of various laser beam profiles and their characteristics is essential for performing laser safety analyses in the next section.

\subsection{Single Transverse Mode Beam}
\label{idealGaussian}
\begin{figure}[t!]
 \centering
\includegraphics[width=0.6\textwidth]{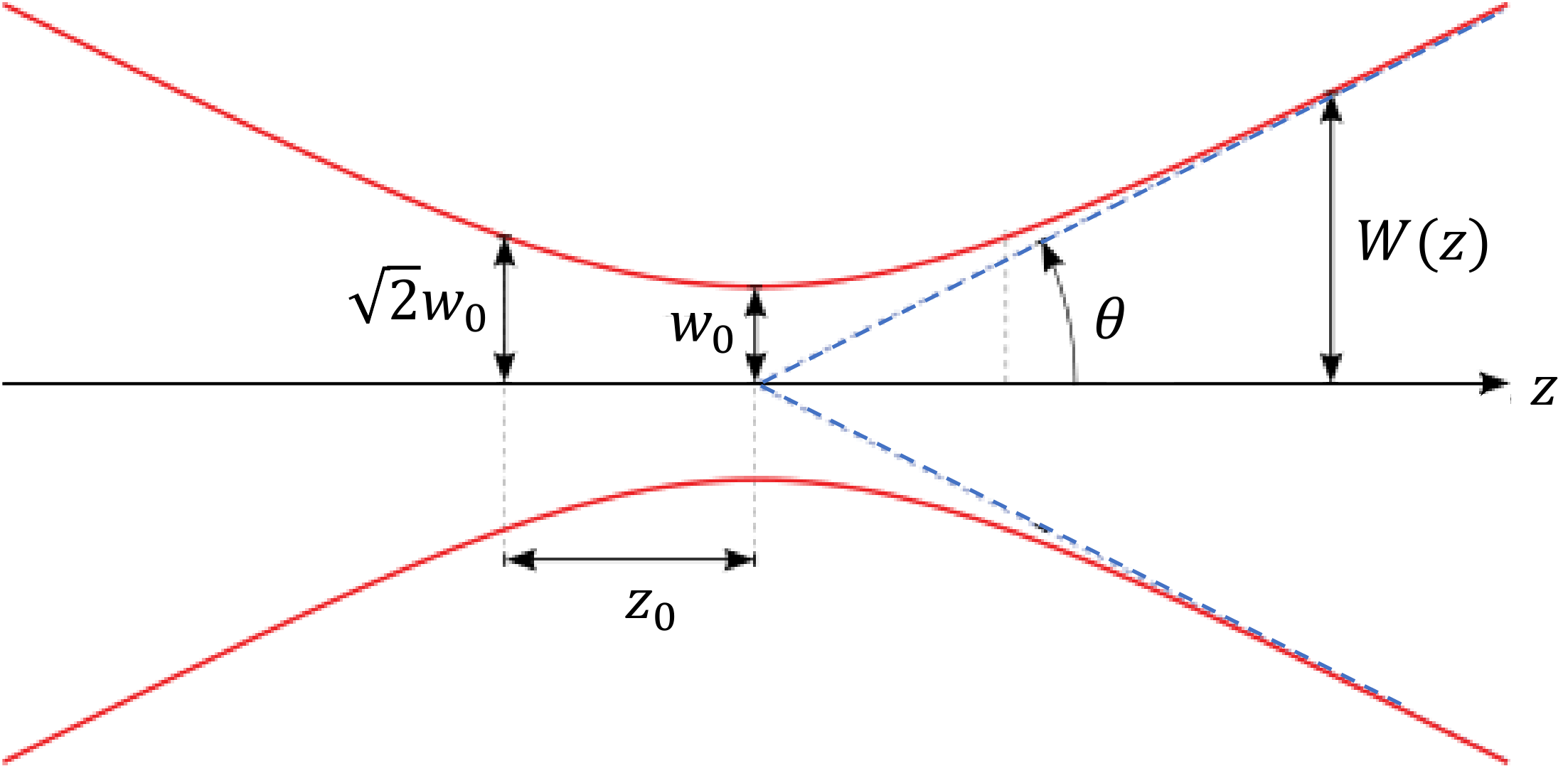} \caption{Gaussian beam geometry. }
\label{Fig-GaussainBeam}
\end{figure}

An ideal single-mode Gaussian beam has an intensity distribution given by the following  equation \cite{saleh2019fundamentals}:
\begin{equation}
	I(r,z) = \frac{2P_\mathrm{t}}{\pi W^2(z)}{\exp{\left( -\dfrac{2r^2}{W^2(z)} \right )}},
	\label{Eq:2_1}
\end{equation}
where $P_\mathrm{t}$ is the optical power carried by the beam, and $r$ and $z$ are the radial and axial positions, respectively.
For ideal Gaussian beams at a distance of $z$, the spot radius\footnote{Spot radius or spot size is the radius at which the intensity values fall to $1/e^2$ of their axial values, in the transverse plane at position $z$ along the beam axis.} and the wavefront radius of the curvature are respectively expressed as below \cite{saleh2019fundamentals}:
\begin{subequations}
\label{W_z}
\begin{align}   
    &W(z)=w_0\sqrt{1+\left(\frac{z}{z_0}\right)^2},\\ &R(z)=z\left(1+\left(\frac{z_0}{z}\right)^2\right),
\end{align}
\end{subequations}
where $z_0$ is the Rayleigh range and is given by:
\begin{equation}
    z_0=\frac{\pi w_0^2}{\lambda}
\end{equation}
and $\lambda$ is the wavelength of the laser source. 
The complex $q$-parameter of the Gaussian beam is defined as $q(z)=z+jz_0$ where $j=\sqrt{-1}$. The reciprocal of $q(z)$ contains the radius of the curvature and the spot radius in its real and imaginary parts, respectively \cite{saleh2019fundamentals}:
\begin{equation}
    \frac{1}{q(z)}=\frac{1}{R(z)}-j\frac{\lambda}{\pi W^2(z)}.
    \label{qparameter}
\end{equation}
The beam divergence angle\footnote{In this paper, the term ``beam divergence angle" specifies the half-angle divergence as shown in Fig.~\ref{Fig-GaussainBeam}. In some studies, beam divergence angle stands for full-angle divergence.}, as shown in Fig.~\ref{Fig-GaussainBeam}, is defined as:
\begin{equation}
    \theta=\lim_{z \to \infty}\tan^{-1}\left(\frac{W(z)}{z}\right).
    \label{eq_DivergenceAngle}
\end{equation}
In the paraxial case (i.e, very small $\theta$), the beam divergence (in radians) is approximated as:
\begin{equation}
\label{thetaFWHMEq}
    \theta \approx \frac{\lambda}{\pi w_0}.
\end{equation}

\begin{table}[t]
\renewcommand{\arraystretch}{1.5}
\centering
\caption[blah]{Commonly used measures for the beam spot radius.}
\label{TableBeamRadiusMetric}
\begin{tabular}{|l||c|c|c|}
\hline
Measure    & Power ratio  & Spot radius                 & Divergence angle            \\ \hline
$1/e^2$    & $86\%$       &  $W(z)$                   &$\theta$                     \\ \hline
$1/e$      & $63\%$       &  $\frac{\sqrt{2}}{2}W(z)$ & $\frac{\sqrt{2}}{2}\theta$  \\ \hline
$1/2$       & $50\%$       &  $\frac{\sqrt{2\ln{2}}}{2}W(z)$     & $\frac{\sqrt{2\ln{2}}}{2}\theta$      \\ \hline
\end{tabular}
\end{table}

The received power at a distance of $z$ through an aperture of radius $r_0$, can be obtained as:
\begin{equation}
\label{intensityEq}
    P_{\rm{r}}=\int_{0}^{r_0}I(r,z)2\pi {r{\rm{d}}}r=P_\mathrm{t}\left(1-\exp{\left( -\dfrac{2r_0^2}{W^2(z)} \right )}\right).
\end{equation}
Therefore, a circle of $r_0=W(z)$ contains $1-e^{-2}$ or $86\%$ of the Gaussian beam's total power. We refer to this as the ``$1/e^2$'' or ``$86\%$ measure'' and use it throughout this paper. Apart from this measure, in other studies $1/e$ and $1/2$ are also used. These commonly used measures are compared in Table~\ref{TableBeamRadiusMetric} in terms of the spot radius and the beam divergence angle. The third row in Table~\ref{TableBeamRadiusMetric} represents the case where the beam spot radius is measured based on the full width at half maximum (FWHM) criterion. In fact, FWHM specifies the beam diameter for which the optical power contained within the beam spot is $1/2$ of the total power.

\subsection{Multimode Laser Beams}
\label{HigherModeBeam}
The Gaussian beam is not the only solution of the paraxial Helmholtz equation \cite{saleh2019fundamentals}. Of practical interest are solutions which have non-Gaussian intensity distributions. In general, non-Gaussian irradiance profiles can be expressed as the incoherent superposition of transverse modes. The total optical intensity of the beam can be represented as:
\begin{equation}
\label{EqSumNonGaussian}
    I =\sum_{l=0}^{L_{\rm max}}\sum_{m=0}^{M_{\rm max}} C_{l,m}|U_{l,m}|^2,
\end{equation}
where $U_{l,m}$ is the complex amplitude of the mode $(l,m)$, hence, $|U_{l,m}|^2$ is the optical intensity of the mode, and $C_{l,m}$ is the mode power coefficient.


Two well known orthogonal sets of transverse modes are Hermite Gaussian modes, which are used mainly in Cartesian coordinate systems, and Laguerre Gaussian modes, which are more suitable for circularly or cylinderically symmetric beams. In the following, these two polynomials are discussed in more detail.

\subsubsection{Hermite-Gaussian Beams} \label{HG}
A Hermite-Gaussian beam of order $(l,m)$ is often represented as ${\rm HG}_{l,m}$, where $l$ and $m$ refer to the mode numbers in transverse $x$ and $y$ directions in the Cartesian coordinate system, respectively. The complex field amplitude of ${\rm HG}_{l,m}$ propagated at distance  $z$ is represented by \cite{saleh2019fundamentals}:
\begin{equation}
\label{HermiteEq}
\begin{aligned}
 U_{l,m}(x,y,z)&=A_{l,m}\left[\frac{w_0}{W(z)}\right] \mathbb{G}_l\left(\frac{\sqrt{2}x}{W(z)}\right) \mathbb{G}_m\left(\frac{\sqrt{2}y}{W(z)}\right)\\ 
 &\times\exp{\left(-jkz-jk\frac{x^2+y^2}{2R(z)}+j(l+m+1)\zeta(z)\right)},
 \end{aligned}
\end{equation}
where $\zeta(z)=\tan^{-1}(z/z_0)$ and $k=2\pi/\lambda$ is the wave number (in radian per meter) for the free-space wavelength of $\lambda$. The mode normalisation coefficient $A_{l,m}$ for Hermite-Gaussian beams is defined as below:
\begin{equation}
\label{HermiteClmEq}
\begin{aligned}
    A_{l,m} =\sqrt{\dfrac{2^{-l-m+1}}{\pi l!m!}}
\end{aligned}
\end{equation}
%
%
In \eqref{HermiteEq}, $\mathbb{G}_l(u)$ is known as the Hermite-Gaussian function of order $l$ and is given as:
\begin{equation}
    \mathbb{G}_l(u)=\mathbb{H}_l(u)\exp{\left(\frac{-u^2}{2}\right)}, \ \ \ l=0,1,2,...
\end{equation}
%
%
The Hermite polynomials, $\mathbb{H}_l(u)$, are defined as:
\begin{equation}
    \mathbb{H}_l(u) = l! \sum_{i=0}^{\left\lfloor \tfrac{l}{2} \right\rfloor} \frac{(-1)^i}{i!(l - 2m)!} (2u)^{l - 2i}. 
\end{equation}
The irradiance of the mode $(l,m)$ at $x,y,z$ is represented by \cite{saleh2019fundamentals}:
\begin{equation}
    \label{HermiteIrradianceEq}
    \begin{aligned}
        |U_{l,m}\left(x,y,z\right)|^2=A_{l,m}^{2} \left[\frac{w_0}{W(z)}\right]^2 \!\!\mathbb{H}_l^2\left(\frac{x\sqrt2}{W_x\left(z\right)}\right)\!\mathbb{H}_m^2\left(\frac{y\sqrt2}{W_y\left(z\right)}\right)\exp{\left(-\frac{2x^2}{W_x^2\left(z\right)}-\frac{2y^2}{W_y^2\left(z\right)}\right)}.
    \end{aligned}
\end{equation}

\begin{figure}[t!]
 \centering
 \subfloat[$\rm {PMN} = 1$ \label{sub1:FigHG}]{%
		\resizebox{.32\linewidth}{!}{\includegraphics{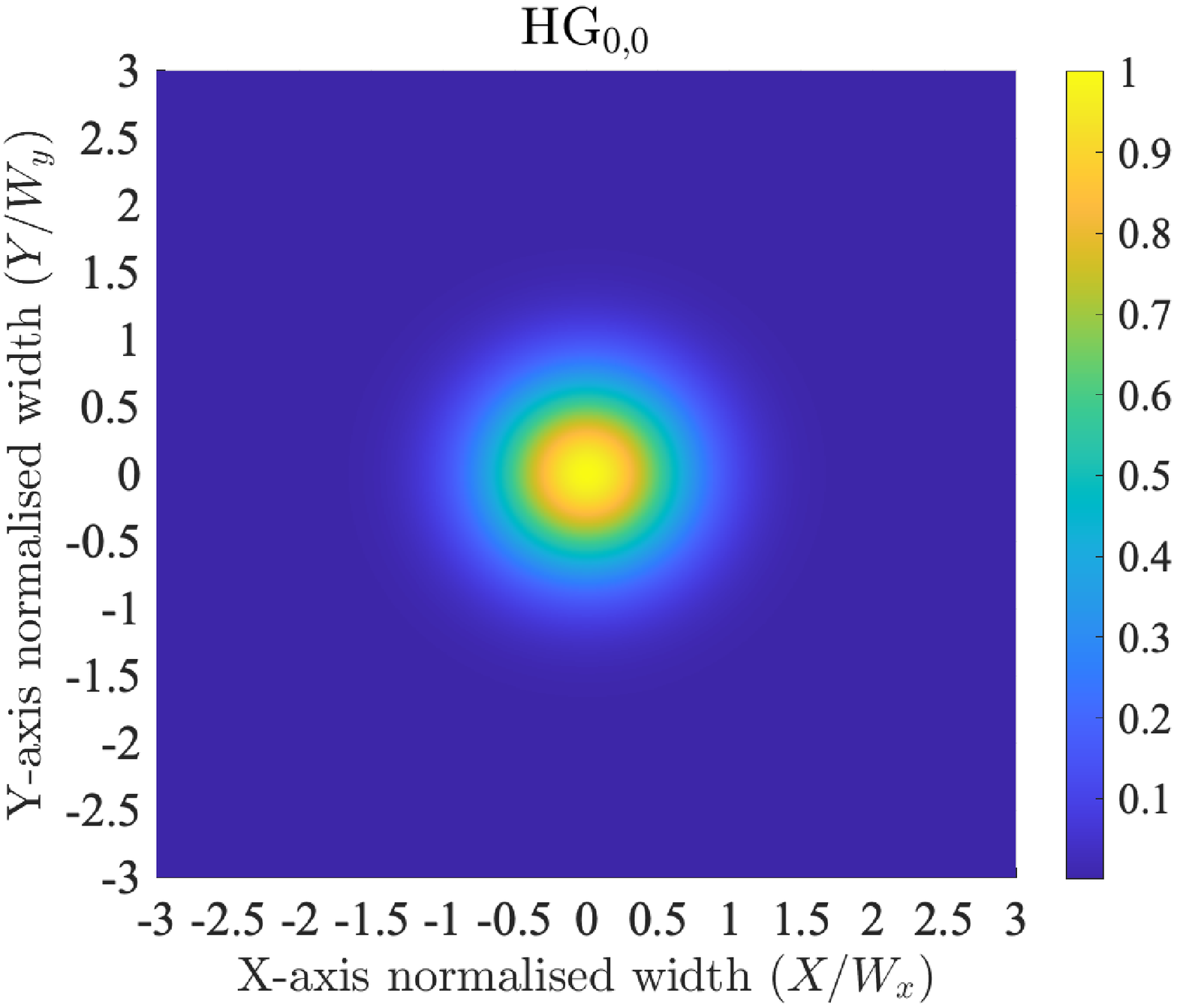}}
	}  
	\subfloat[$\rm {PMN} = 2$ \label{sub2:FigHG}]{%
		\resizebox{.32\linewidth}{!}{\includegraphics{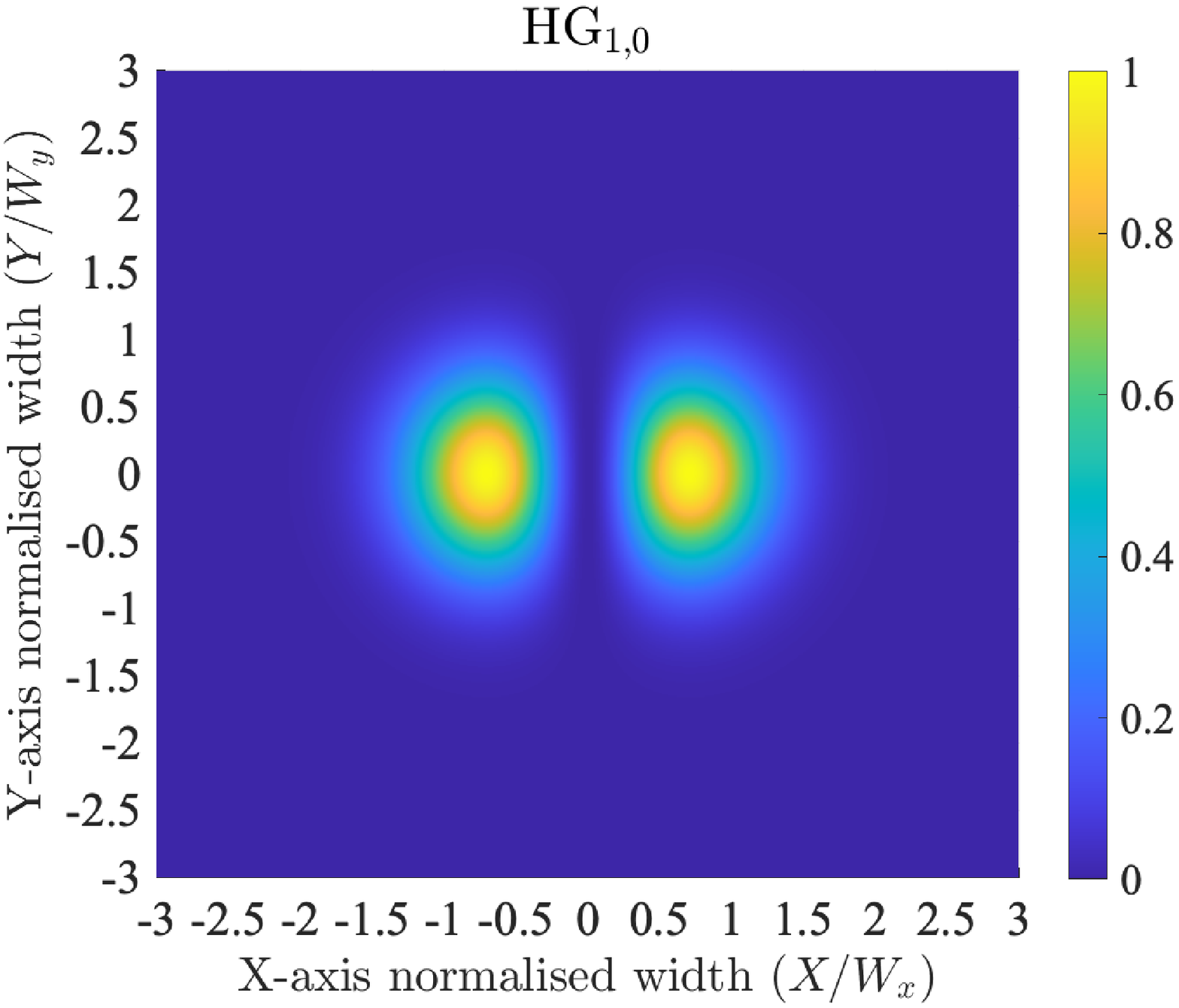}}
	} 
	\subfloat[$\rm {PMN} = 2$ \label{sub3:FigHG}]{%
		\resizebox{.32\linewidth}{!}{\includegraphics{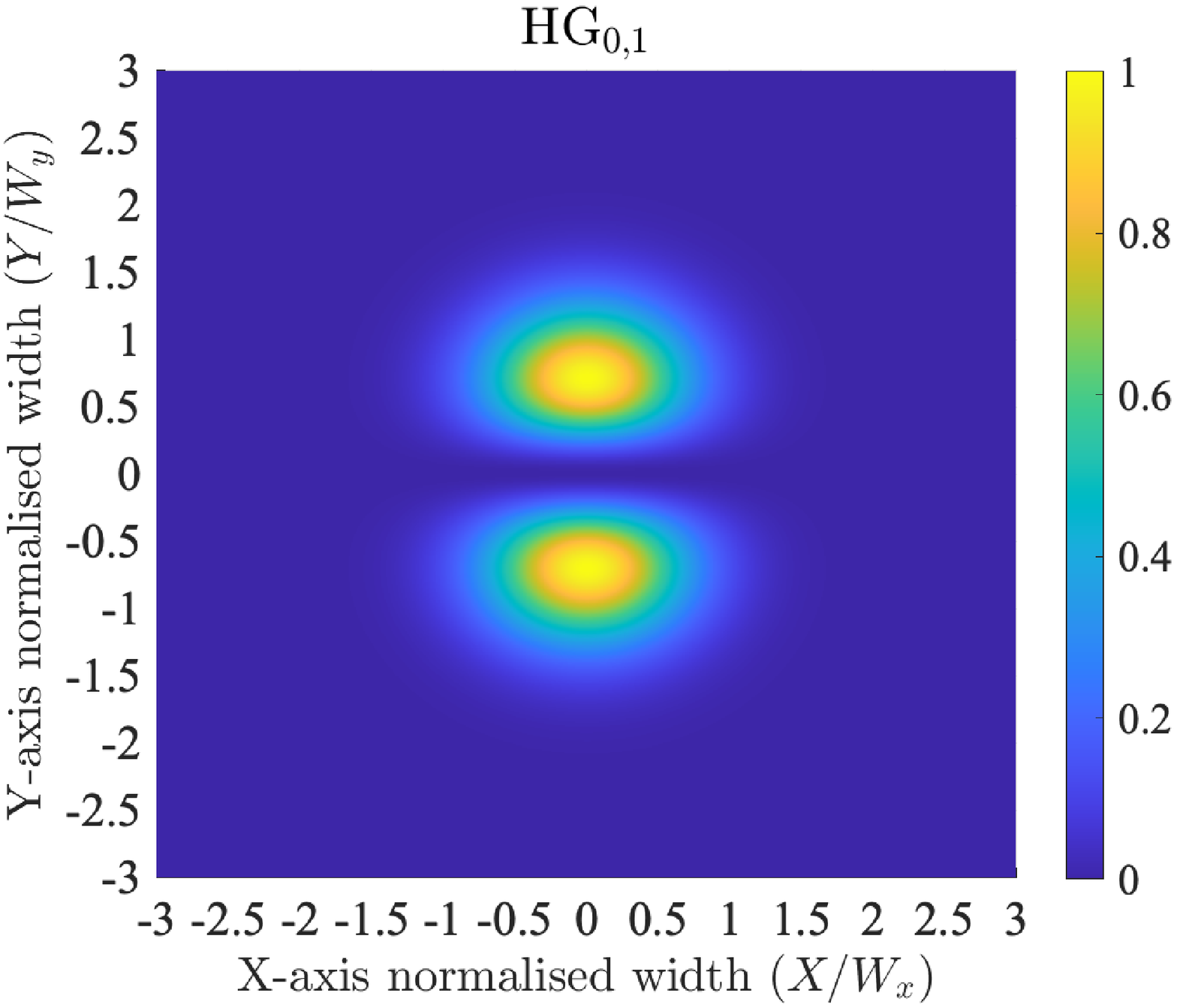}}
	}\\
\subfloat[$\rm {PMN} = 3$ \label{sub4:FigHG}]{%
		\resizebox{.32\linewidth}{!}{\includegraphics{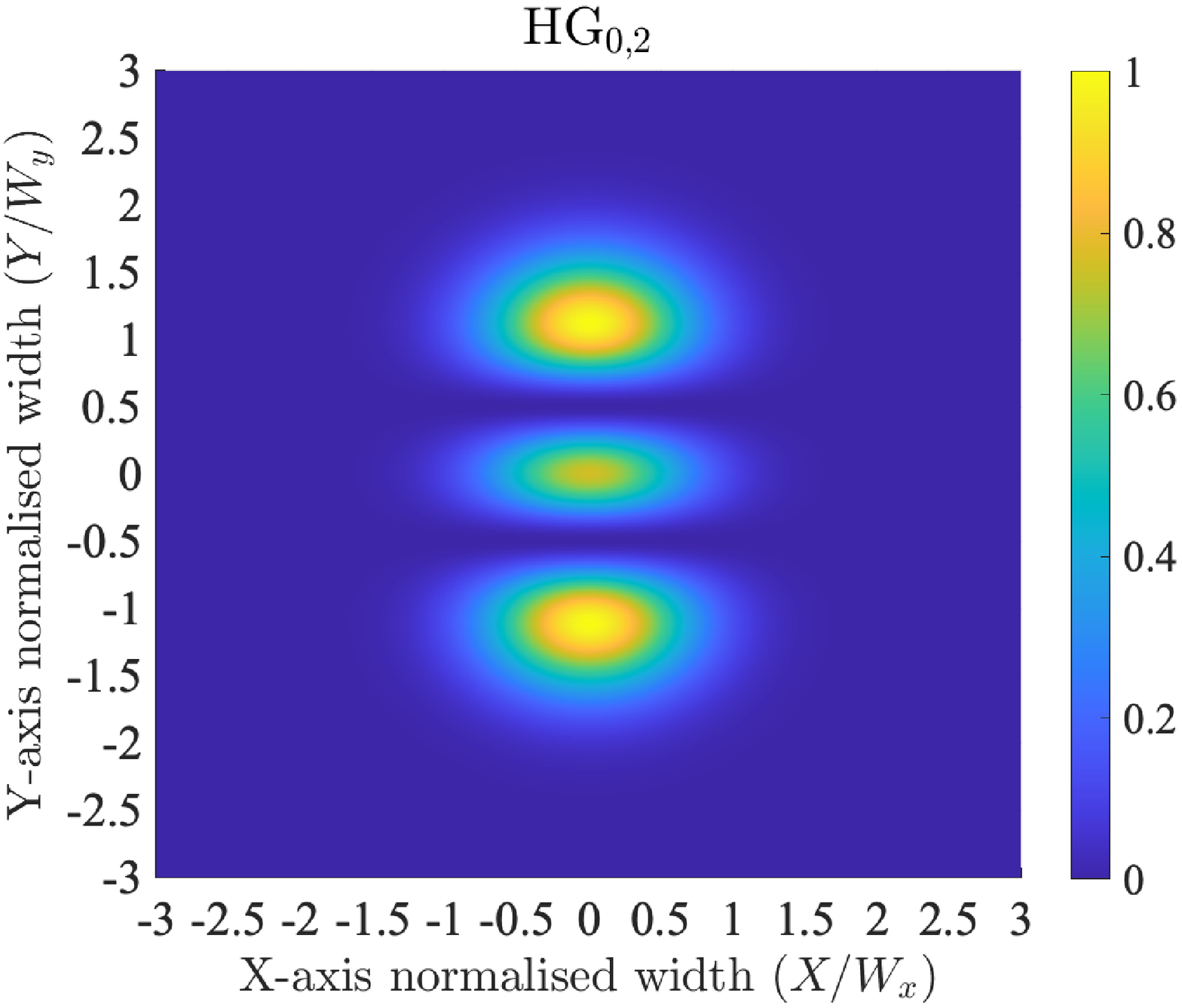}}
	}  
	\subfloat[$\rm {PMN} = 5$ \label{sub5:FigHG}]{%
		\resizebox{.32\linewidth}{!}{\includegraphics{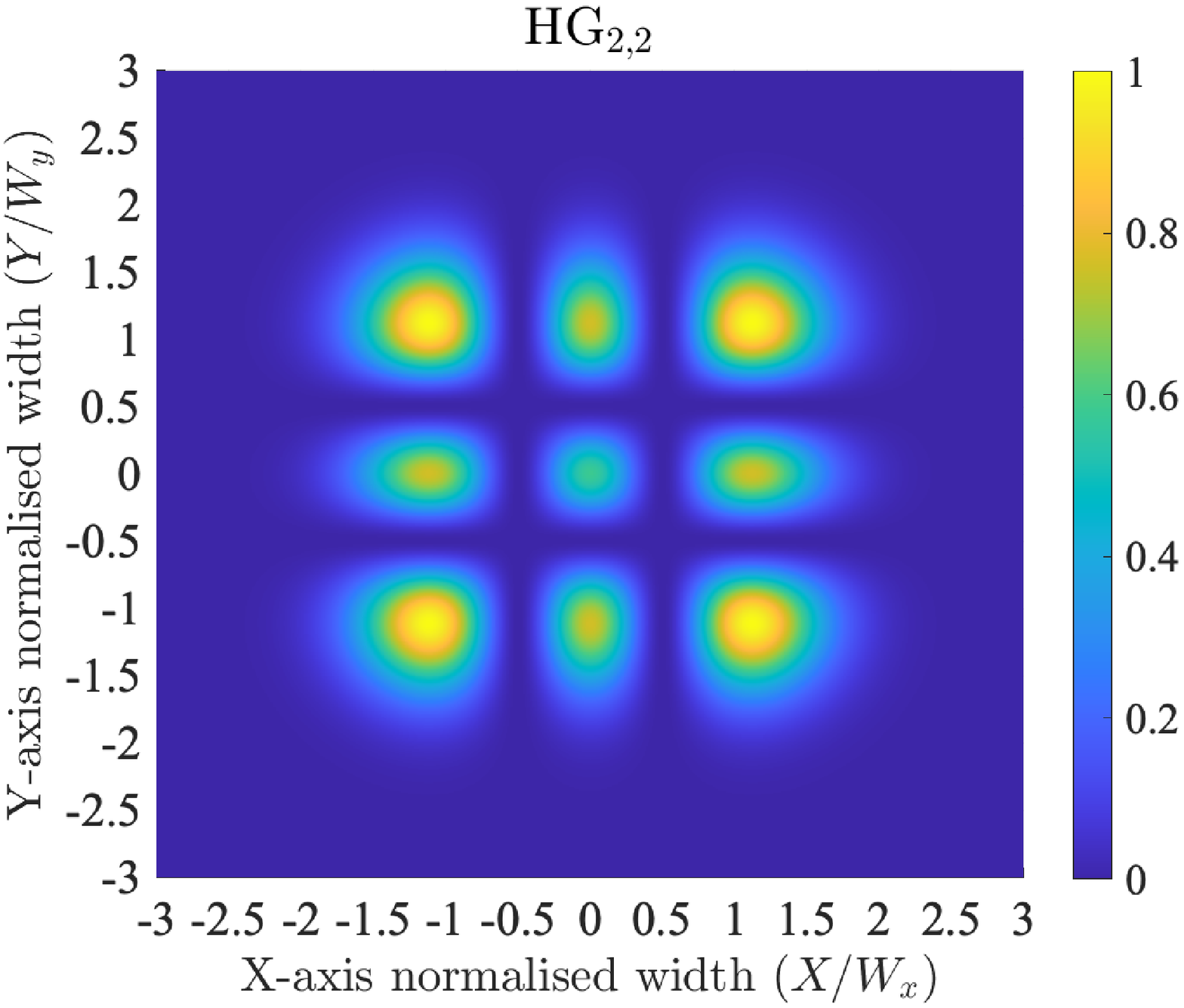}}
	} 
	\subfloat[$\rm {PMN} = 5$ \label{sub6:FigHG}]{%
		\resizebox{.32\linewidth}{!}{\includegraphics{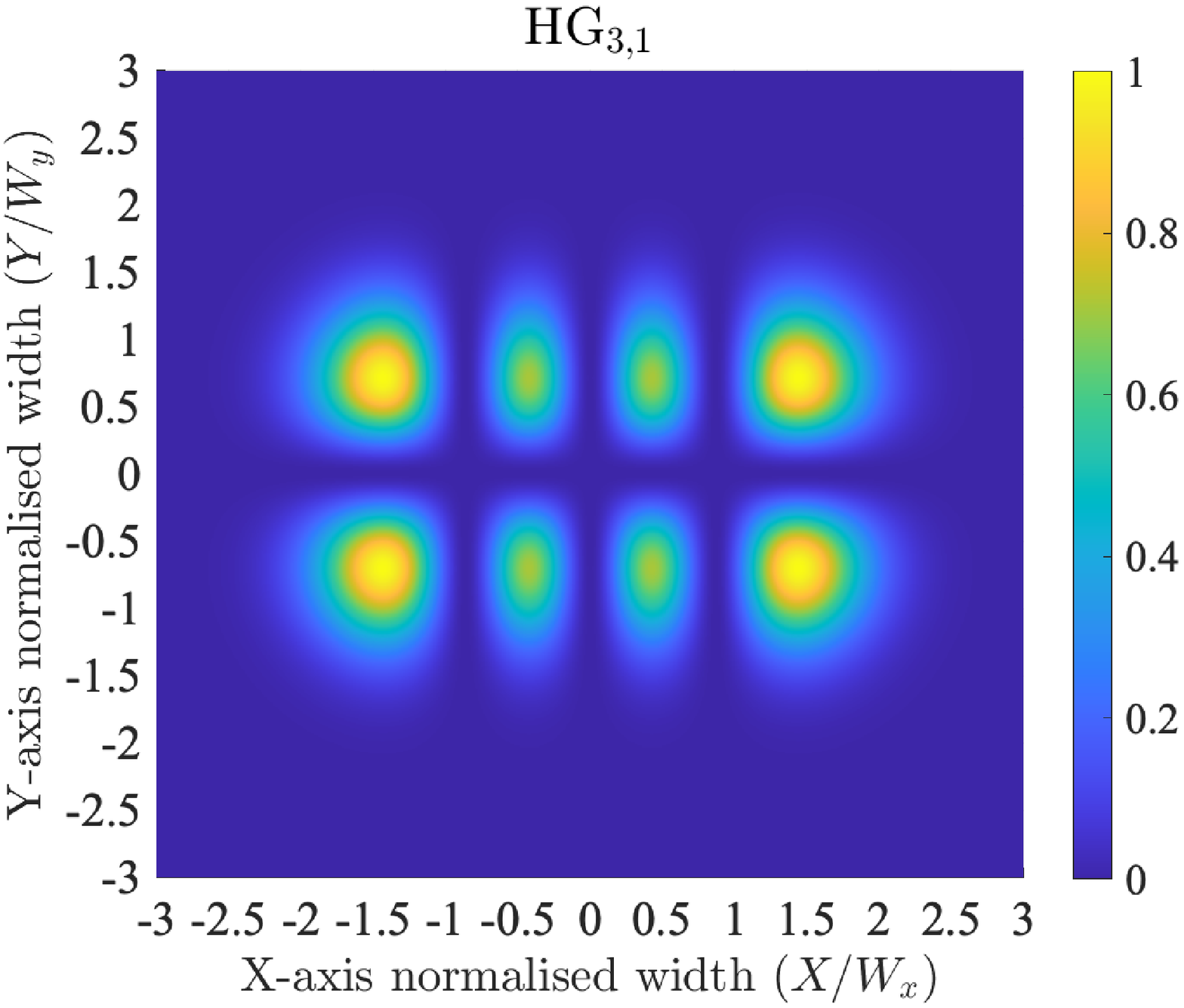}}
	}\\
\caption{Example irradiance profiles (normalised to peak) of Hermite-Gaussian transverse modes ($\lambda = 0.85~\mu$m and $w_0=5~\mu$m).} 
\label{Fig-HG-modes}
\end{figure}

The modes can be grouped into degenerate (i.e. very similar) modes based on their principal mode number (PMN) defined as ${\rm {PMN}}=l+m+1$ for Hermite-Gaussian beams. Fig.~\ref{Fig-HG-modes} presents some examples of the Hermite-Gaussian irradiance profiles for different PMNs. As can be seen in Fig.~\ref{sub1:FigHG}, the Hermite-Gaussian beam of order $(0,0)$, namely ${\rm HG}_{0,0}$, is the ideal Gaussian beam.

\begin{figure}[t!]
 \centering
 \subfloat[Combinations \label{sub1:FigHG-Com}]{%
		\resizebox{.6\linewidth}{!}{\includegraphics{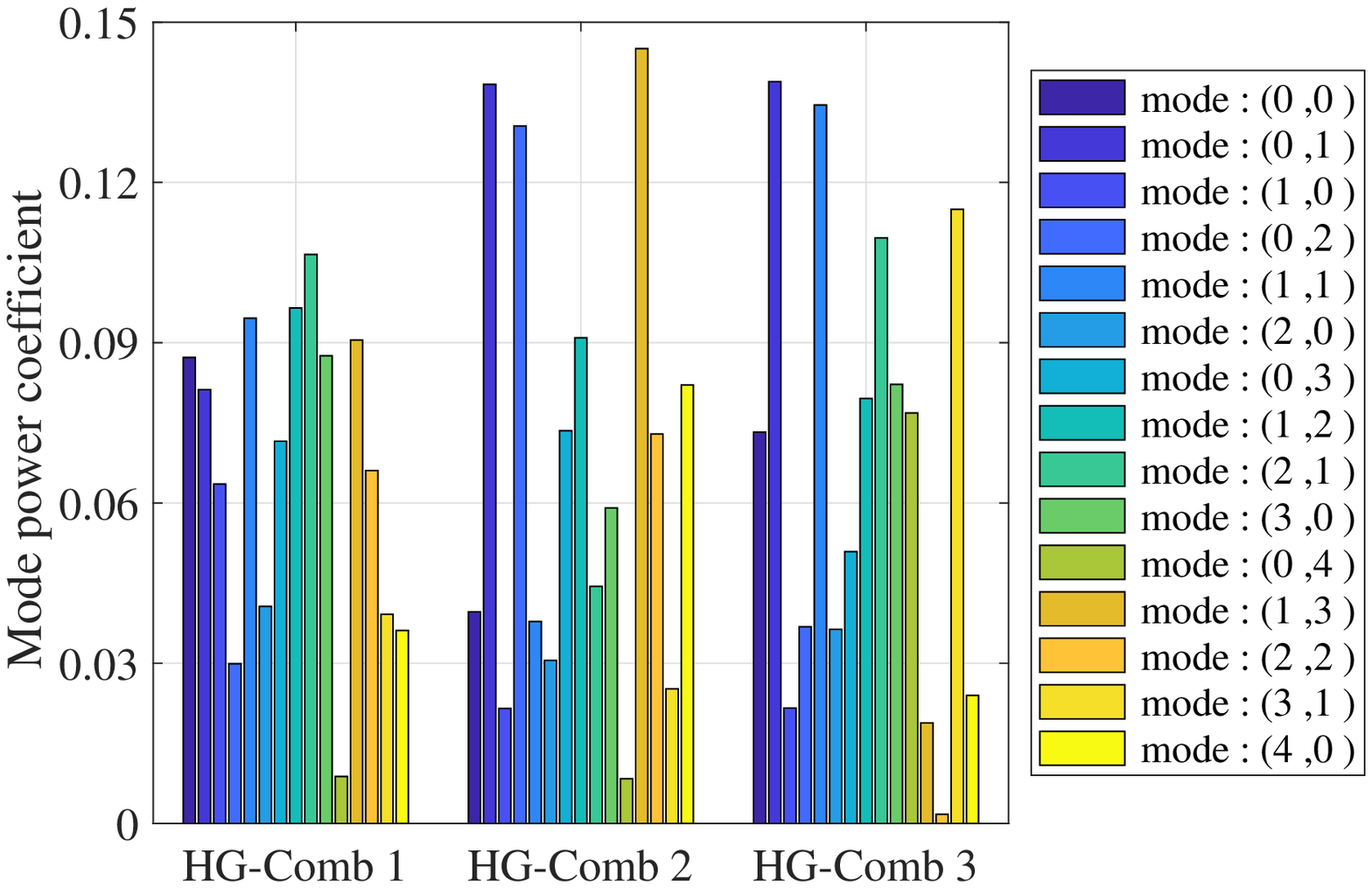}}
	}\\  
	\subfloat[HG-Comb 1 \label{sub2:FigHG-Com}]{%
		\resizebox{.33\linewidth}{!}{\includegraphics{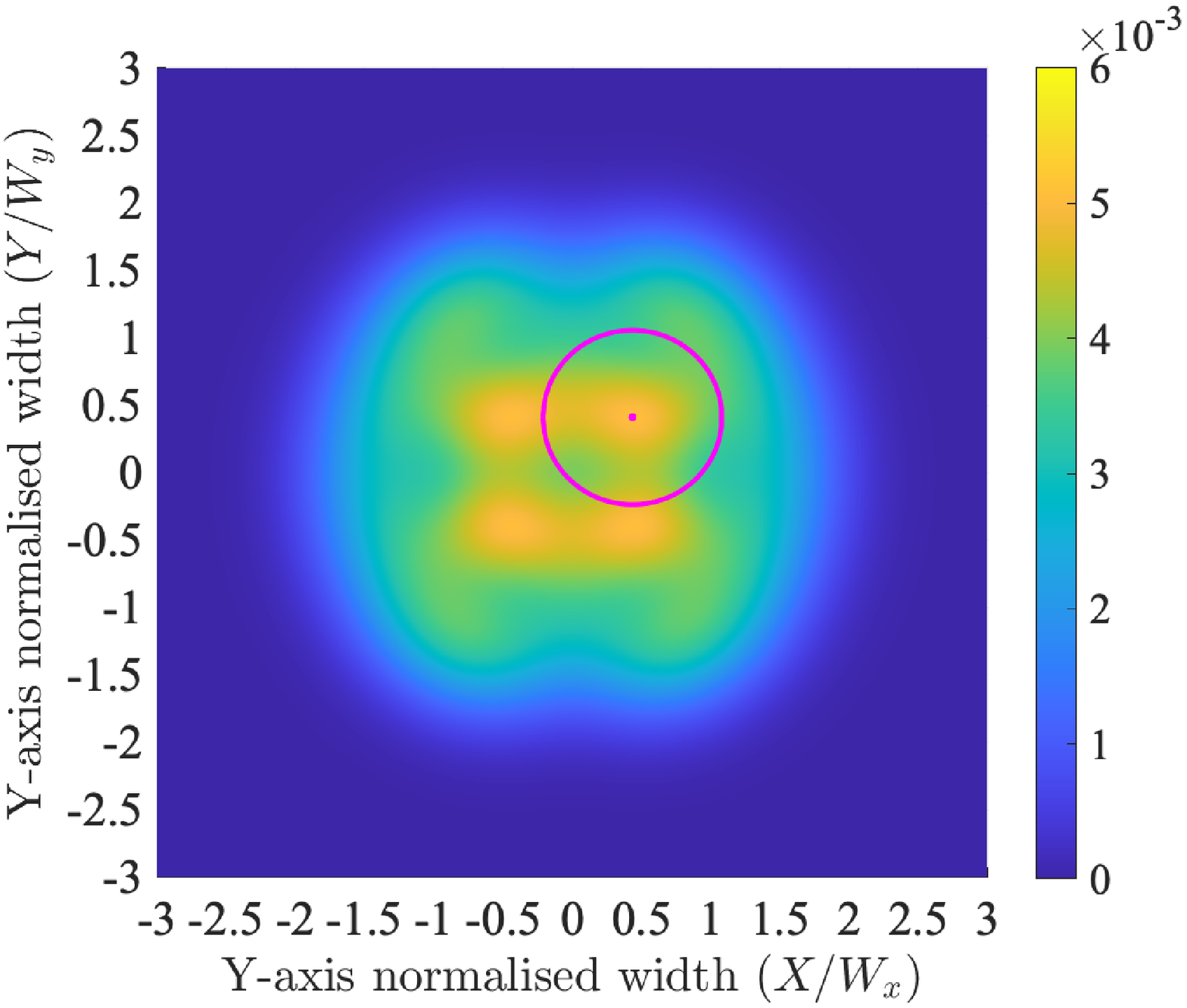}}
	}%
    \subfloat[HG-Comb 2 \label{sub3:FigHG-Com}]{%
		\resizebox{.33\linewidth}{!}{\includegraphics{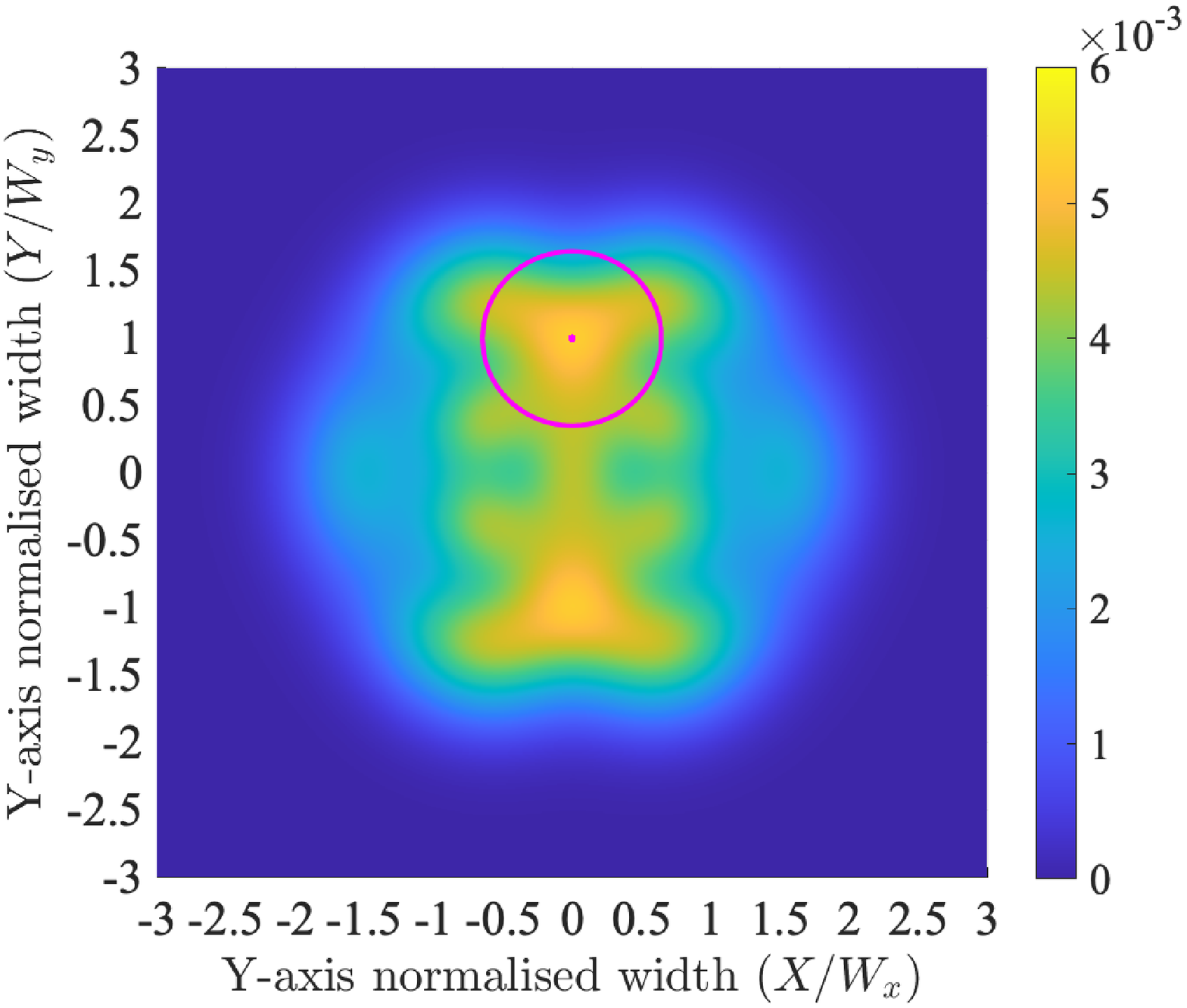}}
	}%
	\subfloat[HG-Comb 3\label{sub4:FigHG-Com}]{%
		\resizebox{.33\linewidth}{!}{\includegraphics{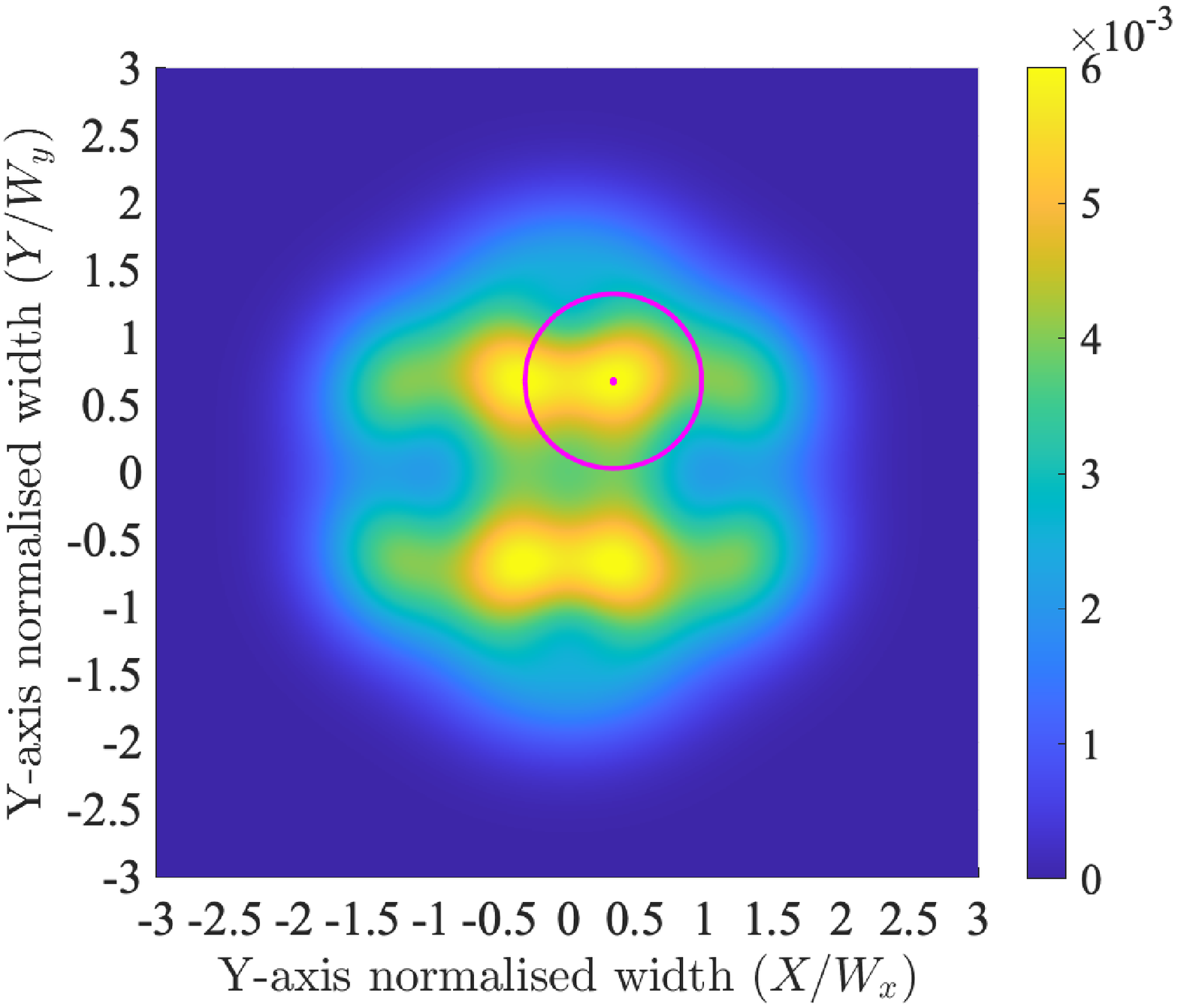}}
	}
\caption{Hermite-Gaussian mode combination: (a) normalized mode coefficients for three different combinations, (b),(c), and (d)  power distribution for each combination.} 
\label{Fig-HG-Com}
\end{figure}

A multimode laser beam can be decomposed into the principal transverse modes as introduced above. The number of modes that a multimode laser supports depends on the device characteristics and operating conditions, while the mode partitioning in the device constitutes a complex dynamic system \cite{252508,643258,720228,1363575}. 
Based on different values of the mode coefficient, $C_{l,m}$, and the mode combinations diverse beam profiles can be generated. Fig.~\ref{Fig-HG-Com} presents an example of three different mode combinations and their corresponding output power profiles. The mode coefficients of each combination are illustrated in Fig.~\ref{sub1:FigHG-Com} and the total power profile of these combinations are depicted in Figs.~\ref{sub2:FigHG-Com}-\ref{sub4:FigHG-Com}. The purple dots indicate the point of maximum irradiance in each case, which is substantial for laser safety analysis. 


\subsubsection{Laguerre-Gaussian Beams} \label{LG}
An alternative set of solutions to the paraxial Helmholtz equation are Laguerre-Gaussian beams. Circularly symmetric beam profiles (e.g., lasers with cylindrical cavities) are often best represented using the Laguerre-Gaussian modal decomposition. Let ${\rm LG}_{l,m}$ denote the Laguerre-Gaussian beam of order $(l,m)$ in the cylindrical coordinate system $(r,\phi,z)$. The integers $l\in \mathbb{Z}^+$ and $m\in \mathbb{Z}^+$ are azimuthal and radial indices, respectively. The complex amplitude of ${\rm LG}_{l,m}$ is expressed as \cite{saleh2019fundamentals}:
\begin{equation}
\label{LaguerreEq1}
    \begin{aligned}
    U_{l,m}(r,\phi,z)&=A_{l,m} \left[\frac{w_0}{W(z)}\right]\left(\frac{\sqrt{2}r}{W(z)}\right)^l \mathbb{L}_m^l\left(\frac{2r^2}{W^2(z)}\right)\exp\left(-\frac{r^2}{W^2(z)}\right)\\
    &\times\exp{\left(-jkz-jk\frac{r^2}{2R(z)}\mp jl\phi+ j(l+2m+1)\zeta(z)\right)},
    \end{aligned}
\end{equation}
%
where $\mathbb{L}_m^l(\cdot)$ denotes generalized Laguerre polynomials expressed as \cite{saleh2019fundamentals}:
\begin{equation}
    \mathbb{L}_m^l(x)=(m+l)!\sum_{i=0}^m\frac{(-x)^i}{i!(m-i)!(l+i)!}, 
\end{equation}

The ideal Gaussian beam corresponds the lowest-order Laguerre-Gaussian beam $\rm LG_{0,0}$. In \eqref{LaguerreEq1}, $A_{l,m}$ is the normalized mode coefficient which is used to ensure that the total power of each mode is equal to $1$ and is given by:
%
\begin{equation}
    A_{l,m}=\sqrt{\frac{2m!}{\pi(m+l)!}} .
\end{equation}
The irradiance of each Laguerre-Gaussian mode, can be expressed as \cite{saleh2019fundamentals}:
\begin{equation}
\label{LaguerreEq}
    \begin{aligned}
    \left|U_{l,m}(r,\phi,z)\right|^2= \Big| A_{l,m} \left[\frac{w_0}{W(z)}\right] \left(\frac{\sqrt{2}r}{W(z)}\right)^l \mathbb{L}_m^l\left(\frac{2r^2}{W^2(z)}\right) \exp{\left(-\frac{r^2}{W^2(z)}\right)}\Big|^2.
    \end{aligned}
\end{equation}
%

\begin{figure}[t!]
 \centering
 \subfloat[$\rm{PMN} = 1$ \label{sub1:Fig1}]{%
		\resizebox{.33\linewidth}{!}{\includegraphics{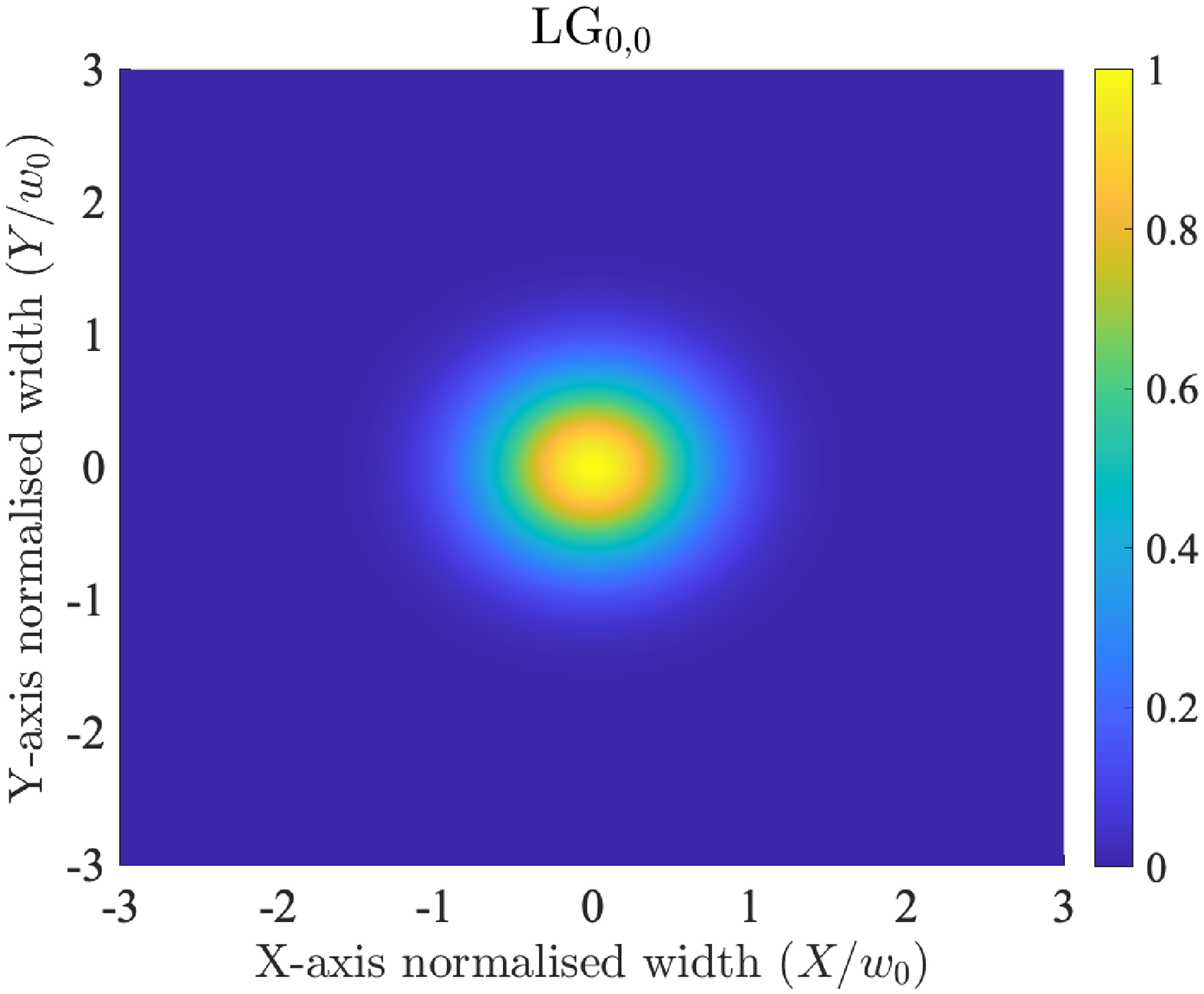}}
	}
	\subfloat[$\rm{PMN} = 4$ \label{sub2:Fig2}]{%
		\resizebox{.33\linewidth}{!}{\includegraphics{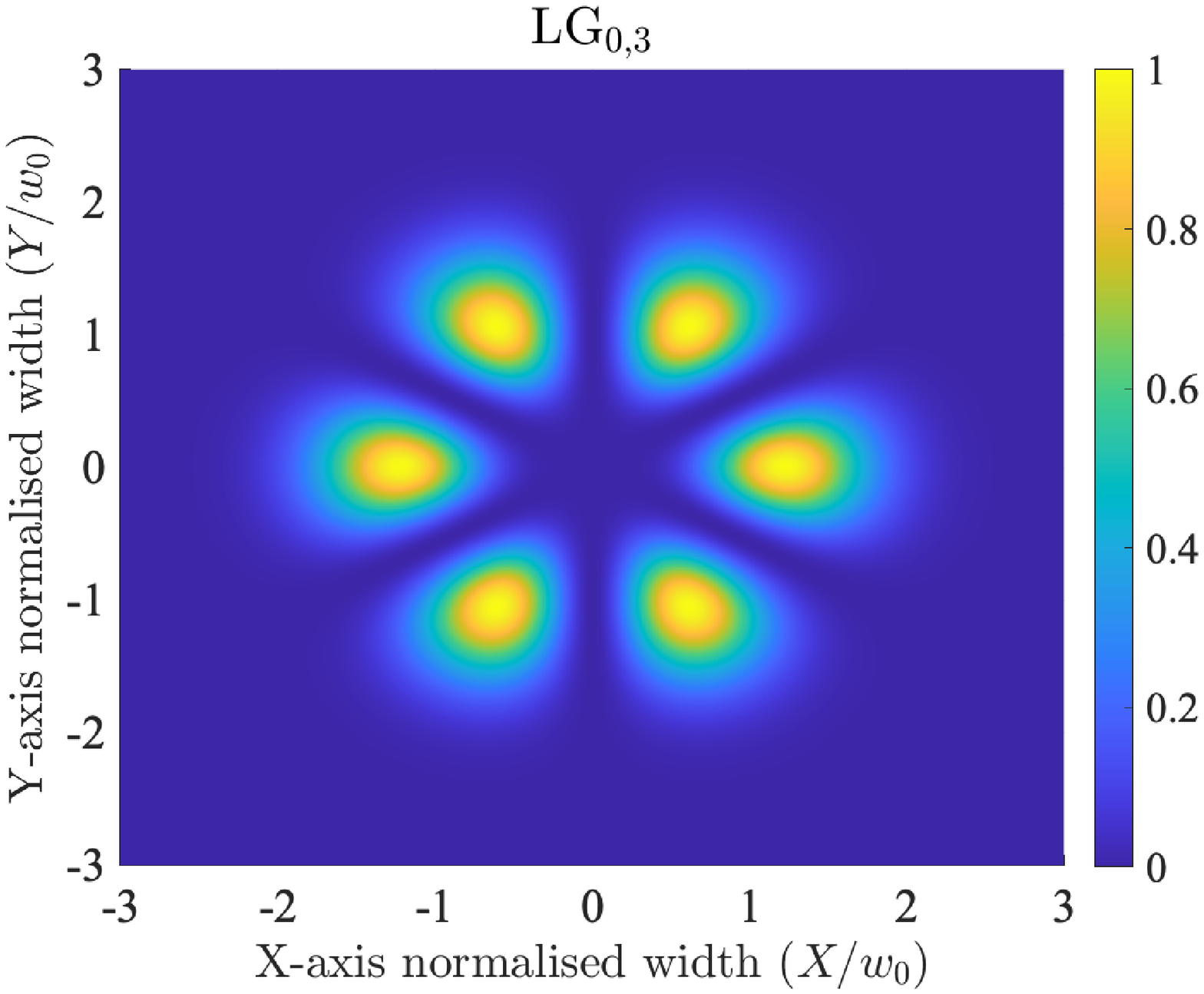}}
	}
	\subfloat[$\rm{PMN} = 4$ \label{sub2:Fig2}]{%
		\resizebox{.33\linewidth}{!}{\includegraphics{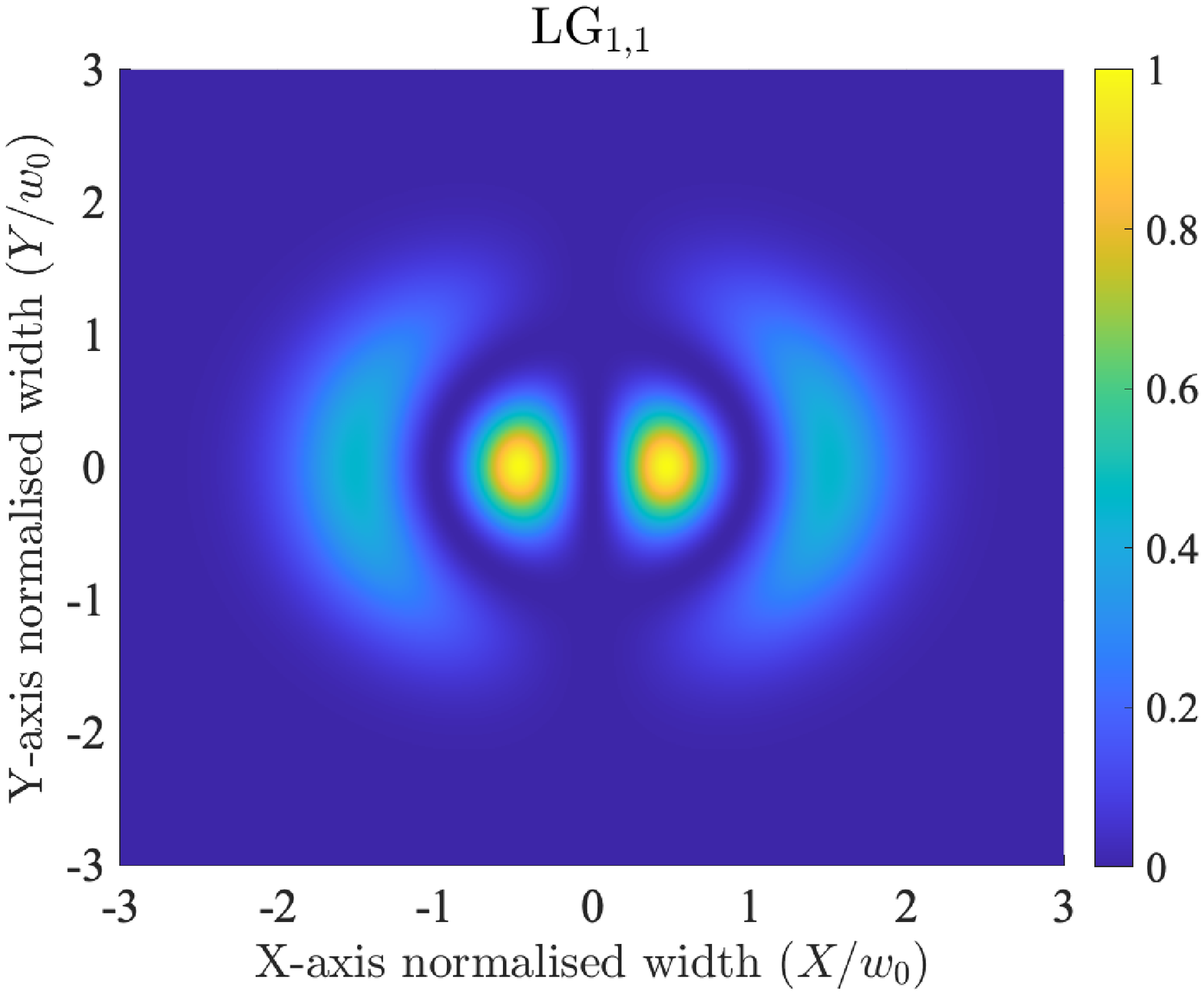}}
	}
\caption{Example irradiance profiles (normalised to peak) of VCSEL Laguerre-Gaussian transverse modes ($\lambda = 0.85~\mu$m and $w_0=5~\mu$m).}
\label{Fig-LG1}
\end{figure}

For Laguerre-Gaussian beams, PMN is defined as ${\rm {PMN}} = l + 2m +1$. 
Fig.~\ref{Fig-LG1} shows some examples of irradiance profiles of low order Laguerre-Gaussian beams.

\subsubsection{$M$-Squared Factor}
\label{MsquaredFactor}
In practice, rather than decomposing the profile of the actual laser beam to Hermite or Laguerre Gaussian modes, the propagation of the laser beam is described approximately by an ideal Gaussian beam but with a modified beam radius $W_{\rm R}(z)$ given by:
\begin{equation}
    W_{\rm R}(z)=w_{0\rm R} \sqrt{1+\left(\frac{z\lambda M^2}{\pi w_{0\rm R}}\right)^2}, 
\end{equation}
where the degree of deviation of the practical laser beam from the ideal Gaussian beam is quantified by the quality factor $M^2$, often known as the \textit{$M$-squared factor} or the \textit{beam propagation factor} \cite{kardosh2004beam,214507} and defined as \cite{Siegman98Laser,TechReport}:
\begin{equation}
    M^2=\frac{w_{0\rm R}\theta_{\rm R}}{w_0\theta}= \frac{\pi w_{0\rm R}\theta_{\rm R}}{\lambda},
    \label{eqMsquared}
\end{equation}
In~(\ref{eqMsquared}), $w_{0\rm R}$ and $\theta_{\rm R}$ are the beam waist and far-field divergence angle of the practical beam, respectively, and $w_0$ and $\theta$ are those of the benchmark ideal Gaussian beam, respectively. Note that for an ideal Gaussian beam, we have $w_0 \theta =\lambda/\pi$. 

%
%
%
%

The $M$-squared factor reflects how well a collimated beam can be focused to a small spot, or how well a divergent beam can be collimated. For an ideal Gaussian beam, $M^2$ is equal to $1$. Lower values of $M^2$ represent a better beam quality. Practical laser beams have factors greater than one. The $M^2$ factor of VCSELs can be as high as $20$ \cite{kardosh2004beam}. This is due to their transverse multimode emission which results from their large active diameters.
%
%
%
In \cite{Hall_2007}, an experimental methodology for measuring the beam propagation factor $M^2$ has been developed based on the technique described in the ISO 11146 series of standards \cite{ISO11146-1:2005,ISO11146-2:2005,ISO11146-3:2004}. 

\subsubsection{VCSEL Beam Decomposition Using Experimental Measurements}
\label{Sec:ExperimentalData}
Far field measurements of VCSEL devices allow the estimation of mode power coefficients based on the ideal models described in Sections~\ref{HG} and~\ref{LG}.
Such measurements consist of recording the far field beam profile at a given distance from the laser source. Although the exact mode power coefficients of multimode VCSELs depend on the laser dynamics and the operating conditions, a common approach is to use a set of pre-defined mode combinations for system design and/or the performance analysis.
As an example, in \cite{Pepeljugoski03}, a set of Laguerre-Gaussian mode combinations is presented. Table~\ref{Table-ExpData2} provides the mode power coefficients of these combinations and Fig.~\ref{Fig-ExpData6} illustrates the normalized radial power profile assuming $\lambda = 850$~nm, $w_0=5~\mu$m, and $z=0$. In this figure, the red dashed line indicates the radial position where the power falls into the $1/{e}^2$ of the maximum value.
For a single-mode Gaussian beam, this position is known as the spot radius (see Section~\ref{idealGaussian}). This definition of the spot radius can also be used for non-Gaussian modes and multimode beams. The spot radius can be found analytically or numerically from the mode power profiles. Once the spot radius is known, the divergence angle can be calculated according to (\ref{eq_DivergenceAngle}). Let $\theta_{l,m}$ denote the divergence angle of LG$_{l,m}$. Table~\ref{Table-Beamdivergence} presents the spot radii and divergence angles of the first six Laguerre-Gaussian transverse modes. The approximation values in this table correspond to the small value of ${\lambda}/{(\pi w_0)}$.

\setlength\doublerulesep{0.1cm}



%
In Table~\ref{Table-ExpData2}, the ratio $\theta_{\rm R} / \theta$ is provided for each combination, where $\theta_{\rm R}$ represents the divergence angle of each combination and $\theta$ is that of LG$_{0,0}$ (assuming that $\lambda/(\pi w_0)$ is small). The $M^2$ factor can be found according to this ratio and therefore, it can help simplify the eye safety analysis. This approximation method has been used in Section~\ref{SafetyAnalysisMultiMode} and it is compared to the exact analytical technique.

\noindent\begin{minipage}{\linewidth}
\centering
\captionof{table}{Mode power coefficients for various combinations of the first six Laguerre-Gaussian transverse modes \cite{Pepeljugoski03}.}
\label{Table-ExpData2}
{%
\begin{tabular}{|c|c||c|c|c|c|c|c|c|}
\cline{3-9} 
\multicolumn{2}{l|}{\multirow{2}{*}{}} & \multicolumn{7}{c|}{Mode combinations} \\ \cline{3-9} 
\multicolumn{2}{l|}{}                  
     & \begin{tabular}[c]{@{}c@{}} LG-\\Comb~1 \end{tabular}   
     & \begin{tabular}[c]{@{}c@{}} LG-\\Comb~2 \end{tabular} 
     & \begin{tabular}[c]{@{}c@{}} LG-\\Comb~3 \end{tabular} 
     & \begin{tabular}[c]{@{}c@{}} LG-\\Comb~4 \end{tabular} 
     & \begin{tabular}[c]{@{}c@{}} LG-\\Comb~5 \end{tabular} 
     & \begin{tabular}[c]{@{}c@{}} LG-\\Comb~6 \end{tabular} 
     & \begin{tabular}[c]{@{}c@{}} LG-\\Comb~7 \end{tabular}\\ 
    \hhline{|==|:=======|}
\multirow{6}{*}{\rotatebox[origin=c]{90}{Modes}}        
& 1  & $1$ & $0.5$ & $0$    & $0$    & $0.125$   & $0$      & $0.06$\\ \cline{2-9} 
& 2  & $0$ & $0.5$ & $0.5$  & $0$    & $0.75$    & $0.125$  & $0.44$\\ \cline{2-9} 
& 3  & $0$ & $0$   & $0.25$ & $0.25$ & $0.0625$  & $0.375$  & $0.22$\\ \cline{2-9} 
& 4  & $0$ & $0$   & $0$    & $0.25$ & $0$       & $0.0625$ & $0.03$\\ \cline{2-9} 
& 5  & $0$ & $0$   & $0.25$ & $0.25$ & $0.0625$  & $0.375$  & $0.22$\\ \cline{2-9} 
& 6  & $0$ & $0$   & $0$    & $0.25$ & $0$       & $0.0625$ & $0.03$\\ \hhline{|==|:=======|} 

\multicolumn{2}{l|}{\multirow{2}{*}{}} & \multicolumn{7}{c|}{ $\theta_{\rm R}/{\theta} = M^2 ({w_0}/w_{0\rm R})$} \\  \cline{3-9}
\multicolumn{2}{l|}{}   &               
$1$ &  $1.3248$ &  $1.7346$ &  $1.9451$ &  $1.5432$ &  $1.7511$ & $1.7237$ \\\cline{3-9}
\end{tabular}}


\vspace{10pt}
\centering
\captionsetup{type=figure}
\includegraphics[width=0.8\textwidth]{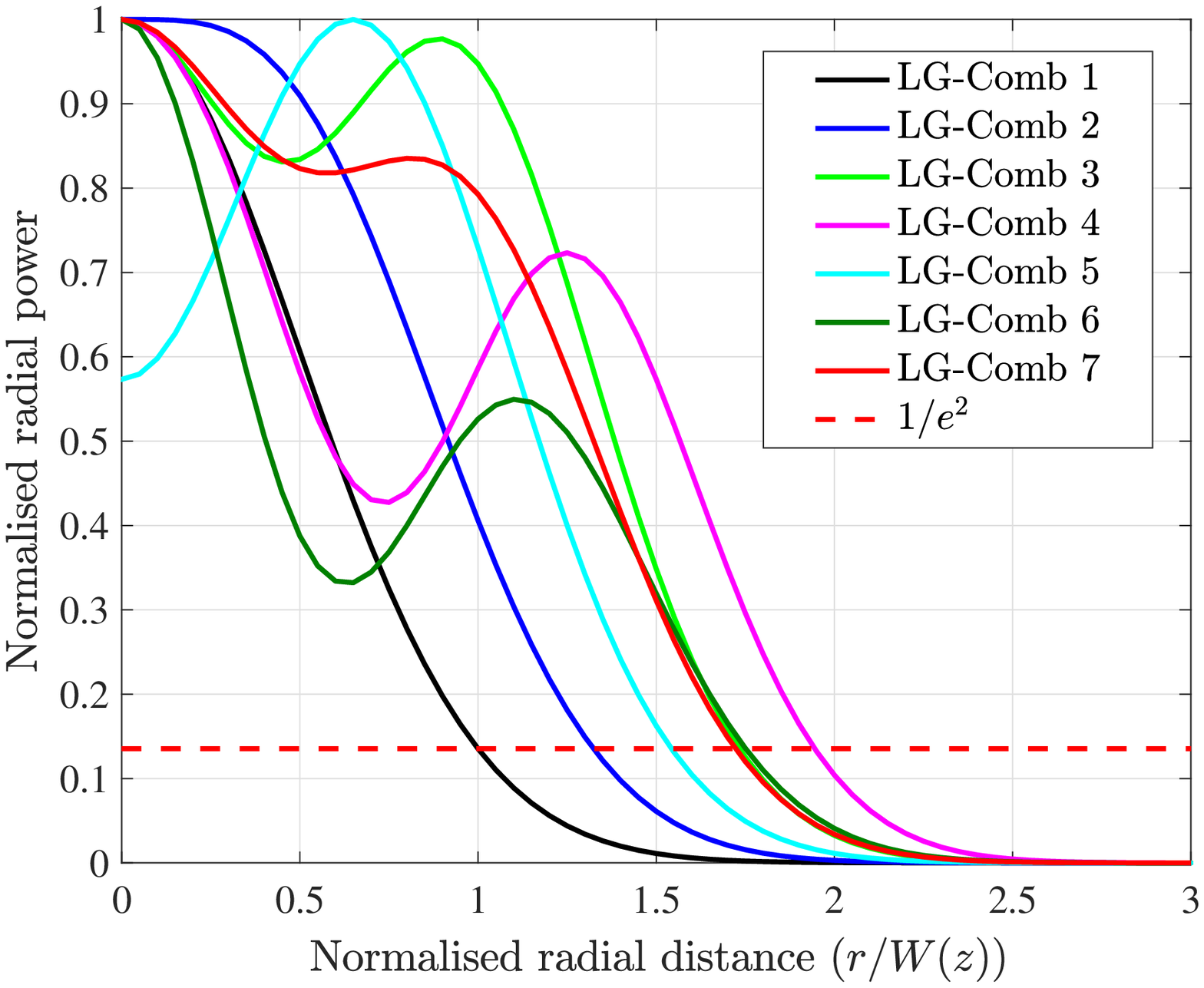} 
\captionof{figure}{Normalised radial power for various mode combinations given in Table~\ref{Table-ExpData2} for $\lambda = 850$~nm, $w_0=5~\mu$m, and $z=0$.}
\label{Fig-ExpData6}
\end{minipage} 

\begin{table}[ht!]
\renewcommand{\arraystretch}{1.6}
\centering
\caption[blah]{Spot radii and divergence angles of the first six Laguerre-Gaussian transverse modes.}
\label{Table-Beamdivergence}
\begin{tabular}{|c|c|c|c|c|}
\hline
Mode & $(l,m)$ & PMN & Spot radius  & Divergence angle, $\theta_{l,m}$  \\ \hline\hline
$1$ & $(0,0)$ & $1$ & $W(z)$ & $\theta_{0,0}=\theta_0\approx\dfrac{\lambda}{\pi w_0}$         
\\ \hline 
$2$ & $(1,0)$ & $2$ & $1.500W(z)$.  & $\theta_{1,0}\approx 1.5\theta_0$   \\ \hline
$3$ & $(2,0)$ & $3$ & $1.774W(z)$ & $\theta_{2,0}\approx 1.774\theta_0$ \\ \hline
$4$ & $(0,1)$ & $3$ & $1.502W(z)$ & $\theta_{0,1}\approx 1.502\theta_0$ \\ \hline
$5$ & $(3,0)$ & $4$ & $1.999W(z)$ & $\theta_{3,0}\approx 1.999\theta_0$ \\ \hline
$6$ & $(1,1)$ & $4$ & $2.008W(z)$ & $\theta_{1,1}\approx 2.008\theta_0$ \\ \hline
\end{tabular}
\end{table}

\subsection{Beam Propagation in Lens Systems}
\label{BeamAfterLens}
\begin{figure}[t]
		\centering  
	\subfloat[\label{sub1:Fig-LensSystem}]{%
		\resizebox{.45\linewidth}{!}{\includegraphics{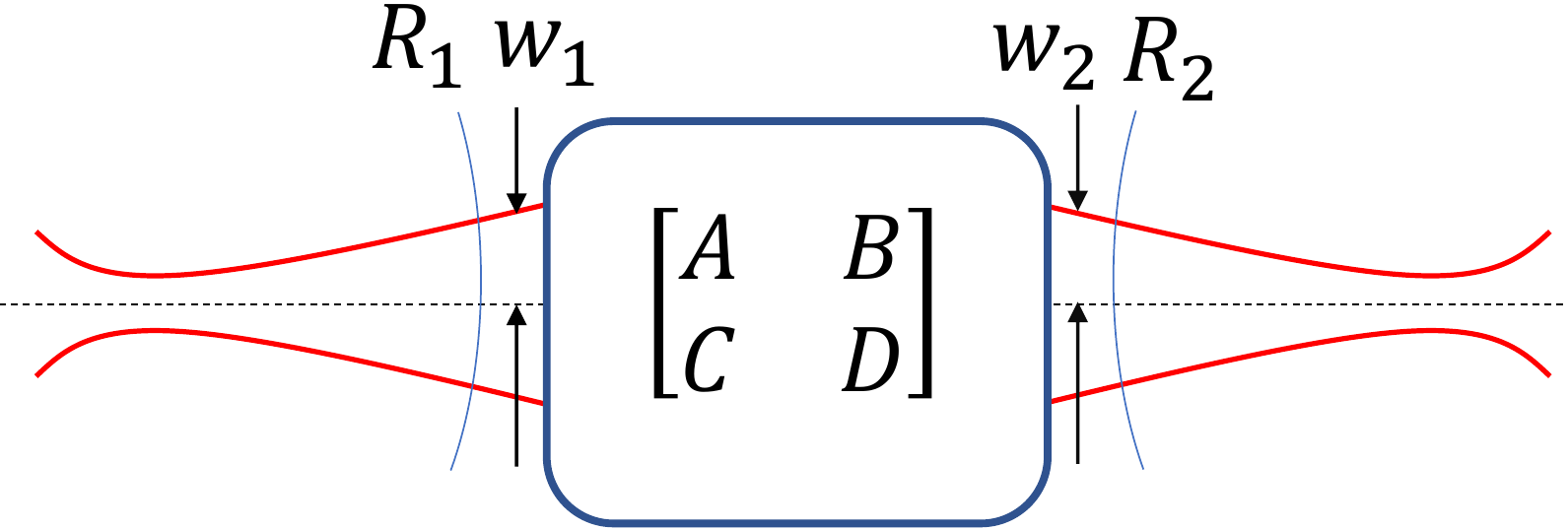}}
	} \quad\ \ 
	\subfloat[\label{sub2:ThinLens}]{%
		\resizebox{.45\linewidth}{!}{\includegraphics{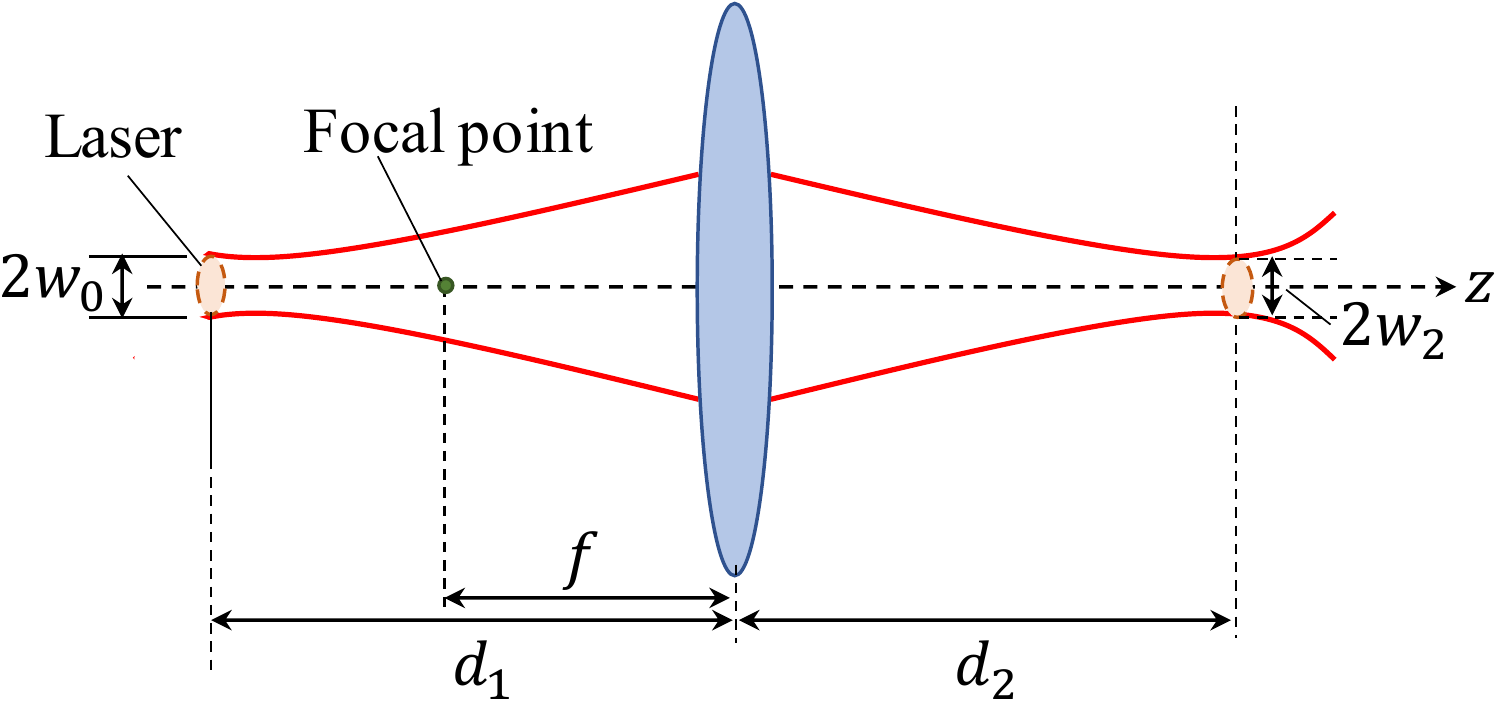}}
	}\\
	\caption{(a) Transformation of a Gaussian beam by an arbitrary paraxial system described by an $ABCD$ matrix. Here, $R_1$ and $R_2$ are the radii of curvature of the input and output wavefronts, respectively. (b) Gaussian beam propagation after a thin lens. }
	\label{Fig-LensSystem}
\end{figure}

A Gaussian beam remains Gaussian after passing through a lens (if there are no aberrations). However, the parameters of the output Gaussian beam are different from those of the input beam. The \textit{ray transfer method} is a simple method which relates the input and output beam parameters of any optical system using ABCD matrices \cite{kogelnik1966laser}. This method is schematically shown in Fig.~\ref{sub1:Fig-LensSystem}. A $2\times 2$ matrix associates the position and angle of paraxial rays at the input and output of the optical system by means of simple linear algebraic equations. This method is described briefly in the following paragraphs.

Let us assume that the input and output beam waists corresponds to the object and image, respectively. The object and/or the image may be real or virtual, depending on the type of lens and the location of the beam waist with respect to lens. Let $q$ and $w_0$ be the parameters of the input Gaussian beam, and $q_2$ and $w_2$ be the corresponding parameters of the output Gaussian beam. The relation between the complex beam parameters of the input and the output, $q$ and $q_2$, are given as follows:
\begin{equation}
    q_2=\frac{Aq+B}{Cq+D},
\end{equation}
where the matrix elements are expressed as \cite{hecht1998hecht}:
\begin{subequations}
\begin{align}
&A=1+\frac{\rho(n-1)}{n\upsilon_1},\\
&B=\frac{\rho}{n},\\
&C=(n-1)\left(\frac{1}{\upsilon_1}-\frac{1}{\upsilon_2}\right)-\frac{\rho(n-1)^2}{n\upsilon_1\upsilon_2},\\
&D=1-\frac{\rho(n-1)}{n\upsilon_2}.
\end{align}
\end{subequations}
Here, $n$ is the refractive index of the lens and $\rho$ shows the lens thickness. The parameters $\upsilon_1$ and $\upsilon_2$ are the radii of the input and output surfaces of the lens, respectively. 

As an example, consider a thin lens of focal length $f$ and let $d_1$ and $d_2$ denote the distances of the input and output beam waists from the lens, respectively, as illustrated in Fig.~\ref{sub2:ThinLens}. The parameters $d_2$ and $w_2$ are given as \cite{saleh2019fundamentals}:
\begin{equation}
    d_2=\frac{\dfrac{1}{f}-(1-\dfrac{d_1}{f})d_1(\dfrac{\lambda}{\pi w_0^2})^2}{\dfrac{1}{f^2}+(1-\dfrac{d_1}{f})^2(\dfrac{\lambda}{\pi w_0^2})^2},
    \label{beamwaistloc}
\end{equation}
\begin{equation}
    w_2=\frac{\lambda f}{\pi w_0\sqrt{1+(1-\dfrac{d_1}{f})^2(\dfrac{\lambda f}{\pi w_0^2})^2}}.
    \label{w0AfterLens}
\end{equation}
The beam divergence after the lens is also given by:
\begin{equation}
\label{BeamDivergenceLens}
    \theta_2=\frac{\theta}{\kappa},
\end{equation}
where $\kappa=w_2/w_0$ is the magnification factor of the lens.

\subsection{Beam Propagation for Array of VCSELs}
\label{BeamPropArray}
\begin{figure}[t]
		\centering  
	\subfloat[\label{sub1:Fig-ArraySystemModel}]{%
		\resizebox{.65\linewidth}{!}{\includegraphics{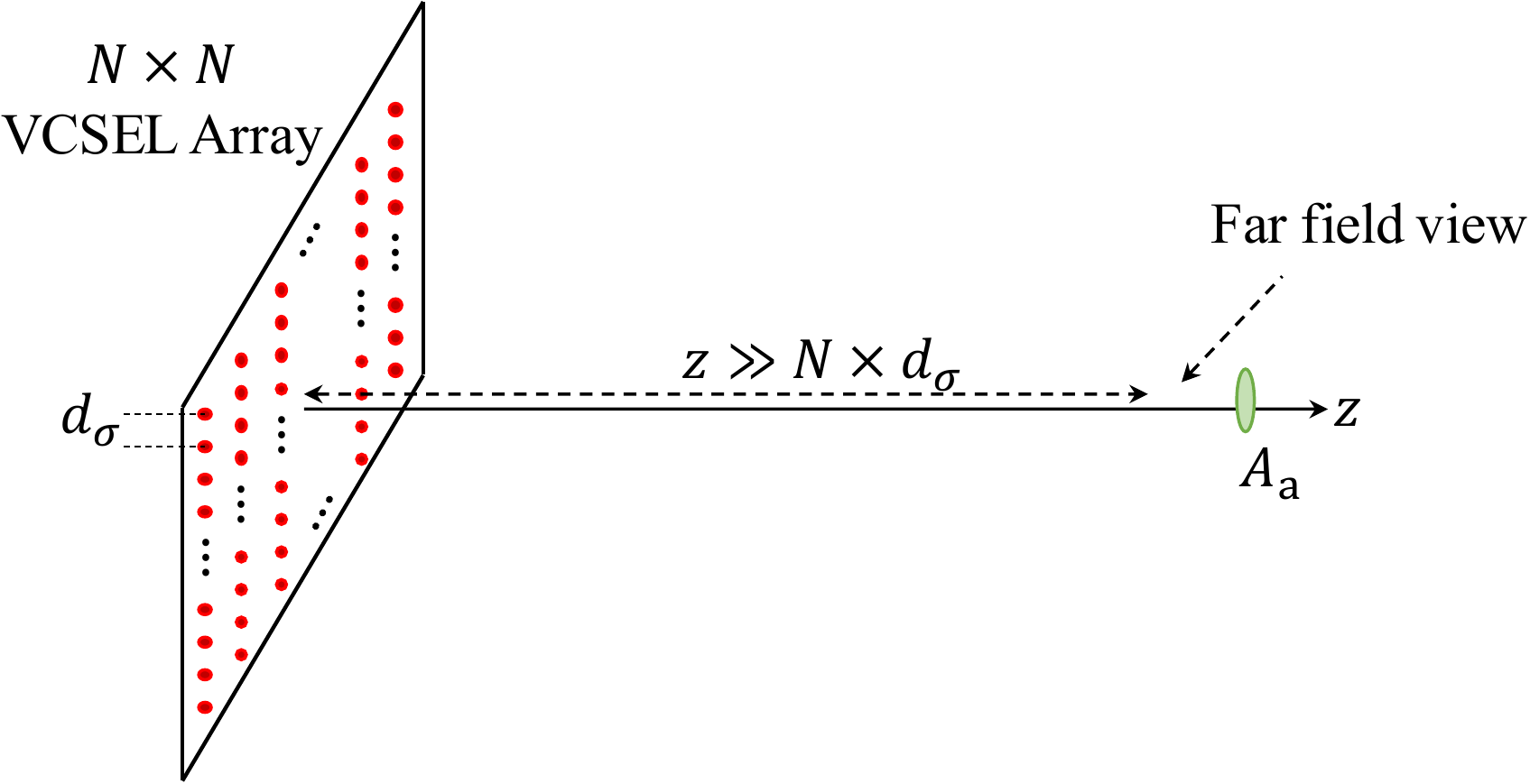}}
	} \quad\ \ 
	\subfloat[\label{sub2:Fig-ArrayCollimated}]{%
		\resizebox{.15\linewidth}{!}{\includegraphics{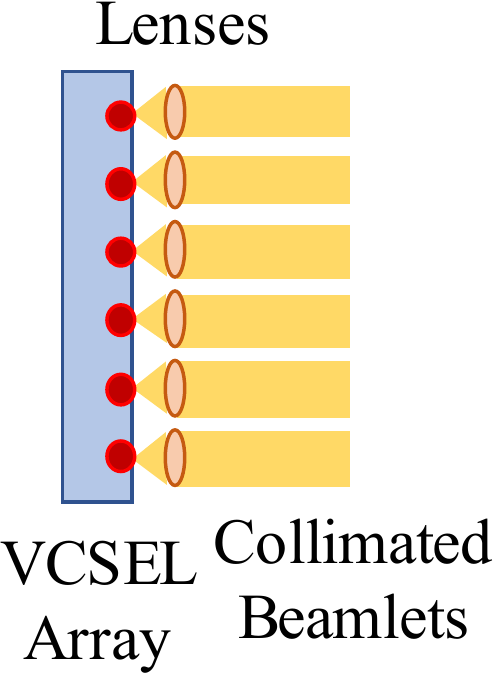}}
	}\\
	\caption{(a) An array of $N\times N$ VCSELs with the pitch of $d_{\sigma}$. (b) Collimation of the output of a VCSEL array.}
	\label{Fig-ArraySystemModel}
\end{figure}

An $N\times N$ two dimensional ($2$D) VCSEL array is illustrated in Fig.~\ref{sub1:Fig-ArraySystemModel}.
The total output power of a VCSEL array is simply the sum of the output powers of all single VCSEL emitters. Practically, the emission per VCSEL might have to be somewhat decreased if the marginal constraints of the cooling system are reached. It should be noted that the total dissipated power of a VCSEL array and the thermal power density can be considerable (often several hundreds of Watt per cm$^2$). The cooling can be made simpler by increasing the spacing between the emitters, however, this does decrease the beam quality. This spacing may be referred to as pitch distance in some studies and is denoted by $d_{\sigma}$ in Fig.~\ref{sub1:Fig-ArraySystemModel}.

\label{BeamPropagationArray}

Assuming that all VCSELs in the array are identical, the received power collected by an aperture with an area of $A_{\rm a}$ which is located at $(x_{\rm a},y_{\rm a})$ is given by:
\begin{equation}
\label{ReceivedPowerArray}
    P_{\rm r}=\sum_{i=1}^{N^2}\iint\limits_{A_{\rm a}}\frac{2P_{\rm t}}{\pi W(z)^2}\exp{\left(-2\frac{(x_i-x_{\rm a})^2+(y_i-y_{\rm a})^2}{W(z)^2}\right)} {\rm d}x{\rm d}y ,
\end{equation}
where $(x_i,y_i)$ is the location of $i$th VCSEL of the array.
In the far-field region, i.e., $z \gg N\times d_{\sigma}$, \eqref{ReceivedPowerArray} can be approximated as:
\begin{equation}
    P_{\rm r}\approx  N^2\iint\limits_{A_{\rm a}}\frac{2P_{\rm t}}{\pi W(z)^2}\exp{\left(-2\frac{(x-x_{\rm a})^2+(y-y_{\rm a})^2}{W(z)^2}\right)} {\rm d}x{\rm d}y \approx N^2P_{\rm r,single},
\end{equation}
where $P_{\rm r,single}$ is the received power of a single VCSEL at the considered position. 

It should be noted that, in some applications, a lens might be used in front of the VCSEL arrays to separate beam spots at the receiver plane (see Figs. 3b, 3d and 3f of \cite{9217158}). In such cases, the total received power at the photodetector can be approximated by the received power of a single VCSEL. On the other hand, some applications necessitate the reduction of beam divergence by the use of a microlens array, where a separate lens is in front of each VCSEL element. A set of collimated beamlets are shown in Fig.~\ref{sub2:Fig-ArrayCollimated}. In such a case, the total received power depends on the area of receiver \cite{9217368}.

\begin{figure}[t]
		\centering  
		\subfloat[\label{sub1:FigArray-BeamProfile6}]{%
        \resizebox{.48\linewidth}{!}{\includegraphics{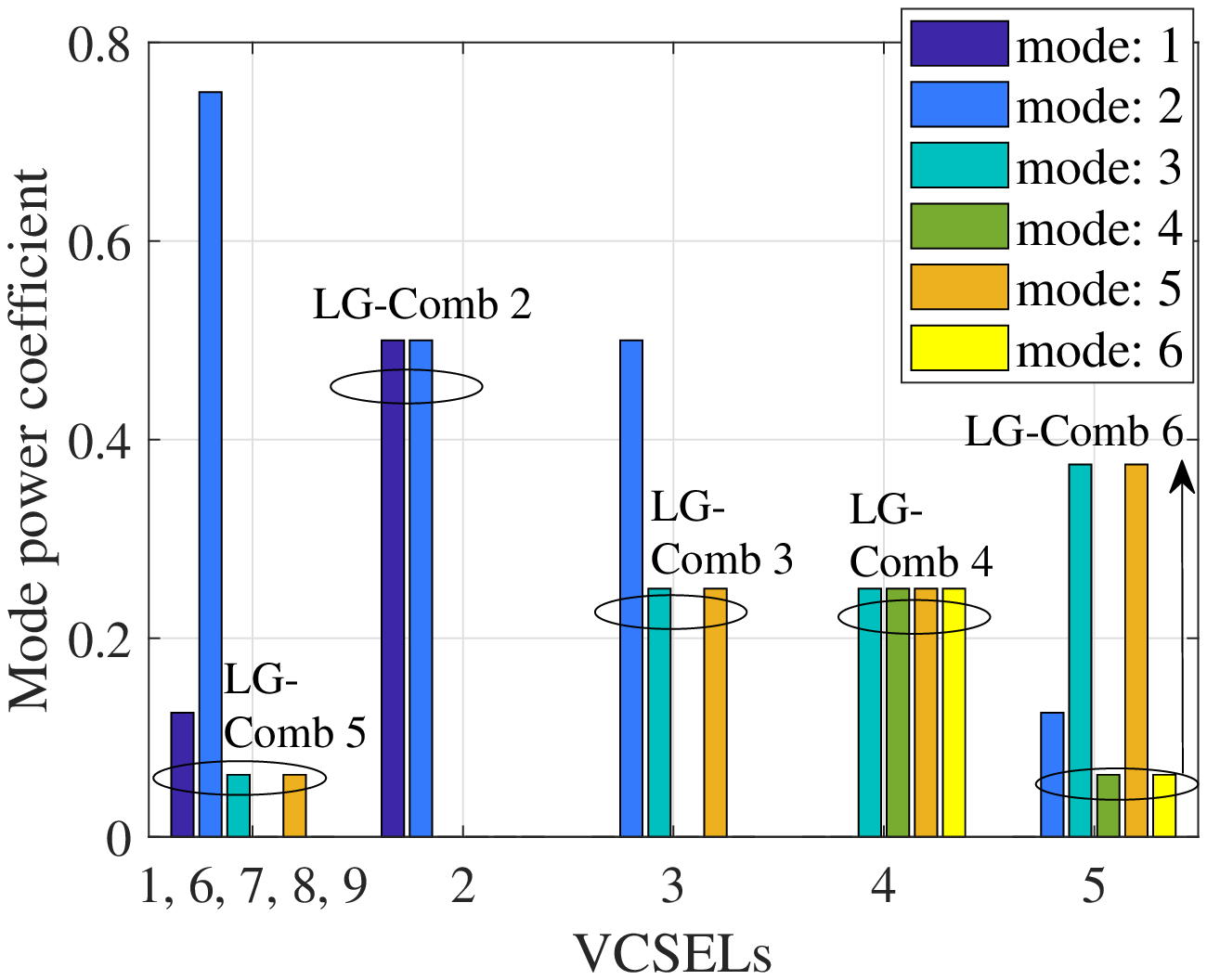}}
	}
	\subfloat[$z=0.1$ cm\label{sub1:FigArray-BeamProfile8}]{%
		\resizebox{.47\linewidth}{!}{\includegraphics{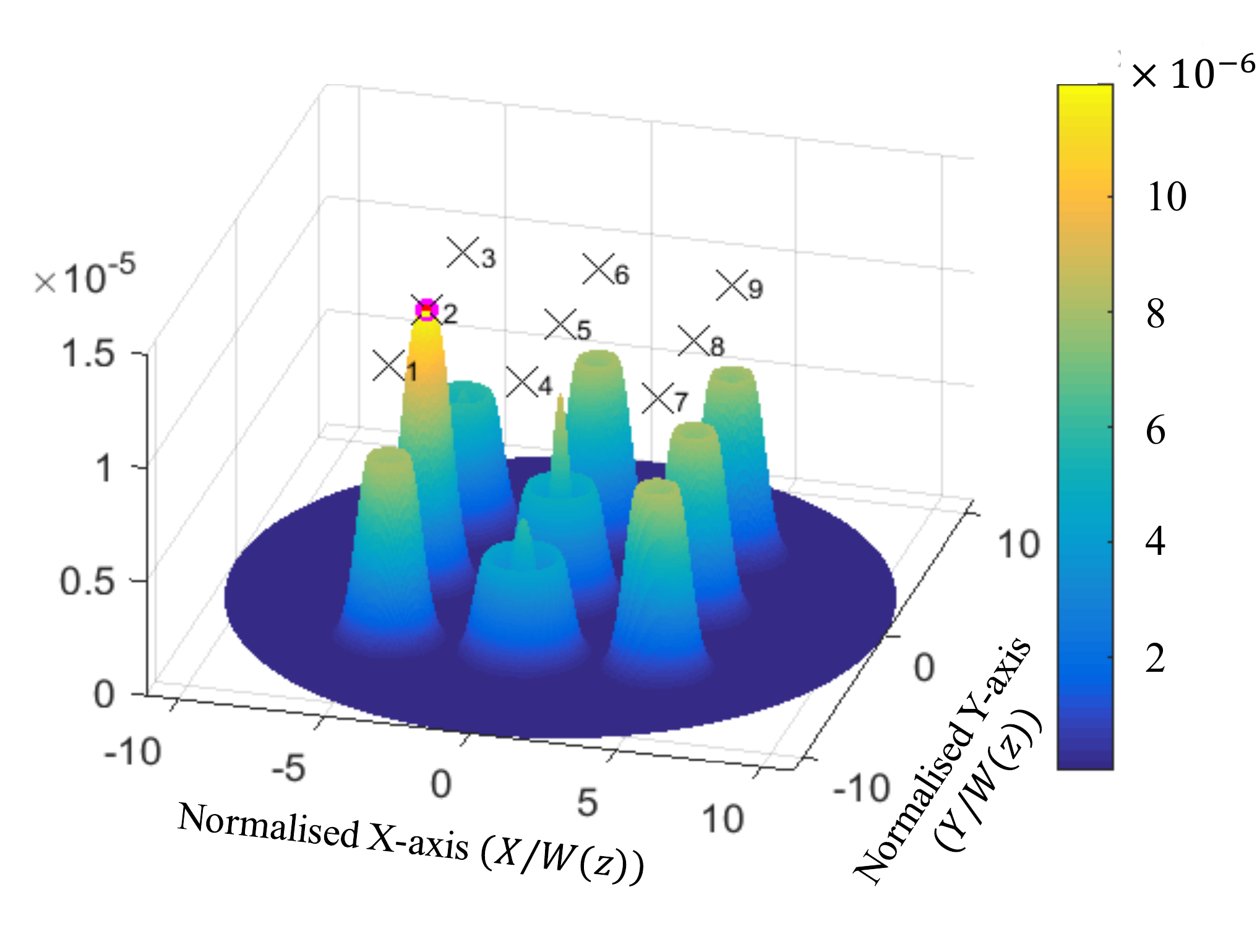}}
	} \\ 
	\subfloat[$z=1$ cm\label{sub2:FigArray-BeamProfile9}]{%
		\resizebox{.45\linewidth}{!}{\includegraphics{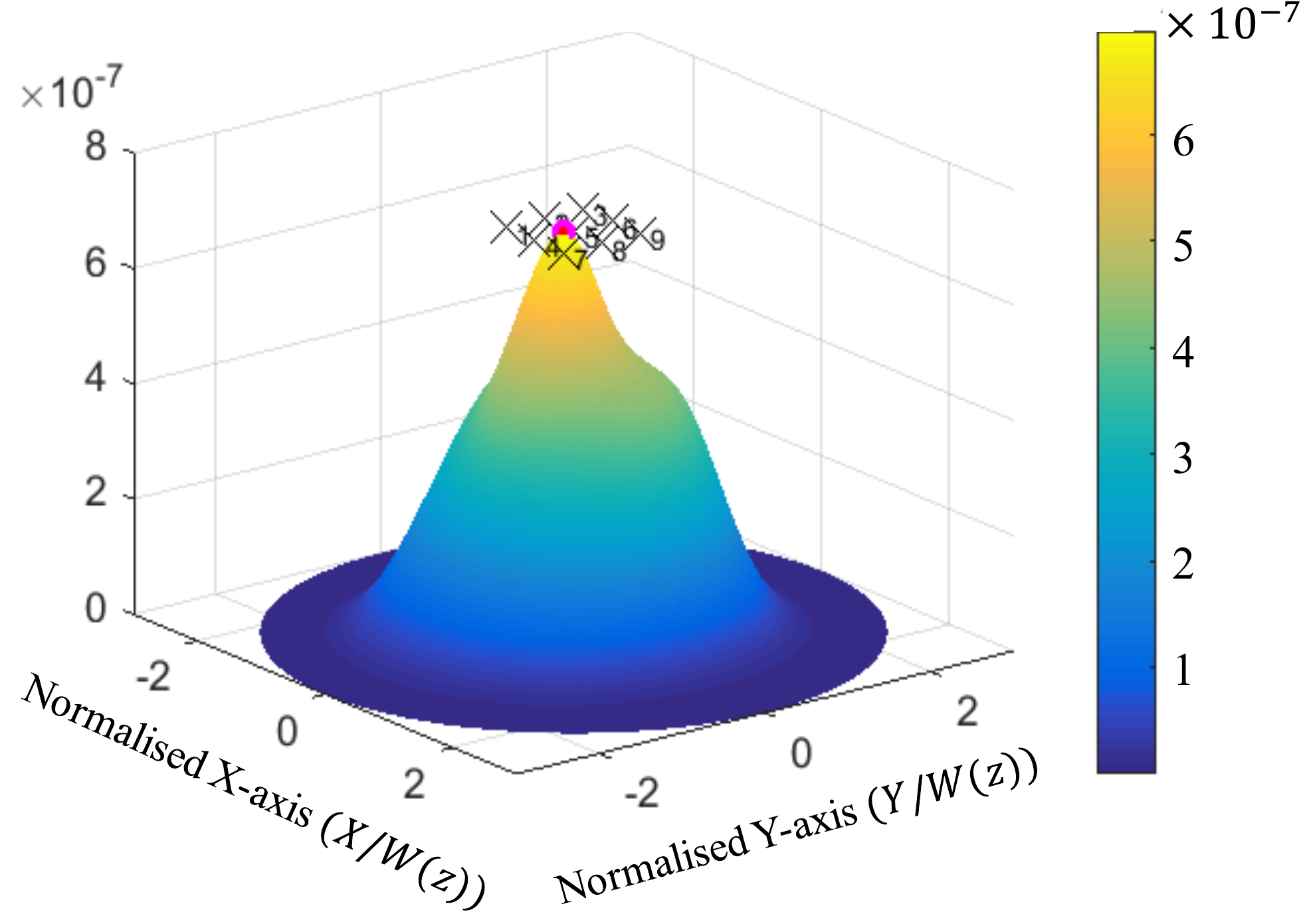}}
	}
		\subfloat[$z=10$ cm\label{sub1:FigArray-BeamProfile10}]{%
		\resizebox{.47\linewidth}{!}{\includegraphics{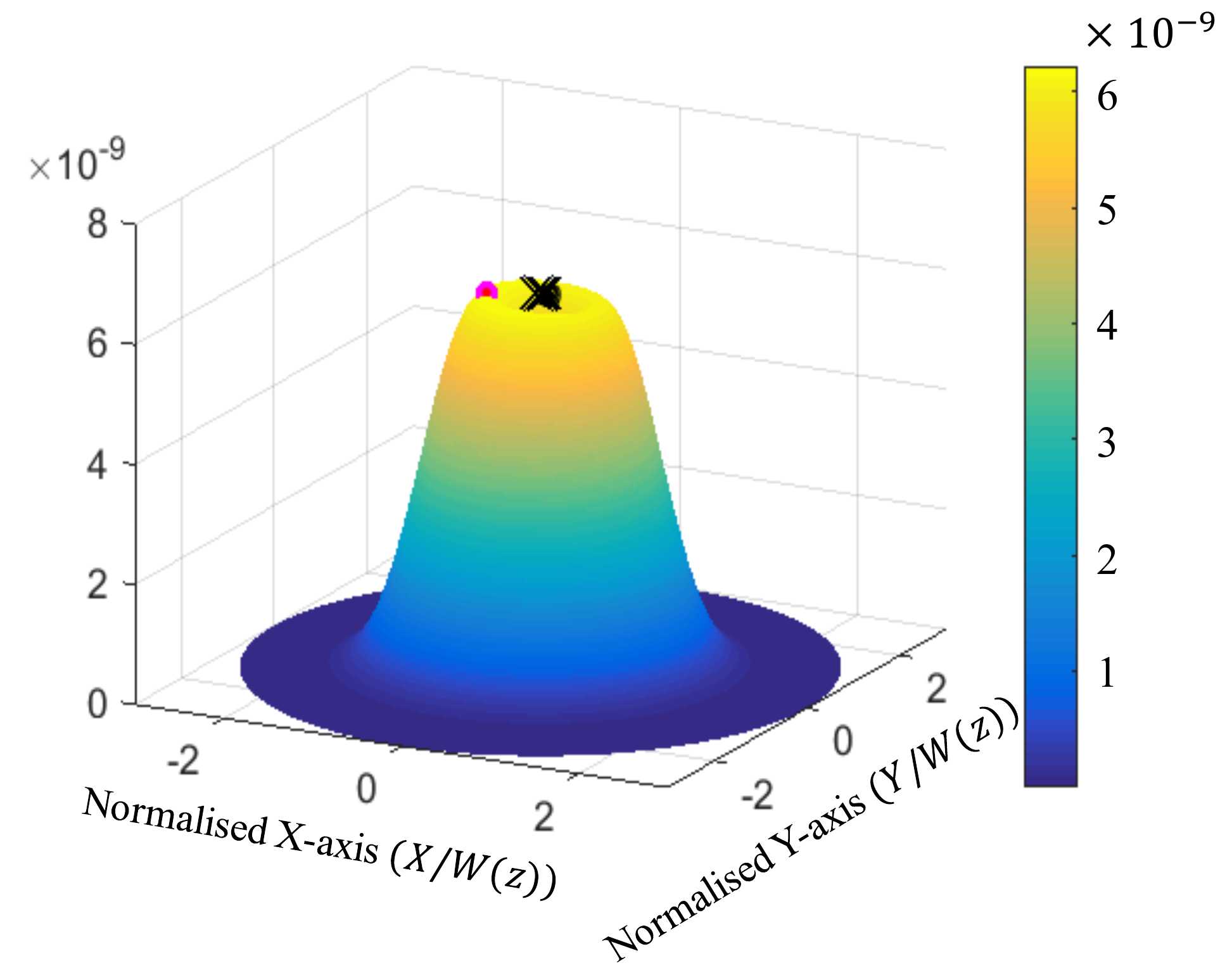}}
	} \quad\ 
	\\
	\caption{Power profiles of a $3\times 3$ multimode VCSEL array at different distances for $w_0=5~\mu$m and $\lambda=850$~nm. The VCSELs are assumed to transmit the same output power. The mode power coefficients of the multimode VCSEL for each VCSEL is shown in the first sub-figure.}
	\label{Fig-Array-BeamProfile2}
\end{figure}



In Fig.~\ref{Fig-Array-BeamProfile2}, a $3 \times 3$ multimode VCSEL array is considered. For these results, the mode power coefficients of the different VCSELs are varied, while their output power is assumed to be the same. The beam waist of each VCSEL and the pitch distance are assumed to be $w_0=5 \ \mu$m and $d_{\sigma}=250 \ \mu$m, respectively. The total output power at different $z$-distances ($z=0.1, 1, 10$ cm) and the point of maximum irradiance are found and indicated with a purple dot in the plots. 

\begin{figure}[t!]
 \centering
\includegraphics[width=0.8\textwidth]{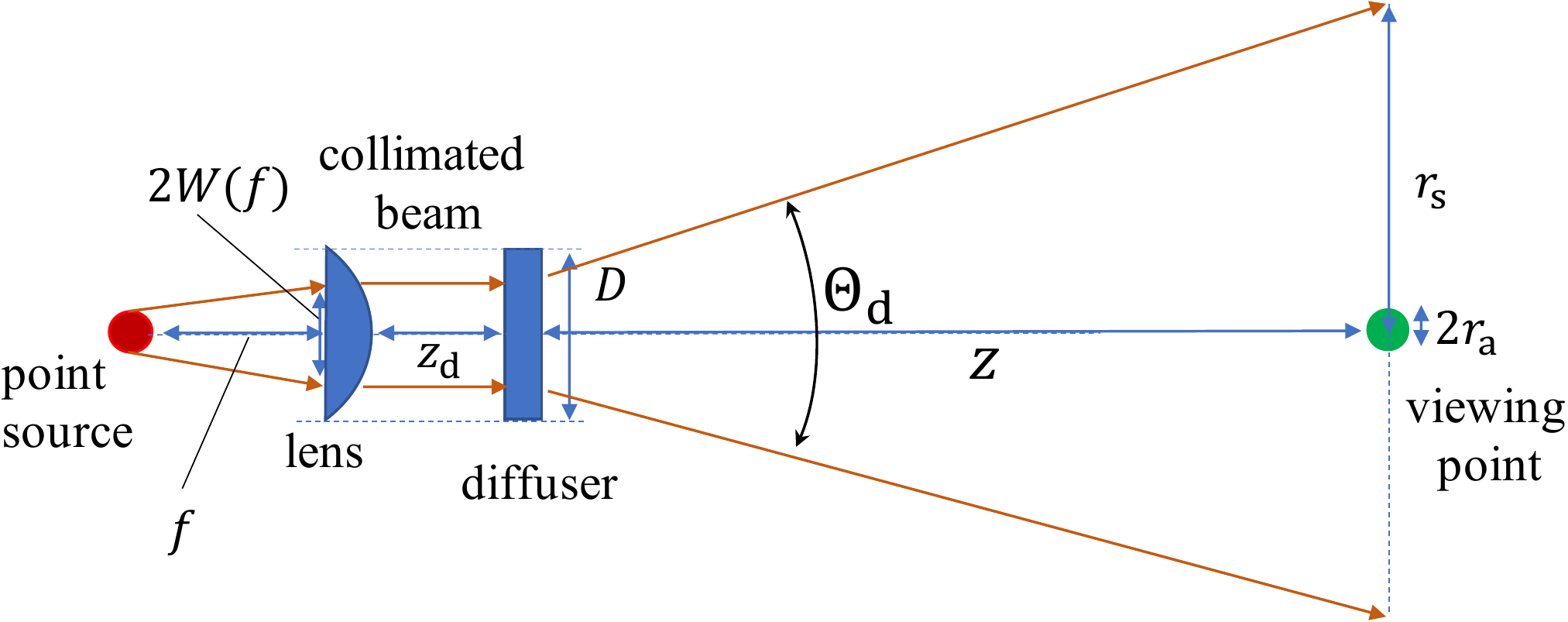} 
\caption{System model of a diffuser.}
\label{Fig-diffuser}
\end{figure}

\subsection{Beam Propagation through Diffusers}
A diffuser can be used to alter the intensity pattern of a laser beam. In this paper, we will consider two types of diffusers: the Lambertian diffuser, and the uniform diffuser\cite{DataSheetLambertian,DataSheetUniform}.
The system model of a diffuser is shown in Fig.~\ref{Fig-diffuser}. The input beam of a diffuser should be a collimated beam, therefore, a lens is required before the diffuser. The laser source is located at the focal point of the lens and a collimated beam is generated. The distance between the lens and the diffuser is denoted by $z_{\rm d}$. The diameter and the FWHM angle\footnote{The FWHM angle is often reported in the datasheets for practical diffusers.} of the diffuser are denoted by $D$ and $\Theta_{\rm d}$, respectively. 
The received power at a distance of $z$ from the diffuser can be obtained as:
\begin{equation}
\label{GeneralReceivedPower}
    P_{\rm r}=\iint\limits_{\Omega_{\rm p}} I(\theta,\phi)\sin(\theta){\rm d}\theta{\rm d}\phi,
\end{equation}
where $I(\theta,\phi)$ is the intensity in W/Sr (i.e. Watt per steradian)\footnote{The steradian (Sr) is the unit of a solid angle which has its vertex at the center of a sphere. The solid angle cuts off an area of the surface of the sphere equal to that of a square with side lengths equal to the radius of the sphere.}, $\theta$ and $\phi$ are the polar and azimuthal angles, respectively, in the spherical coordinate system. Note that the integral is taken over the solid angle, $\Omega_{\rm p}$, defined by the area of relevant aperture at a given distance from the source. In the following, we study the Lambertian and uniform diffusers.

\subsubsection{Lambertian Pattern Engineered Diffusers}
\hfill\\
The Lambertian diffusers are available with various diffusing angles (from $0.5^{\circ}$ to $50^{\circ}$ at FWHM angle) to control the diffuse area of illumination and increase the transmission efficiency. 
 It is important to note that diffusing angles are given for a collimated input beam and angular divergence will vary for different incidence angles.
The intensity of a Lambertian diffuser depends only upon the azimuthal angle, as it is symmetric around the polar angle$,\phi$, that is $I(\theta,\phi)$ reduces to $I(\theta)$. 
The radiation pattern of a generalized Lambertian diffuser of order $m$ is given by \cite{Kahn97WirelessInfrared,DimitrovJSAC2009}:
\begin{equation}
\label{LambertianIntensity}
    I(\theta)=\frac{m+1}{2\pi}\cos^m(\theta)P_{\rm t},
\end{equation}
where $P_{\rm t}$ is the transmit optical power. Substituting \eqref{LambertianIntensity} into \eqref{GeneralReceivedPower} and calculating the integral, the received power is obtained as:
\begin{equation}
\label{EqReceivedPowerLambertianDiffuser}
P_{\rm r}=P_{\rm t}\int_{0}^{2\pi}\int_{0}^{\psi_{\rm c}} \frac{m+1}{2\pi}\cos^m(\theta)\sin(\theta){\rm d}\theta{\rm d}\phi=(1-\cos^{m+1}(\psi_{\rm c}))P_{\rm t},
\end {equation}
where, $\psi_{\rm c}=\tan^{-1}(\frac{r_{\rm{a}}}{z})$ where $z$ is the viewing distance and $r_{\rm{a}}$ is the radius of the aperture (see Fig.~\ref{Fig-diffuser}).

\begin{figure}[t]
		\centering  
	\subfloat[$\Theta_{\rm d}=20^\circ$\label{sub1:Fig1}]{%
		\resizebox{.46\linewidth}{!}{\includegraphics{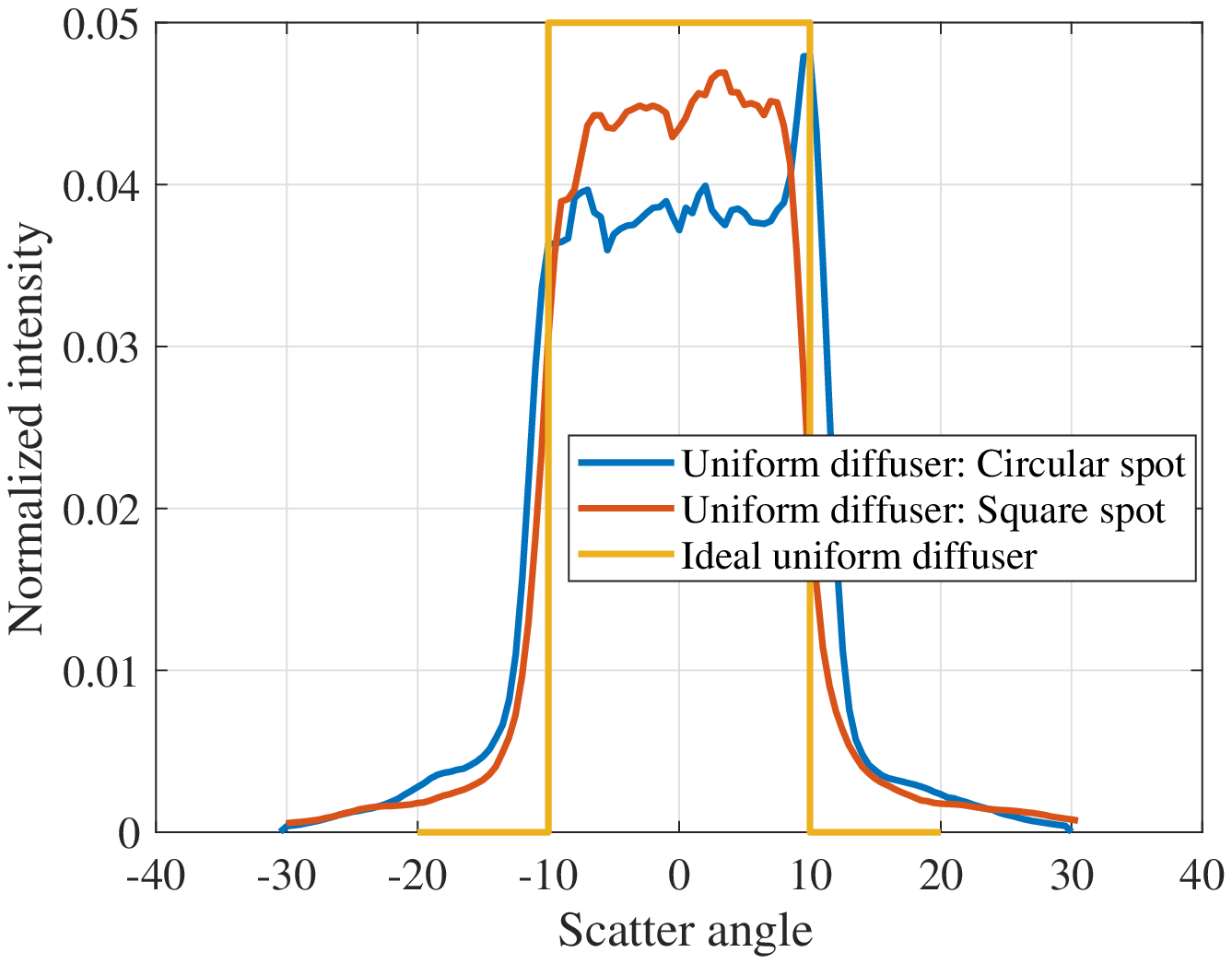}}
	} \quad\ \ 
	\subfloat[$\Theta_{\rm d}=50^\circ$\label{sub2:Fig2}]{%
		\resizebox{.45\linewidth}{!}{\includegraphics{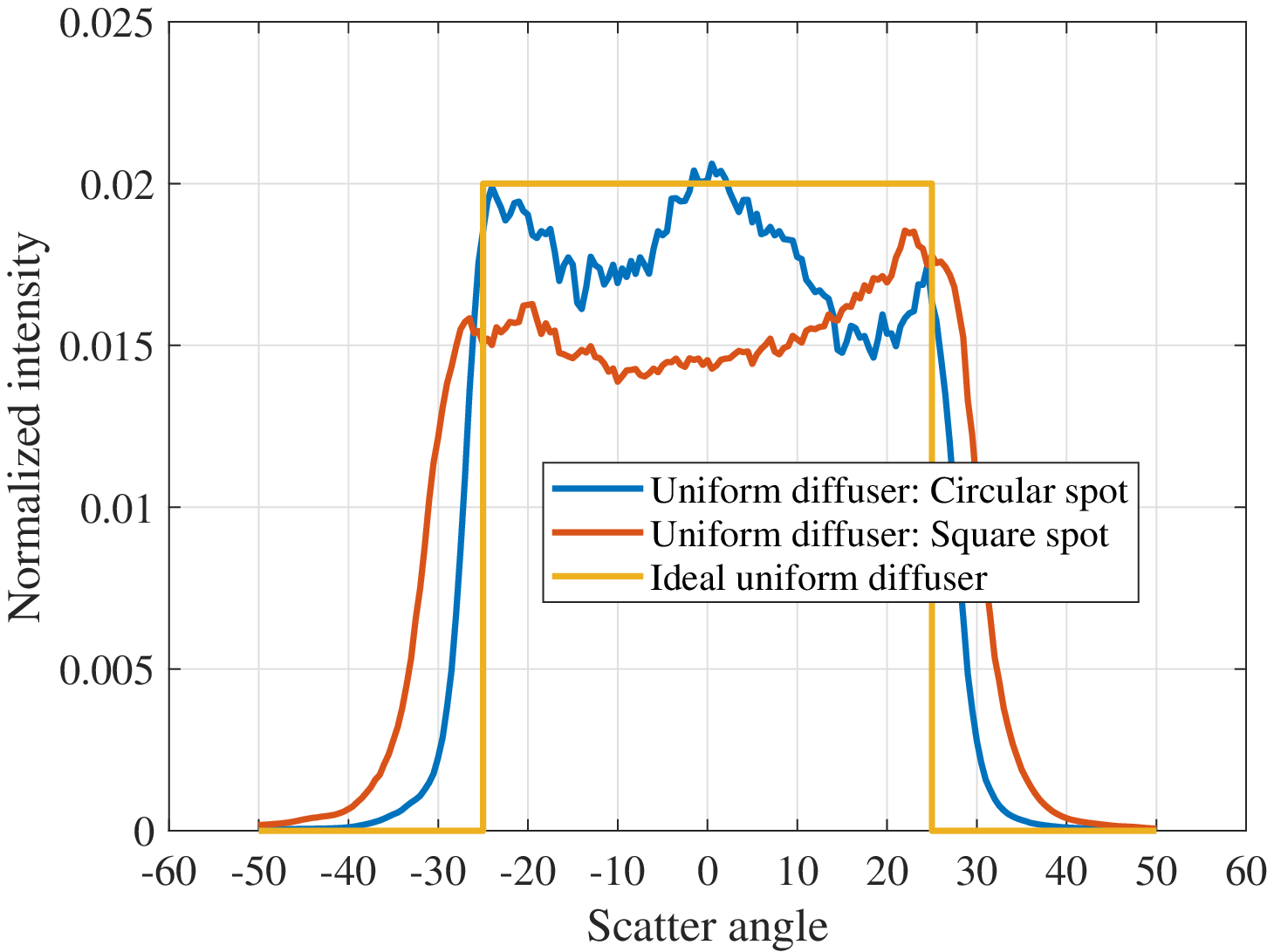}}
	}\\
	\caption{Normalized intensity versus scatter angle of uniform diffusers with $\lambda = 785$ nm \cite{DataSheetUniform}. The normalization is based on the total intensity.}
	\label{Fig-Diffuser-measurement}
\end{figure}

\subsubsection{Uniform Pattern Engineered Diffusers}
\hfill\\ 
Uniform diffusers are designed to create non-Gaussian intensity distributions typically in circular, square or hexagonal shapes. 
The FWHM, $\Theta_{\rm d}$, of practical diffusers ranges from several milliradian to $1.4$ rad ($80^\circ$).  
As an example, the normalized irradiance profile of two diffusers with of $\Theta_{\rm d}=20^{\circ}$ and $\Theta_{\rm d}=50^{\circ}$ are shown in Fig.~\ref{Fig-Diffuser-measurement} and compared with an ideal diffuser \cite{DataSheetUniform}. As it can be seen, the normalized intensity, particularly for square pattern diffusers, is almost the same for all scatter angles. Hence, we can assume that the irradiance profile is independent of the scatter angle. Another point is that the maximum transmit power based on the assumption of an ideal diffuser is the worst case, and therefore, the eye safety constraint on the transmit power for practical diffusers (with the irradiance profile shown in Fig.~\ref{Fig-Diffuser-measurement}) would be relaxed to some degree. 

The solid angle of a cone whose cross-section subtends the angle $\Theta_{\rm d}$ is $\Omega=2\pi(1-\cos(\Theta_{\rm d}/2))$. Therefore, for a circularly uniform diffuser with a FWHM of $\Theta_{\rm d}$, the irradiance is given as:
\begin{equation}
\label{IntensityUniform}
    I=\frac{P_{\rm t}}{2\pi(1-\cos(\Theta_{\rm d}/2))}. 
\end{equation}
It can be seen that the irradiance presented in the above equation is independent of the polar and azimuthal angles\footnote{It can be readily justified that $\int_{0}^{2\pi}\int_{0}^{\Theta_{\rm d}/2} I\sin(\theta){\rm d}\theta{\rm d}\phi=P_{\rm t}$.}. 
Substituting the above equation in \eqref{GeneralReceivedPower} and calculating the integral over $\phi$ and $\theta$ from $0$ to $2\pi$ and from $0$ to $\psi_{\rm c}$, respectively, the received power in the pupil of the eye is obtained as:
\begin{equation}
\label{PrUniformDiffuser}
    P_{\rm r}=\frac{P_{\rm t}}{{(1-\cos(\Theta_{\rm d}/2))}}(1-\cos(\psi_{\rm c})).
\end{equation}
In the next Section, we present eye safety evaluations for the discussed beam propagation models.

\section{Eye Safety Analysis}
\label{Sec: eye safety analysis}
The hazards from laser sources can best be classified into four categories: optical radiation hazards to the eye and skin; chemical hazards; electrical hazards; and casualty hazards\footnote{Chemical hazards are essentially due to vaporization of material, metal oxides and other fumes. Electrical hazards refer to electrical shock and possibly resulting in electrocution. Casualty hazards might be due to several reasons like causing an accidental fire by means of lasers.}. Eye hazards are by far the most significant \cite{sliney2013safety}. Hence, we mainly focus on the analysis of eye safety under different laser beam propagation conditions (single mode, multimode, with lens, array of VCSELs and with diffusers) and skin safety  will be discussed at the end of this section.

\subsection{Safety Analysis for Single Mode laser}
For an ideal single-mode (Gaussian) beam, the maximum received power is at the center of the beam. Therefore, the maximum transmit power to ensure eye safety is calculated based on the received power into the eye pupil (the pupil can be as large as 6–7 mm when it is wide open, which translates into the maximum physical aperture \cite{atchison2000optics}) located in the center of the beam at the most hazardous distance from the laser source (see Fig.~\ref{Fig-SubtenseAngle}). 

%
Substantially, the maximum transmit power can be obtained based on the MPE defined by ANSI standard \cite{ANSI2014}. In order to fulfill eye safety regulations, the following condition should be met:
\begin{equation}
\label{irradianceCondition}
    \frac{P_{\rm r}}{\pi r_{\rm p}^2}<{\rm MPE},
\end{equation}
where $P_{\rm r}$ is the fraction of the power ($86\%$\footnote{This metric here is more conservative in comparison to the $63\%$ by \cite{IEC_Std608251} and $72\%$ adopted in \cite{henderson2003laser}}) into the eye at the MHP, $z_{\rm haz}$ and $r_{\rm p}$ is the cornea radius. Based on \eqref{intensityEq}, the maximum transmit is given as:
\begin{equation}
    P_{\rm t,max}=\frac{1}{\eta}{\pi r_{\rm p}^2}\times{\rm MPE},
\end{equation}
where $\eta=\left(1-\exp{\left( -\dfrac{2r_{\rm p}^2}{W^2(z_{\rm haz})} \right )}\right)$.

The subtense angle is required in order to specify the MPE. The subtence angle can be determined based on the distance between the eye and the source, and the beam waist of the laser. Let denote the distance for which the $86\%$ of the power passes into a $7$ mm pupil by $d_{86}$. If $d_{86}$ is less than $0.1$ m, then the MHP is set to $0.1$ m. For this case, we have $\alpha=2\tan^{-1}({w_0}/{z_{\rm{haz}}})$. Otherwise, $\alpha=2\tan^{-1}({w_0}/{d_{86}})$. Next, the MPE can be calculated according to the operating wavelength, exposure duration and the calculated subtense angle from Table~\ref{table_eye_safety}. The approach to obtain the maximum transmit power that satisfies eye safety regulations is summarized in algorithm~\ref{Algo-singlemodelaser}. It should be mentioned that this algorithm covers the analysis for both point and extended sources. There is only one consideration in reading the MPE from the Table~\ref{table_eye_safety}. The MPE can be distinguished based on the subtense angle for point and extended sources.

\begin{algorithm}[t]
 \caption{Determining maximum transmit power of a single mode laser source.}
 \label{Algo-singlemodelaser}
 \begin{algorithmic}[1]
 \renewcommand{\algorithmicrequire}{\textbf{Input}:}
 \renewcommand{\algorithmicensure}{\textbf{Output}:}
 \REQUIRE  Denote the wavelength of the laser source by $\lambda$; the beam waist by $w_0$;
 the radius of the pupil by $r_{\rm p}$; exposure duration by $t_{\rm{ex}}$.    
 \ENSURE  Maximum transmit power, $P_{\rm{t,max}}$ of the laser source.
 \STATE Compute $d_{86}=\dfrac{\pi w_0}{\lambda}\sqrt{\dfrac{-2r_{\rm p}^2}{\ln{(1-\varepsilon_{86})}}}$, where $\varepsilon_{86}=0.86$
 \IF{$2\theta>92$ mrad AND $d_{86}<0.1$}
 \STATE Set $z_{\rm{haz}}=0.1$
 \STATE Compute $\alpha=2\tan^{-1}\left(\dfrac{w_0}{z_{\rm{haz}}}\right)=2\tan^{-1}(10w_0)$
 \ELSE
 \STATE Compute $\alpha=2\tan^{-1}\left(\dfrac{w_0}{d_{86}}\right)$
 \IF{$\alpha<\alpha_{\rm min}$}
 \STATE Set $z_{\rm{haz}}=\dfrac{w_0}{0.0021}$
 \ELSE
 \STATE Set $z_{\rm{haz}}=\dfrac{9.2\pi w_0}{2\lambda}$
 \ENDIF
 \ENDIF
 \STATE Compute $\eta=1-\exp{\left(\dfrac{-2r_{\rm p}^2(\pi w_0^2)^2}{(\lambda z_{\rm{haz}})^2}\right)}$
 \STATE Calculated the MPE based on the given wavelength, exposure duration, $t_{\rm{ex}}$, and the calculated $\alpha$ from Table~\ref{table_eye_safety}. 
 \STATE Calculate $P_{\rm{t,max}}=\dfrac{1}{\eta}({\rm MPE})\pi r_{\rm p}^2$ using the calculated MPE
\RETURN $P_{\rm{t,max}}$ as the maximum transmit power of the laser source
\end{algorithmic}
\end{algorithm}

Fig.~\ref{sub1:Fig1} shows the maximum transmit power versus exposure duration for a point source. Two wavelengths of $\lambda=850$ nm and $\lambda=950$ nm are considered and it is assumed that $w_0=5\ \mu$m. As shown in the results, the maximum transmit power increases as the wavelength of the laser source increases, generating an increment of $1.14$ mW between operating wavelengths of $850$ nm and $950$ nm for exposure durations higher than $10$ s. As shown in these simulations, $P_{\rm t,max}$ highly depends on exposure durations of less than $10$ s while it is constant for exposure durations of higher than $10$ s.  Fig.~\ref{sub2:Fig2} illustrates $P_{\rm t,max}$ versus beam divergence angle, $\theta$, of a laser source with $w_0=5\ \mu$m assuming an exposure duration greater than $10$ s. As shown in this set of results, the maximum transmit power is constant up to the divergence angle of $2.6^{\circ}$ and then after that, the maximum transmit power increases as the beam divergence increases. This is due to the reduction of the amount of power passing into the eye as the beam divergence increases, which leads to a more relaxed condition on the transmit power limitation. 

Fig.~\ref{Fig-PtmaxversusLambda} presents the results of $P_{\rm t,max}$ versus the wavelength of the laser source assuming an exposure duration greater than $10$ s. 
Results are provided for three practical ranges of the beam waist. It can be seen from these simulation results that as the operating wavelength of a laser source increases, the maximum transmit power increases as well. In particular, the maximum transmit power increases exponentially beyond wavelength $1050$ nm and up to the wavelength $1200$ nm, where the eye safety constraint is much more relaxed in comparison to lower wavelengths. For wavelengths higher than $1200$ nm, $P_{\rm t,max}$ increases linearly as the wavelength increases. The other observation is related to the relation between the beam waist and the $P_{\rm t,max}$. It can be inferred from this set of results that the $P_{\rm t,max}$ reduces as the beam waist increases. This is due to the fact that the laser beam is less divergent for a larger beam waist.

\begin{figure}[t!]
		\centering  
	\subfloat[\label{sub1:Fig1}]{%
		\resizebox{.47\linewidth}{!}{\includegraphics{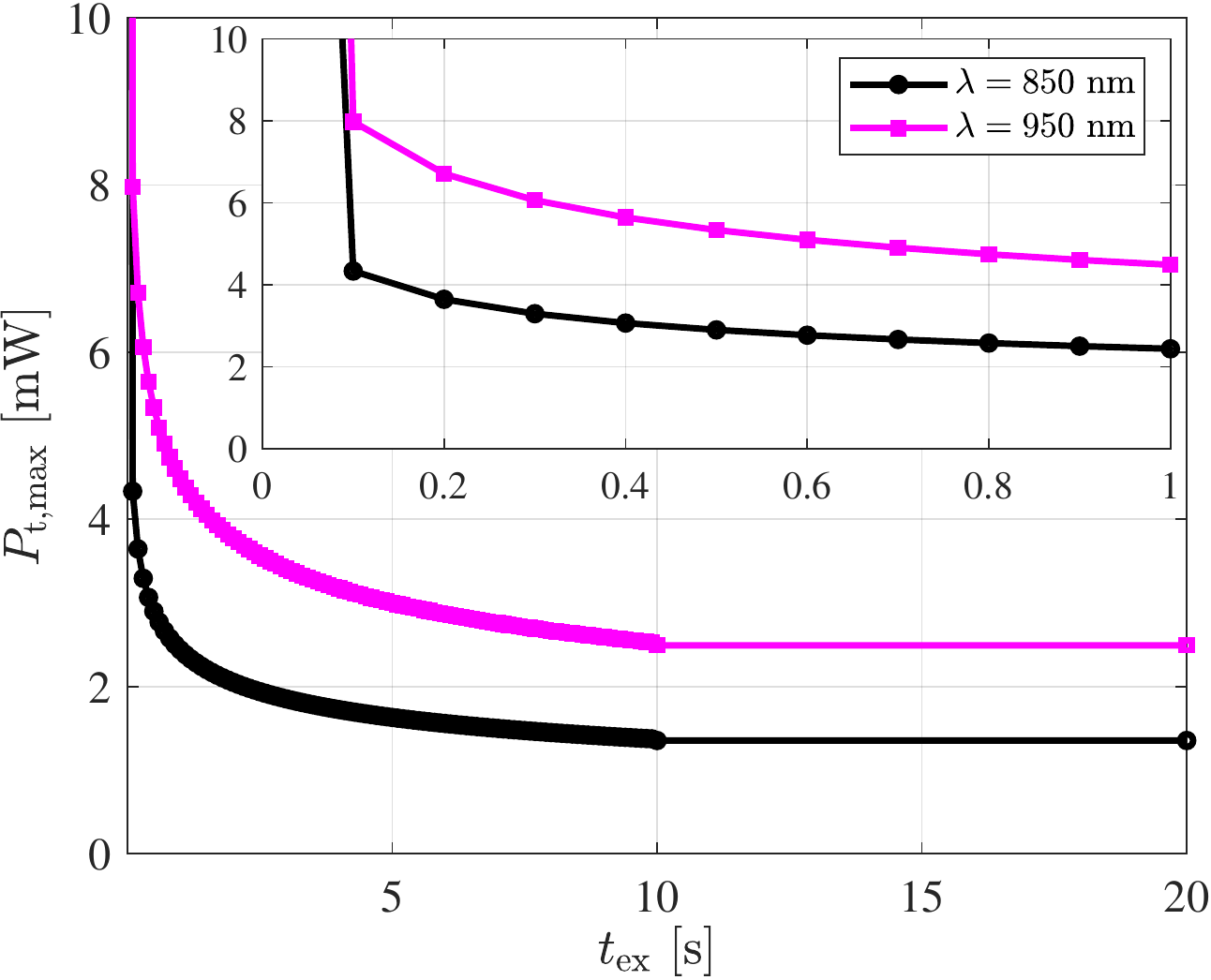}}
	} \quad\ \ 
	\subfloat[\label{sub2:Fig2}]{%
		\resizebox{.47\linewidth}{!}{\includegraphics{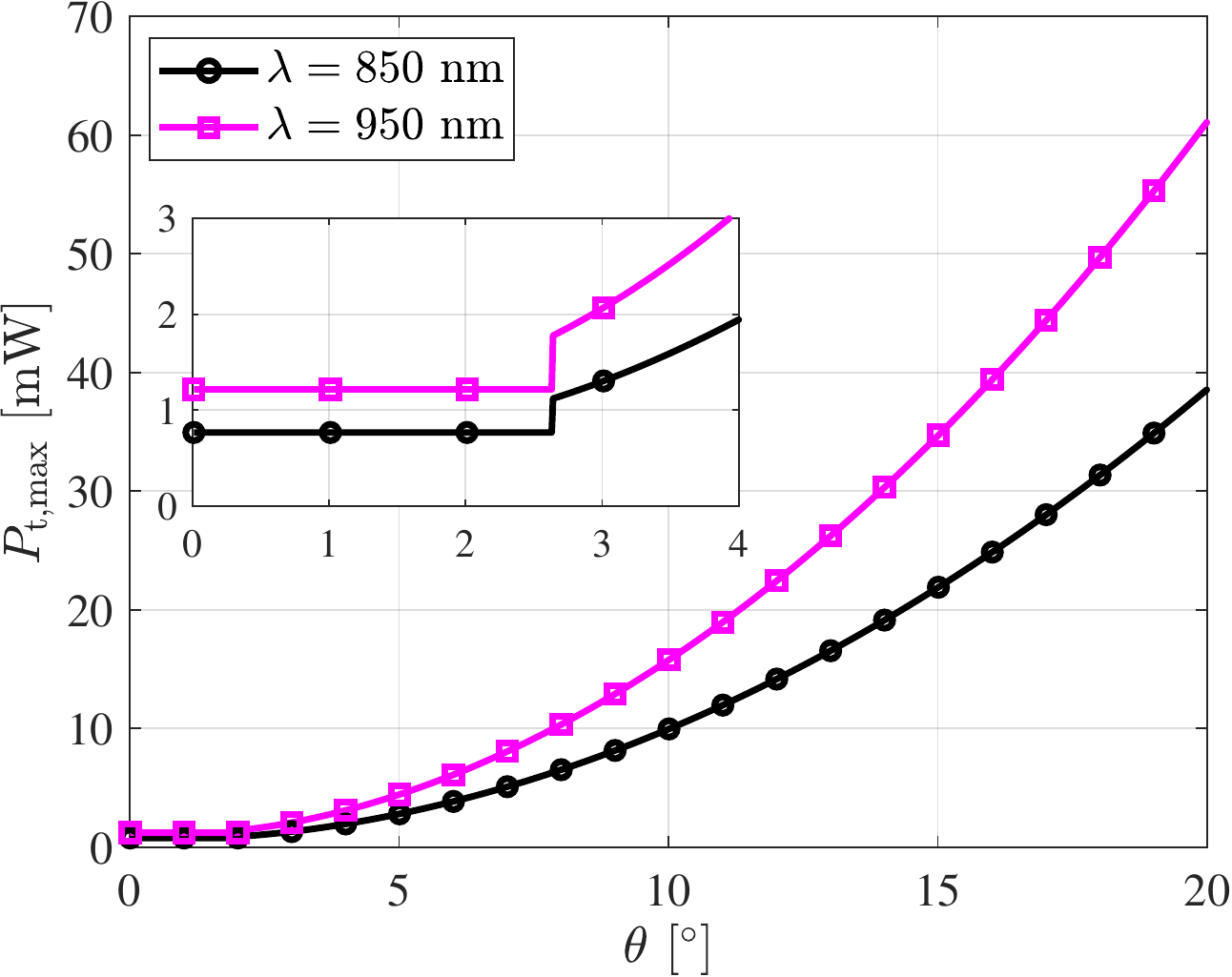}}
	}\\
	\caption{$P_{\rm{t,max}}$ versus (a) exposure duration and (b) divergence angle, $\theta$.}
	\label{Fig-MPE-Exposure}
\end{figure}

\begin{figure}[t!]
 \centering
\includegraphics[width=0.6\textwidth]{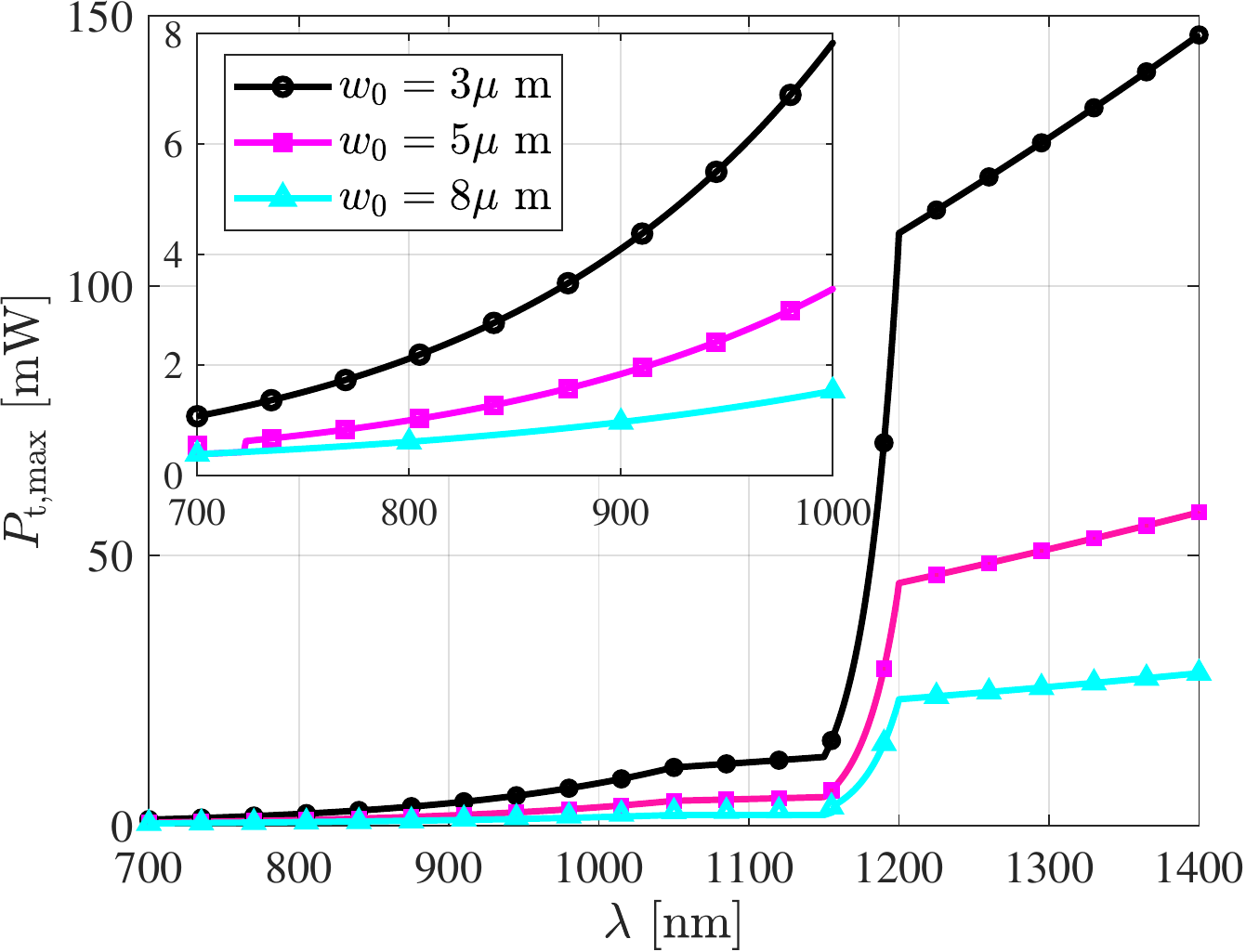} \caption{$P_{\rm t,max}$ versus $\lambda$ assuming an exposure duration greater than $10$ s. }
\label{Fig-PtmaxversusLambda}
\end{figure}

\subsection{Safety Analysis for Multimode Lasers}
\label{SafetyAnalysisMultiMode}
Eye safety analysis of multimode laser sources is normally a complicated task. An accurate approach is based on the beam profile of such laser, which can be obtained through experimental measurements. Once, the beam profile is measured, the maximum power that passes through an aperture of diameter $7$ mm can be specified. Then, based on this calculated irradiance, the maximum transmit power to ensure eye safety can be determined. 
Evaluating the received power passes through an aperture of radius $r_\mathrm{p}$ centered at ($x_0$, $y_0$) is a case of calculating the appropriate integral of the intensity. Assuming that the multimode laser beam is decomposed into a set of Hermite-Gaussian modes, the power that passes through the considered aperture can be specified as: 
\begin{equation}
\label{HermiteGaussianPrEq}
\begin{aligned}
P_\mathrm{r}\left(x,y,z\right)&=P_{\rm t} \sum_{l=0}^{L_{\rm max}}\sum_{m=0}^{M_{\rm max}}C_{l,m}\int_{x_0-r_\mathrm{p}}^{x_0+r_\mathrm{p}}\int_{y_0-\sqrt{r_\mathrm{p}^2-\left(x-x_0\right)^2}}^{y_0+\sqrt{r_\mathrm{p}^2-\left(x-x_0\right)^2}}|U_{l,m}\left(x,y,z\right)|^2\ {\rm d}x\ {\rm d}y .
\end{aligned}
\end{equation}
For the purposes of the eye safety analysis, this double integral should be calculated at the point at which the maximum power is gathered by the aperture. 
We assume that the integral evaluated around the position of peak irradiance ($x_0$, $y_0$) will provide the maximum (or close to the maximum) received power. The coordinates of the maximum point can be found by solving the following set of equations at $z=z_{\rm haz}$: 
\begin{equation}
\label{HermiteGaussianMaxX}
\begin{aligned}
\frac{\partial}{\partial x}{\left(\sum_{l=0}^{L_{\rm max}}\sum_{m=0}^{M_{\rm max}}C_{l,m}|U_{l,m}\left(x,y,z\right)|^2\right)}=0 \\
\frac{\partial}{\partial y}\left(\sum_{l=0}^{L_{\rm max}}\sum_{m=0}^{M_{\rm max}}C_{l,m}|U_{l,m}\left(x,y,z\right)|^2\right)=0 .
\end{aligned}
\end{equation}
Setting $r_x=\frac{x\sqrt2}{W_x\left(z\right)}$ and $r_y=\frac{y\sqrt2}{W_y\left(z\right)}$, the coordinates of the peak irradiance will be the roots of the polynomials defined by the following equations: 
\begin{subequations}
\label{HermiteGaussianRootsX}
\begin{align}
\label{HermiteGaussianRootsX-a}
\sum_{l=0}^{L_{\rm max}}\sum_{m=0}^{M_{\rm max}}C_{l,m}A_{l,m}\mathbb{H}_m^2\left(r_y\right)\mathbb{H}_l\left(r_x\right)\left[2l\mathbb{H}_{l-1}\left(r_x\right)-r_x\mathbb{H}_l\left(r_x\right)\right]=0 \\
\label{HermiteGaussianRootsX-b}
\sum_{l=0}^{L_{\rm max}}\sum_{m=0}^{M_{\rm max}}C_{l,m}A_{l,m}\mathbb{H}_l^2\left(r_x\right)\mathbb{H}_m\left(r_y\right)\left[2m\mathbb{H}_{m-1}\left(r_y\right)-r_y\mathbb{H}_m\left(r_y\right)\right]=0 .
\end{align}
\end{subequations}
These roots can be calculated numerically if the mode weights (i.e., $C_{l,m}$) are known, and accordingly the received power can be evaluated around the peak intensities to calculate the maximum permissible transmit power.

As an example, let's consider the results shown in Fig.~\ref{Fig-HG-Com}. The coordinates of the maximum point are calculated based on \eqref{HermiteGaussianRootsX} and depicted by a purple dot in the subsets of Fig.~\ref{Fig-HG-Com}. The purple circle corresponds to an aperture of diameter $7$ mm. The incident power fraction passed through this circle located $10$ cm away from the laser source is calculated analytically for mode combinations $\#1-3$ to be respectively $0.14$, $0.144$ and $0.149$ of the total emitted power. The beam waist is assumed to be $5\ \mu$m. Since $\alpha=2\tan^{-1}\left(\frac{w_0}{z_{\rm haz}}\right)=0.2$ mrad less than $\alpha_{\rm min}$, then, the laser is treated as a point source. The MPE for such a laser operating at $850$ nm for the exposure duration of higher than 10 s, is $20$ W/m$^2$.
Hence, the maximum transmit powers are $5.5$ mW, $5.3$ mW and $5.2$ mW, respectively.

For a Laguerre-Gaussian beam, the power passes through an aperture of radius $r_\mathrm{p}$ centered at ($r_0$,$\theta_0$) is given by:
\begin{equation}
\label{Lag:Pr:SafetyAnalysis}
    P_\mathrm{r}\left(r,z\right)=P_{\rm t} \sum_{l=0}^{L_{\rm max}}\sum_{m=0}^{M_{\rm max}}C_{l,m} {A_{l,m}\int_{\theta_1}^{\theta_2}\int_{r_1}^{r_2}{\left|U_{l,m}\left(r,z\right)\right|^2r\ {\rm d}r\ {\rm d}\theta}},
\end{equation}
where $\theta_1$ and $\theta_2$ represent the maximum and minimum polar angles that span the aperture. Their values depend on the relative size of the aperture $r_{\rm p}$ with respect to its position $(r_0,\theta_0)$. The two cases and the equations to obtain $\theta_1$, $\theta_2$, $r_1\left(\theta\right)$ and $r_2\left(\theta\right)$ are provided below. \\
If $r_{\rm p}\leq r_0$,
\begin{align*}
r_1\left(\theta\right)&=r_0\cos{\left(\theta-\theta_0\right)}-\sqrt{r_\mathrm{p}^2-r_0^2\sin^2{\left(\theta-\theta_0\right)}}  &\theta_1&=\theta_0+\sin^{-1}{\left(-\frac{r_\mathrm{p}}{r_0}\right)} \\
r_2\left(\theta\right)&=r_0\cos{\left(\theta-\theta_0\right)}+\sqrt{r_\mathrm{p}^2-r_0^2\sin^2{\left(\theta-\theta_0\right)}}  &\theta_2&=\theta_0+\sin^{-1}{\left(\frac{r_\mathrm{p}}{r_0}\right)}. 
\end{align*}
If $r_{\rm p}> r_0$, 
\begin{align*}
r_1\left(\theta\right)&=0  &\theta_1&=0\\
r_2\left(\theta\right)&=r_0\cos{\left(\theta-\theta_0\right)}+\sqrt{r_\mathrm{p}^2-r_0^2\sin^2{\left(\theta-\theta_0\right)}}  &\theta_2&=2\pi. 
\end{align*}

Similarly, to fulfil the laser safety requirement, this double integral should be determined at the point at which the maximum power is gathered by the aperture. 
Again, we assume that the double integral evaluated around the position of peak irradiance will provide the maximum (or close to the maximum) received power. We can find the position of the peak irradiance by solving the following equation:
\begin{equation}
\frac{\partial}{\partial r}\left(\sum_{l=0}^{L_{\rm max}}\sum_{m=0}^{M_{\rm max}}C_{l,m}A_{l,m}|U_{l,m}\left(r,z\right)|^2\right)=0 .
\end{equation}
The roots of this solution can be obtained numerically and are the roots of the polynomial defined by the following equation:
\begin{multline}
\sum_{l=0}^{L_{\rm max}}{\sum_{m=0} ^{M_{\rm max}} 2{C_{l,m}}^2A_{l,m}\left(\frac{2}{W^2\left(z\right)}\right)^l\mathbb{L}_m^l\left(\frac{2r^2}{W^2\left(z\right)}\right)}\\
 \times\left[\mathbb{L}_m^l\left(\frac{2r^2}{W^2\left(z\right)}\right)\left\{lr^{2l-1}-\frac{{2r}^{2l+1}}{W^2\left(z\right)}\right\}-\frac{4r^{2l+1}}{W^2\left(z\right)}\mathbb{L}_{m-1}^{l+1}\left(\frac{2r^2}{W^2\left(z\right)}\right)\right]=0 .
 \end{multline}
Once this peak intensity position is found, then, we can  specify the power collected by an aperture with a radius of $r_{\rm p}$ by means of \eqref{Lag:Pr:SafetyAnalysis}, and comparing this power to the value permitted by the exposure limit defined by the standard.

\begin{algorithm}[t]
 \caption{Determining maximum transmit power of a multi mode laser source.}
 \label{Algo-multimodelaser}
 \begin{algorithmic}[1]
 \renewcommand{\algorithmicrequire}{\textbf{Input}:}
 \renewcommand{\algorithmicensure}{\textbf{Output}:}
 \REQUIRE  Denote the wavelength of the laser source by $\lambda$;\\
 the beam divergence of the actual beam by $\theta_{\rm R}$;\\
the radius of the pupil by $r_{\rm p}$;\\  exposure duration by $t_{\rm{ex}}$;\\
\ENSURE  Maximum transmit power, $P_{\rm{t,max}}$ of the multimode laser source
\STATE Calculate $\theta_{\rm em}$ and $w_{0,\rm em}$ of the embedded Gaussian beam:
\STATE $\theta_{\rm em}=\theta_{\rm R}$ 
\STATE $w_{0,\rm em}=\lambda/(\pi \theta_{\rm em})$
\STATE Use $\theta_{\rm em}$ and $w_{0,\rm em}$ in the equations of algorithm~\ref{Algo-singlemodelaser} instead of $\theta$ and $w_0$, respectively (start from step 1 and continue until step 16)
\RETURN $P_{\rm{t,max}}$ for the multimode laser
\end{algorithmic}
\end{algorithm}

The method described above is analytically complex, however, it can provide an accurate value for the maximum transmit power. 
To deal with such intricacy analyses of the multimode beams, the $M$-squared approximation is suggested as an alternative approach in the literature \cite{kardosh2004beam,Hall_2007,siegman1993defining,siegman1998maybe,ISO2003}. By fitting the multimode beam to the so-called embedded Gaussian beam based on the beam divergence angle, a straightforward framework to analyse eye safety for such practical multimode lasers can then be established. 
The embedded Gaussian beam does not necessarily have any physical reality, i.e., it does not necessarily indicate the mode $(0,0)$ of the actual beam. Once the parameters of the embedded Gaussian beam are obtained (the beam waist and divergence angle), these values can be used in algorithm~\ref{Algo-singlemodelaser} to identify the maximum transmit power for such non-Gaussian beams. The approach to obtain $P_{\rm{t,max}}$ for a multimode laser is summarized in algorithm~\ref{Algo-multimodelaser}. Next, we aim to assess the validity of this approach by comparing its results with the analytical beam decomposition ones.

\begin{figure}[t!]
		\centering  
	\subfloat[\label{sub1:Fig-LG-eyesafety}]{%
		\resizebox{.45\linewidth}{!}{\includegraphics{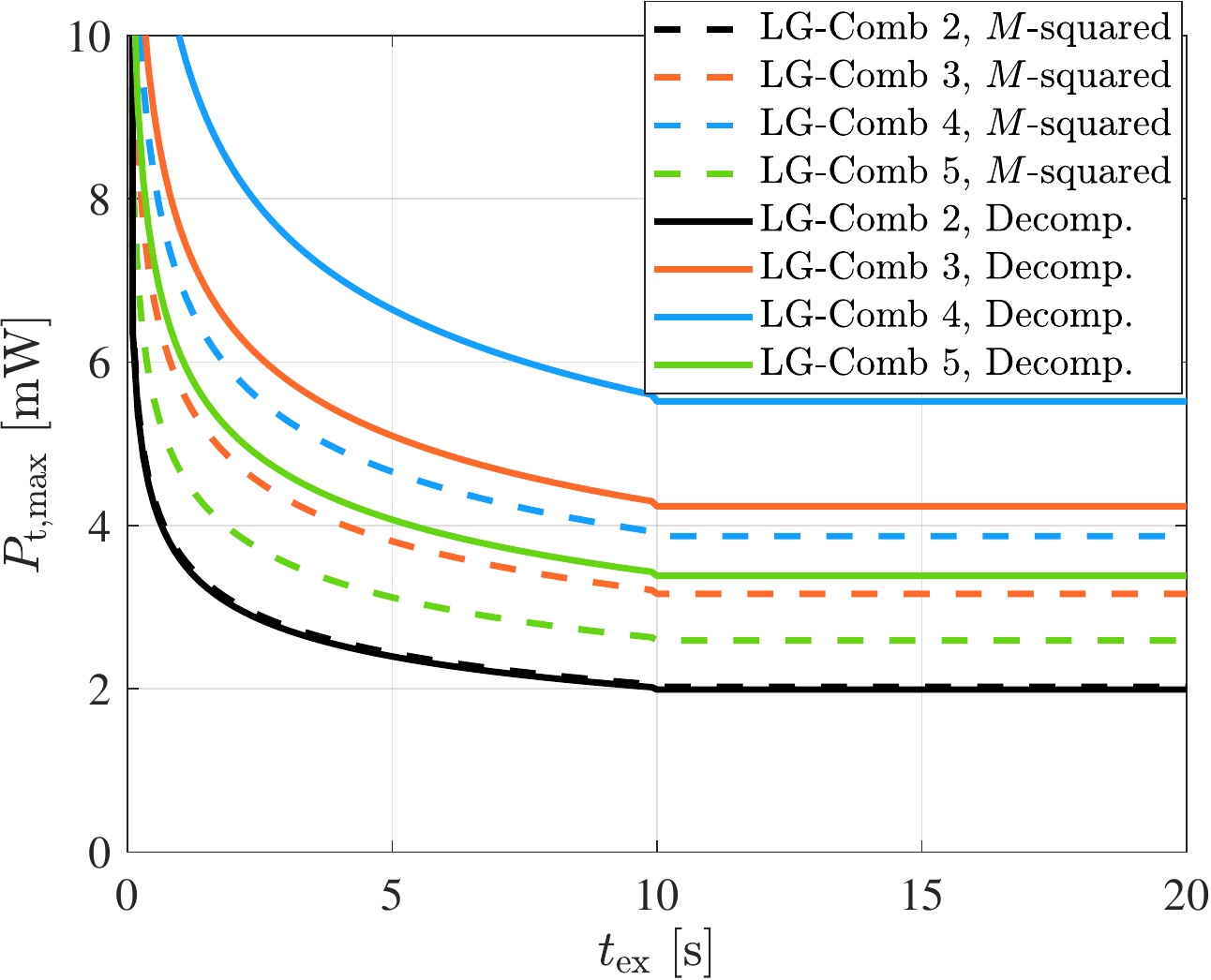}}
	} \quad\ \ 
	\subfloat[\label{sub2:Fig-LG-eyesafety}]{%
		\resizebox{.47\linewidth}{!}{\includegraphics{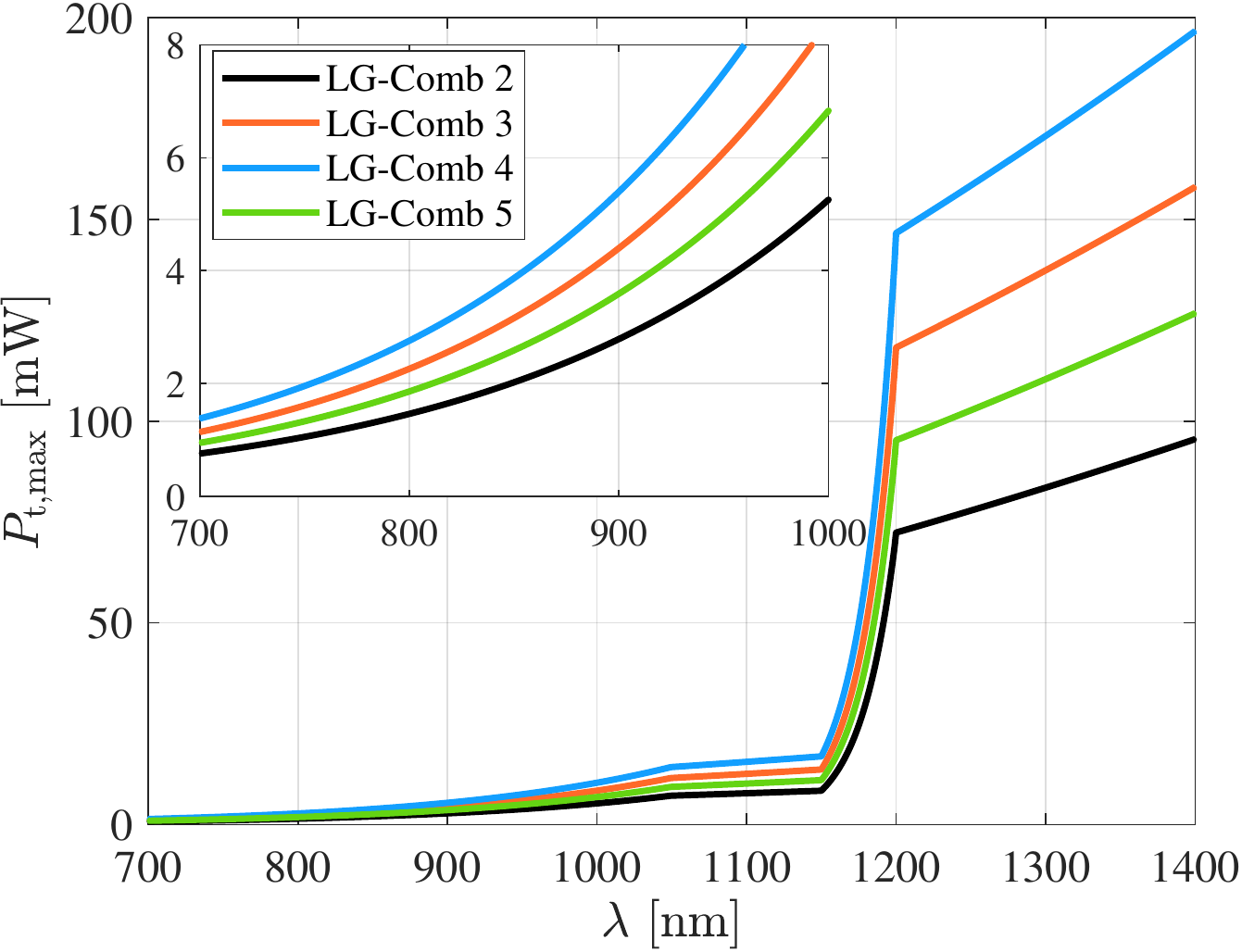}}
	}\\
	\caption{$P_{\rm{t,max}}$ versus (a) exposure duration and (b) wavelength for different modes of a Laguerre-Gaussian laser. The beam waist is assumed to be $w_0=5\ \mu$m.}
	\label{Fig-LG-eyesafety}
\end{figure}

Fig.~\ref{sub1:Fig-LG-eyesafety} shows results of $P_{\rm t,max}$ for a Laguerre-Gaussian laser versus the exposure duration assuming $\lambda=850$ nm and $w_0=5\ \mu$m. The simulation results are presented for different mode combinations of Laguerre-Gaussian beams given in Table~\ref{Table-ExpData2}. In these simulations, LG-Comb 6 and LG-Comb 7 are not chosen since their $P_{\rm t,max}$ are very close to the LG-Comb 3. The results are presented and compared for both $M$-squared and beam decomposition-based methods.  The beam decomposition approach is based on the calculation of the integral given in \eqref{Lag:Pr:SafetyAnalysis} at the maximum point of the beam. For the purpose of short representation, it is denoted as ``Decomp." in the legend of Fig.~\ref{sub1:Fig-LG-eyesafety}.
The outcomes indicate that for LG-Comb $2$ only, $M$-squared and beam decomposition methods perform similarly. For LG-Comb $3$, $4$ and $5$, the difference between the $M$-squared and beam decomposition techniques are $25\%$, $30\%$ and $23\%$, respectively, where $M$-squared is more conservative. In fact, a pure Gaussian beam has more pronounced peak leading to higher relative irradiance than the multimode beams. The values of $P_{\rm t,max}$ calculated by the beam decomposition method are in agreement with the IEC standard for eye safety, which means that the values of $P_{\rm t,max}$ obtained by the $M$-squared-based ensure the IEC regulations as well. However, one can use the precise beam decomposition method to relax the power budget limitations to some degree.
It can be also observed from these results that the fourth mode combination ensures the highest transmit power in comparison to other mode combinations. This is due to the fact that LG-Comb $4$ provides a higher beam divergence at the MHP, therefore, a lower amount of power passes through the eye pupil. 

Fig.~\ref{sub2:Fig-LG-eyesafety} represents the simulation results of $P_{\rm{t,max}}$ versus the wavelength of the laser for the four mode combinations using the $M$-squared based approach. For these results, the exposure duration is set to $10$ s. It is noted that the results given in Table~\ref{Table-ExpData2} are for the wavelength of $850$ nm. A similar ratio of $\theta_{\rm R}/\theta$ is assumed in the simulations of Fig.~\ref{sub2:Fig-LG-eyesafety} for other wavelengths. From these results, the same conclusion, i.e., the LG-Comb $4$ relaxes the transmit power limitation, can be drawn. The other observation is about the gap between $P_{\rm{t,max}}$ of various combinations, which increases for higher wavelengths. This is due to the dependency of the MPE on wavelength, which is larger for higher wavelengths.

\begin{figure}[t!]
\centering
\includegraphics[width=0.8\textwidth]{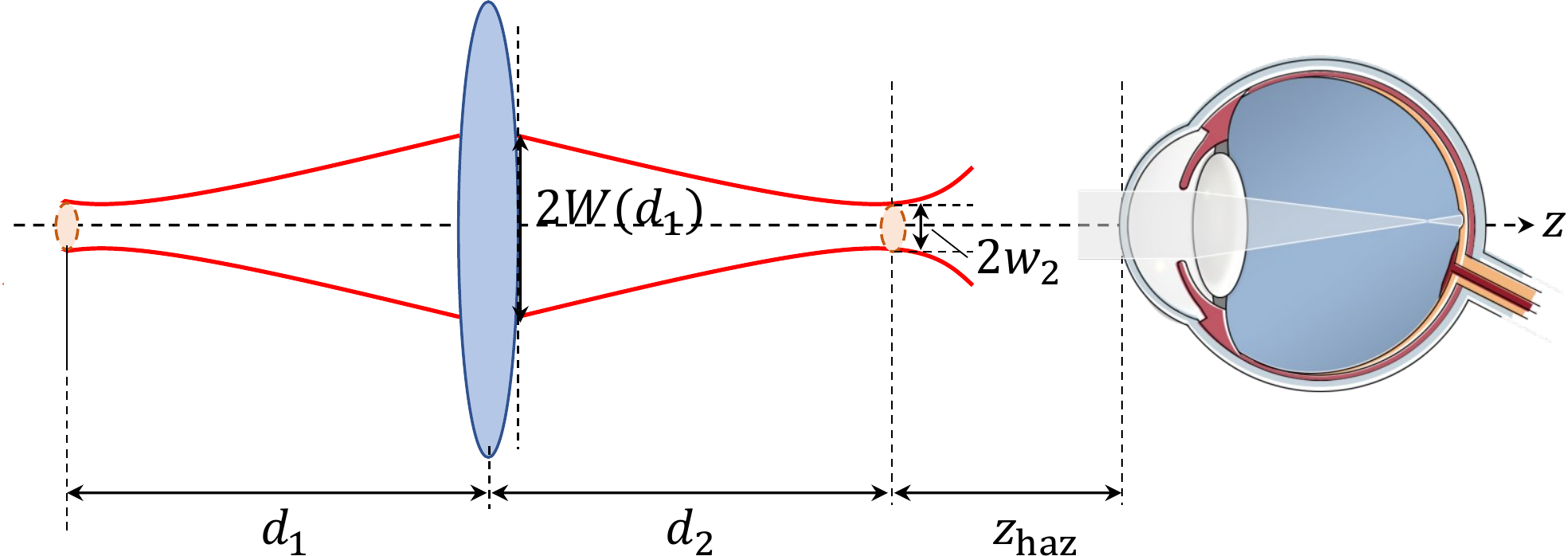} 
\caption{Beam transformation after a thin lens and illustration of the MHP.}
\label{Fig-LensSafetyAnalysis}
\end{figure}

\subsection{Laser Safety Analysis for Lens systems}
In cases where an optical lens is used at the output of a laser source, the eye safety calculations remain the same, but the beam parameters need to be transformed. Assuming the thin lens approximation, the lens will result in the transformation of the output beam according to the $ABCD$ formulation described in Section~\ref{BeamAfterLens}. The beam waist and its location after the thin lens are given in \eqref{beamwaistloc} and \eqref{w0AfterLens}, respectively. In order to determine the MHP, there is one consideration when dealing with such lenses, which is about the location of the image of the beam waist. The location of the image from the lens is denoted by $d_2$ in Fig.~\ref{Fig-LensSafetyAnalysis} and, depending on the relative location of the laser source from the lens and the focal length, it can be either positive or negative. 
Hence, if the location of the image is after the lens ($d_2>0$), the MHP would be considered $z_{\rm haz}$ m away from the location of $w_2$ (or $d_2+z_{\rm haz}$~m away from the lens). Otherwise, the MHP is considered $z_{\rm haz}$ m away from the lens. In this case, the size of the apparent source for the calculation of the subtense angle is the one at the lens location, which is $W(d_1)$. 
The procedure to calculate $P_{\rm{t,max}}$ for a laser with a lens placed in front of it, is described in algorithm~\ref{Algo-Lens}.

\begin{algorithm}[t]
 \caption{Determining maximum transmit power of a laser source with a lens.}
 \label{Algo-Lens}
 \begin{algorithmic}[1]
 \renewcommand{\algorithmicrequire}{\textbf{Input}:}
 \renewcommand{\algorithmicensure}{\textbf{Output}:}
 \REQUIRE  Denote the wavelength of the laser source by $\lambda$;\\
the beam waist by $w_0$;\\
the radius of the pupil by $r_{\rm p}$;\\  exposure duration by $t_{\rm{ex}}$;\\
the distance between laser and lens by $d_1$;\\
the focal length of the lens by $f$.
\ENSURE  Maximum transmit power, $P_{\rm{t,max}}$ 
\STATE Calculate $d_2$, $w_2$ and $\theta_2$ based on \eqref{beamwaistloc}, \eqref{w0AfterLens} and \eqref{BeamDivergenceLens}, respectively
\IF{$d_2>0$ }
\STATE Use $w_2$ and $\theta_2$ as the new beam waist and beam divergence respectively in the equations of algorithm~\ref{Algo-singlemodelaser} 
\STATE Calculate $P_{\rm{t,max}}$ 
\ELSE
\STATE Find the spot radius at the lens location, $W(d_1)$
\STATE Calculate $\alpha$ based on $W(d_1)$
\STATE Use $w_2$ and $\theta_2$, $\alpha$ as the new beam waist, beam divergence and subtense angle respectively in the equations of algorithm~\ref{Algo-singlemodelaser} 
\STATE Calculate $P_{\rm{t,max}}$ 
\ENDIF
\RETURN $P_{\rm{t,max}}$ for the laser with a lens
\end{algorithmic}
\end{algorithm}



\begin{figure}[t]
		\centering  
	\subfloat[\label{sub1:Fig-Lens-eyesafety}]{%
		\resizebox{.47\linewidth}{!}{\includegraphics{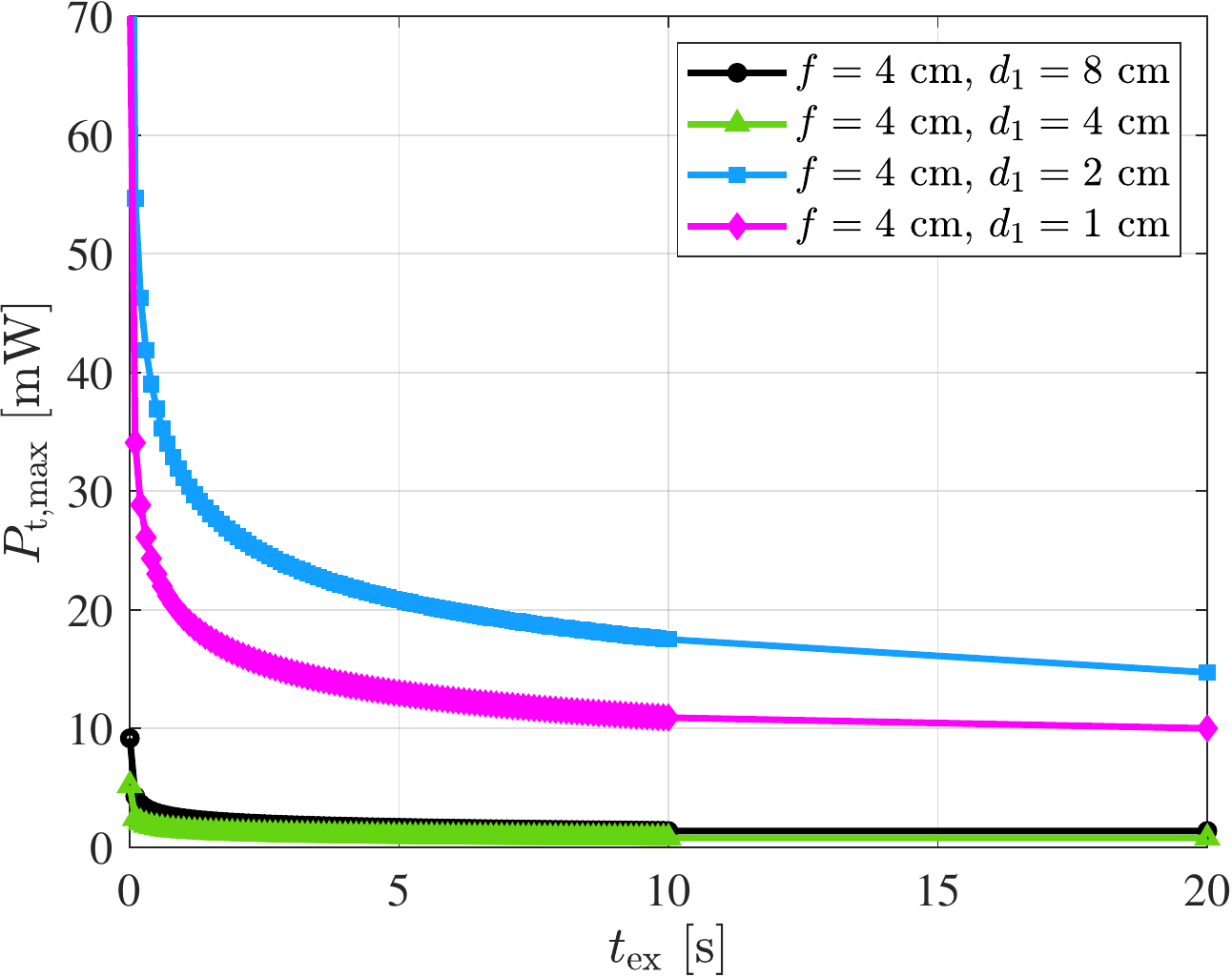}}
	} \quad\ \ 
	\subfloat[\label{sub2:Fig-Lens-eyesafety}]{%
		\resizebox{.47\linewidth}{!}{\includegraphics{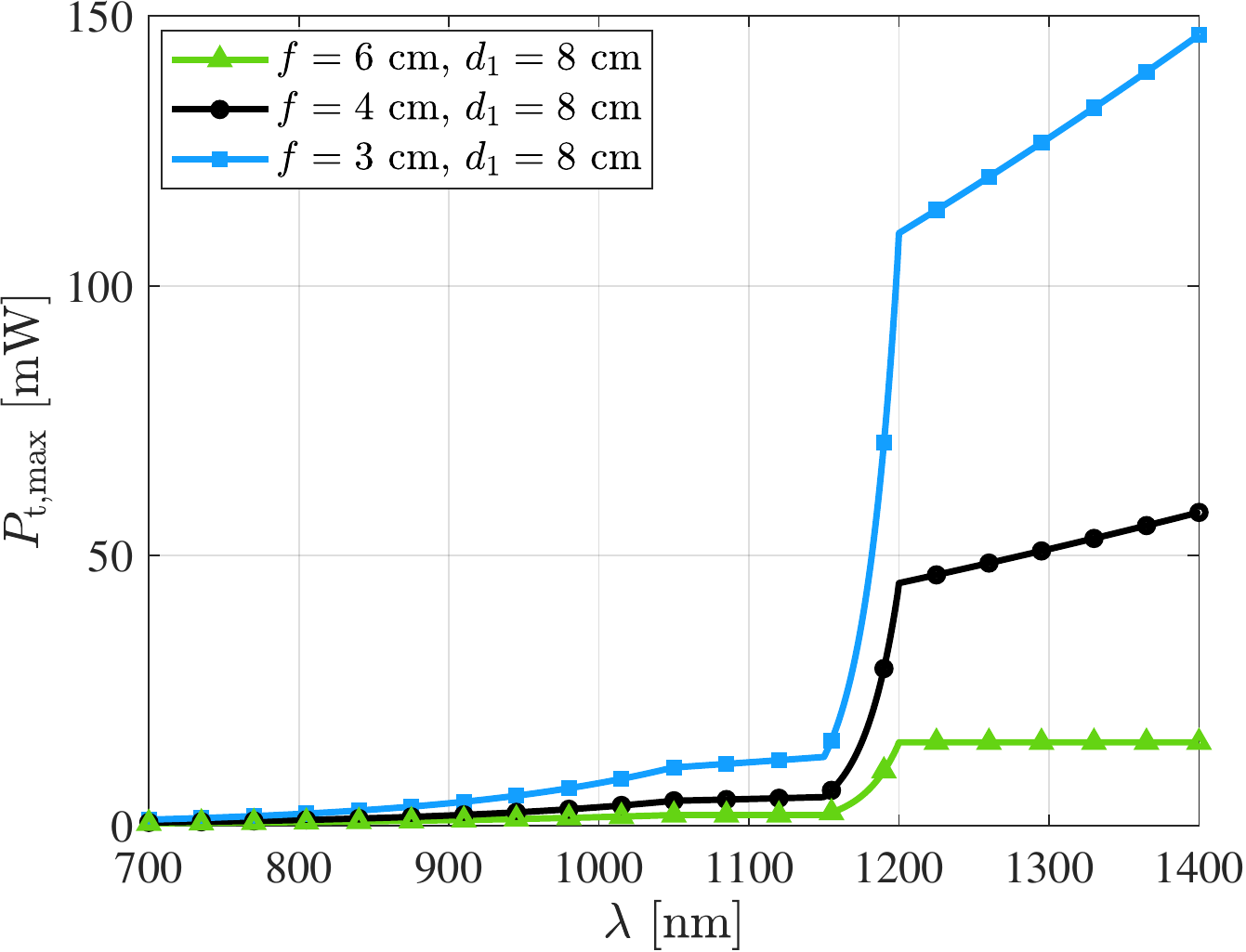}}
	}\\
	\caption{Eye safety analysis for a leaser with a lens: (a) $P_{\rm{t,max}}$ versus $t_{\rm ex}$ for different locations of the laser from the lens, (b) $P_{\rm{t,max}}$ versus $\lambda$ for different focal lengths. }
	\label{Fig-Lens-eyesafety}
\end{figure}

Fig.~\ref{Fig-Lens-eyesafety} depicts the eye safety results for a laser with a thin lens placed in front of it. The beam waist of the laser is assumed to be $w_0=5\ \mu$m in these simulations. Fig.~\ref{sub1:Fig-Lens-eyesafety} represents the results of $P_{\rm{t,max}}$ versus the exposure duration assuming the laser source operates at $\lambda=850$ nm. A lens with a focal length of $4$ cm is considered. Four locations for $d_1$ are studied. When the laser is located 8 cm away from the lens, the beam waist after the lens is $w_2=5\ \mu$m, therefore, the magnification factor is $\kappa=w_2/w_0=1$. Consequently, the results of $P_{\rm{t,max}}$ is the same as the one when there is no lens in front of the laser source (see Fig.~\ref{sub1:Fig1}). 
It can be seen that the most limiting case is related to $d_1=4$ cm where the laser is placed at the focal length of the lens. For this case, a collimated beam will be constructed after the lens, as a result, most of the power passes through the eye pupil.  
If the laser is located between the lens and the focal point, $P_{\rm{t,max}}$ can be increased since the subtense angle in this case is quite large, therefore the parameter $C_6=\alpha/\alpha_{\rm min}$ is large and the MPE value is higher. As the laser becomes closer to the lens, the size of the apparent source $W(d_1)$ becomes smaller, and therefore, $\alpha$ and $C_6$ are smaller, which correspond to lower values for MPE and $P_{\rm{t,max}}$. 
Fig.~\ref{sub2:Fig-Lens-eyesafety} illustrates the results of $P_{\rm{t,max}}$ versus the spectral region from 700 nm to 1400 nm for three different focal lengths. For these three curves, it is assumed that $d_1=8$ cm. Compared to the case without a lens shown in Fig.~\ref{Fig-PtmaxversusLambda}, $P_{\rm{t,max}}$ increases as $\lambda$ increases. However, the case of $f=6$ cm imposes a more limiting condition on the transmit power. When $f=4$ cm, the results are the same as the case without a lens since $w_2=w_0$. Finally, for the case of $f=3$ cm, $P_{\rm{t,max}}$ can be increased more since the beam divergence of the transformed beam is higher and a lower amount of power is collected at the eye pupil.  

\begin{figure}[t!]
\centering
\includegraphics[width=0.7\textwidth]{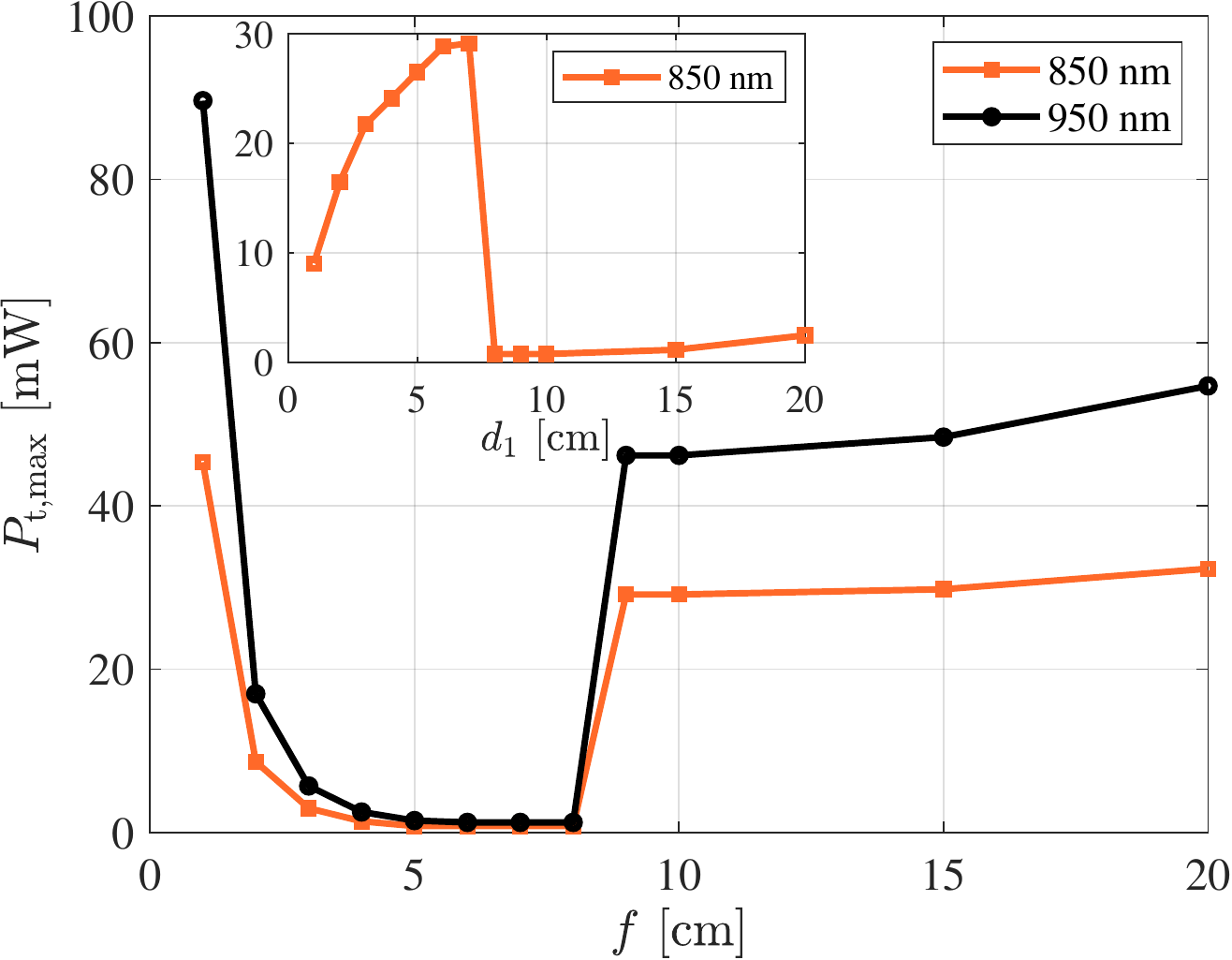} 
\caption{$P_{\rm{t,max}}$ versus focal length, $f$, and the distance between lens and laser, $d_1$.}
\label{Fig-LensFocallength}
\end{figure}

Fig.~\ref{Fig-LensFocallength} shows the impact of geometric parameters such as the focal length of the lens as well as the distance between the laser source and lens on $P_{\rm{t,max}}$. The exposure duration is set to $100$ s for these simulations. Let's first focus on the curves of $P_{\rm{t,max}}$ versus $f$. We set $d_1=8$ cm. The results are shown for two wavelengths of $850$ nm and 950 nm. 
Two different behaviors can be seen in these simulations. When the focal lengths decreases by up to 8 cm, $P_{\rm{t,max}}$ becomes less; while for $f>8$ cm, $P_{\rm{t,max}}$ increases. The minimum value for $P_{\rm{t,max}}$ is related to the focal length of 8 cm where the beam after the lens is collimated and the collected power into the eye is maximum. The inset of Fig.~\ref{Fig-LensFocallength} illustrated the effect of $d_1$ on $P_{\rm{t,max}}$. For this curve, we set $\lambda=850$ nm and $f=8$ cm. The peak value corresponds to $d_1=7$ cm, where the laser source is still between the focal point and the lens, and the apparent size, $W(d_1)$ is near to its maximum value.

\subsection{Laser Safety Analysis for Array of VCSELs}
The beam propagation of a VCSEL array is studied in Section~\ref{BeamPropagationArray}. The general approach to analyze the eye safety for sources that consist of several sources such as arrays or bundles of fibers is based on \textit{combinational} methodology, where in addition to the evaluation of the source as a whole, each individual source needs to be evaluated separately, as well as all combinations of sources. Each individual source or combination is associated with a certain value of $\alpha$, and for each individual source or combination, a particular transmit power level will be measured within the pupil diameter of $r_{\rm p}=7$ mm. 
The pairs of received power values and MPE (depending on $\alpha$) are subsequently compared, and the condition $P_{\rm r}<{\rm MPE}\times \pi r_{\rm p}^2$ should be guaranteed for the individual sources and all combinations. This condition imposes the limitation that the maximum transmit power for each individual source of an array is always less than the maximum transmit power of a single source. 
For an $N\times N$ array with the pitch distance of $d_{\sigma}$, the subtense angle for each combination can be obtained by:
\begin{equation}
\label{alphaArray}
    \alpha^{(i)}=2\tan^{-1}\left(\frac{(i-1)d_{\sigma}+2w_0}{2z_{\rm haz}}\right),
\end{equation}
for $i\in \{1,2,\cdots,N\}$ and $w_0$ is the beam waist of each individual source. The subtence angle for the combinational approach should be limited to a minimum value of $1.5$ mrad and to a maximum value of $100$ mrad.
Therefore, if the pitch distance is small enough that the subtense angle of an array is smaller than $1.5$ mrad, the whole array is treated as one source. The condition on pitch distance to fulfil this requirement is $d_{\sigma}<(0.00015-2w_0)/(N-1)$. 
On the other hand, when spaces between all sources of the array are in such a way that the subtense angle is larger than $100$ mrad, sources are treated as completely independent and not evaluated in a combination (as the corresponding images on the retina can be considered thermally independent) \cite{henderson2003laser}. The constraint $d_{\sigma}>(0.01-2w_0)/(N-1)$ ensures that individual sources can be evaluated independently.

\begin{figure}[t!]
 \centering
\includegraphics[width=0.6\textwidth]{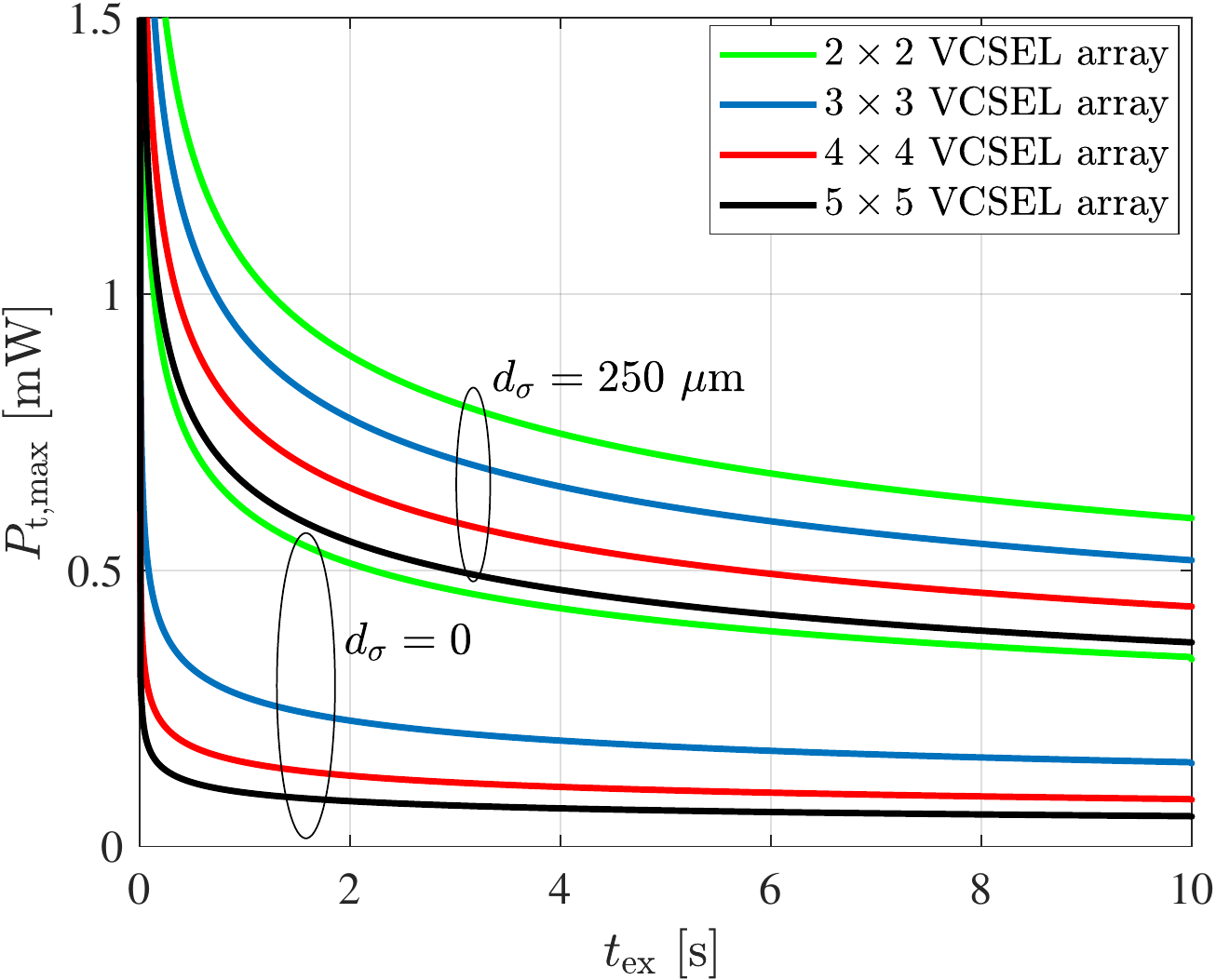} \caption{Eye safety analysis for a VCSEL array with $\lambda =850$ nm and $w_0=5\ \mu$m.}
\label{Fig-Array-eyesafety}
\vspace{-0.5cm}
\end{figure} 

Fig.~\ref{Fig-Array-eyesafety} illustrates the results of the maximum transmit power per laser source versus the exposure duration assuming that the laser source operates at $\lambda=850$ nm with a beam waist of $5\ \mu$m. The results are given for different array sizes and for two pitch distances of $d_\sigma=0$ and $d_\sigma=250\ \mu$m. Obviously, as the size of the array increases, the maximum transmit power for each individual source decreases. The reason for this is that the output power of each individual VCSEL will be summed up at the eye pupil, particularly when the distance between the array and eye is large enough in comparison to the size of the array. As the pitch distance increases, the subtence angle also increases, which means that the source is categorized as an extended one, for which the MPE value and $P_{\rm t,max}$ increase as well. 
We can see from these results that an increase of $7$ times in the maximum transmit power of a $5\times5$ VCSEL array is possible when the pitch distance increases from $0$ to $250\ \mu$m.
Another important note regarding these results is that the maximum transmit power for each individual source of an array is less than the maximum transmit power of a single source (see results shown in Fig.~\ref{sub1:Fig1}). It is noted that the values of $P_{\rm t,max}$ for $10<t_{\rm ex}<3\times10^4$ s is the same as the values of $P_{\rm t,max}$ for $t_{\rm ex}=10$, which are not represented in this figure.

\subsection{Laser Safety Analysis Considering Diffusers}
\subsubsection{Lambertian Pattern Diffusers}
\hfill\\
For a diffuser in front of a laser source, the condition  described in \eqref{irradianceCondition} should be fulfilled to guarantee the eye safety regulation. Hence, the maximum transmit power can be obtained by substituting \eqref{EqReceivedPowerLambertianDiffuser} into \eqref{irradianceCondition}, which should be evaluated at $z=z_{\rm{haz}}$ and for an aperture radius equals to the radius of pupil, i.e., $r_{\rm a}=r_{\rm p}$. Therefore, we have:
\begin{equation}
\label{MaximumTransitPower}
    P_{\rm{t,max}}=\frac{({\rm{MPE}})\pi r_{\rm{p}}^2}{1-\cos^{m+1}(\psi_{\rm{haz}})},
\end{equation}
where $\psi_{\rm{haz}}=\tan^{-1}(\frac{r_{\rm{p}}}{z_{\rm{haz}}})$ calculated at the hazardous distance, $z_{\rm{haz}}$.
To find out the $\rm MPE$, we require to calculate the subtense angle. Let's denote $\alpha_{\rm haz}=2\tan^{-1}(\frac{D}{2z_{\rm{haz}}})$, if $\alpha_{\rm haz}>\alpha_{\rm{min}}$, which corresponds to the regions close to the diffuser, the MPE of an extended source  should be adopted to be used in (\ref{MaximumTransitPower}). The values of MPE are provided in Table~\ref{table_eye_safety}. Let's denote $z_{\rm ps}=\frac{D}{2\tan(\frac{\alpha_{\rm{min}}}{2})}$ as the location after which the source is viewed as a point source. If $z_{\rm{haz}}<z_{\rm ps}$ (or equivalently $\alpha_{\rm haz}<\alpha_{\rm{min}}$) the MPE of a point source should be used in \eqref{MaximumTransitPower}.  

Practical diffusers work more efficiently with collimated beams \cite{Stefansson_2017}. In order to provide a collimated beam, a lens is required. By placing the point source at the focal point of the lens, a collimated beam will be generated after the lens. This beam passes through the diffuser and diverges according to the divergence angle of the diffuser. The geometry of a collimated beam using a lens is shown in Fig.~\ref{Fig-diffuser}. 
In general, and depending on the focal length of the lens, denoted as $f$, the diameter of the spot on the diffuser might be less than the diameter of the diffuser. In this case, we have:
\begin{equation}
    \alpha_{\rm{haz}}=2\tan^{-1}\left(\frac{W(f)} {z_{\rm{haz}}}\right),
\end{equation}
where based on (\ref{W_z}a), $W(f)=w_0\sqrt{1+\left(\dfrac{f}{z_0}\right)^2}$. 

Practical lasers have a unique beam divergence angle, $\theta$, up to several degrees. 
If we want to ensure the condition $2W(f)=D$, $f$ should be:
\begin{equation}
    f=\frac{D}{2\tan(\theta)},
\end{equation}
which means that for a diffuser with $D=2.5$ cm and a laser with beam divergence angle of $1^\circ$, the focal length should be $70$ cm.
Therefore, we usually have $W(f)<D$ for practical lenses. On the other side, $W(f)$ cannot be smaller than the \textit{minimum beam requirement} of a diffuser, which is specified in a datasheet. In fact, the diameter of the beam that hits a diffuser should be higher than the minimum beam requirement. This value for practical diffusers changes from $1$~mm to $5$ mm. 

\subsubsection{Uniform Pattern Engineered Diffusers}
\hfill\\ 
For uniform diffusers with a specific beam divergence angle of $\Theta_{\rm d}$, the geometry of the beam is shown in Fig.~\ref{Fig-diffuser}. Using \eqref{PrUniformDiffuser} and \eqref{irradianceCondition}, the maximum transmit power that meets the eye safety requirement is expressed as:
\begin{equation}
\label{PtmaxUniformPatternEq}
    P_{\rm t,max}={\rm MPE}\times\pi r_{\rm p}^2\times \frac{1-\cos(\Theta_{\rm d}/2)}{1-\cos(\psi^{'}_{\rm haz})},
\end{equation}
where $\cos(\psi^{'}_{\rm{haz}})=\frac{z^{'}_{\rm{haz}}}{\sqrt{{z^{'}}_{\rm{haz}}^2+r_{\rm p}^2}}$, where $z^{'}_{\rm{haz}}=z_{\rm{haz}}+\frac{W(f)} {\tan(\Theta_{\rm d}/2)}$.

\begin{algorithm}[t]
 \caption{Determining maximum transmit power of a laser source with a diffuser}\label{Algo-diff}
 \begin{algorithmic}[1]
 \renewcommand{\algorithmicrequire}{\textbf{Input}:}
 \renewcommand{\algorithmicensure}{\textbf{Output}:}
 \REQUIRE  Denote the diameter of the diffuser by $D$; the radius of the pupil by $r_{\rm p}$; the most hazardous distance by $z_{\rm{haz}}$; the focal length of the lens by $f$ and the divergence angle at FWHM of a diffuser by $\Theta_{\rm d}$.
 \ENSURE  Maximum transmit power, $P_{\rm{t,max}}$ of the laser source
 \STATE Compute $\alpha$ at $z=z_{\rm{haz}}$, as $\alpha_{\rm{haz}}=2\tan^{-1}\left(\frac{W(f)} {z_{\rm{haz}}}\right)$
 \STATE Calculate the associated MPE to the $\alpha_{\rm{haz}}$ from Table~\ref{table_eye_safety}
 \IF {Diffuser is Lambertian one} 
 \STATE Obtain $\psi_{\rm{haz}}=\tan^{-1}(\frac{r_{\rm{p}}}{z_{\rm{haz}}})$
 \STATE Calculate $P_{\rm{t,max}}=\frac{({\rm{MPE}})\pi r_{\rm{p}}^2}{1-\cos^{m+1}(\psi_{\rm{haz}})}$
 \ELSIF{Diffuser is uniform one}
 \STATE Obtain $\cos(\psi^{'}_{\rm{haz}})=\frac{z^{'}_{\rm{haz}}}{\sqrt{{z^{'}}_{\rm{haz}}^2+r_{\rm p}^2}}$, where $z^{'}_{\rm{haz}}=z_{\rm{haz}}+\frac{W(f)} {\tan(\Theta_{\rm d}/2)}$
 \STATE Calculate $P_{\rm t,max}={\rm MPE}\times\pi r_{\rm p}^2\times \frac{1-\cos(\Theta_{\rm d}/2)}{1-\cos(\psi^{'}_{\rm haz})}$
 \ENDIF
\RETURN $P_{\rm{t,max}}$ as the maximum transmit power of the laser source
\end{algorithmic}
\end{algorithm}

The approach for determining the maximum transmit power for a laser source with a diffuser has been summarized in algorithm~\ref{Algo-diff}. Next, we provide some results of the maximum transmit power. The enclosed extended source is assumed to be constructed with a diffuser which is placed in front of a point source. Furthermore, no limitations have been considered on the minimum beam requirement of a diffuser in the following simulations (this assumption indeed imposes more constraint on the maximum transmit power, particularly for smaller beam divergence angles).

\begin{figure}[t]
		\centering  
	\subfloat[$m=1$\label{sub1:Fig-Diffuser-eyesafety}]{%
		\resizebox{.47\linewidth}{!}{\includegraphics{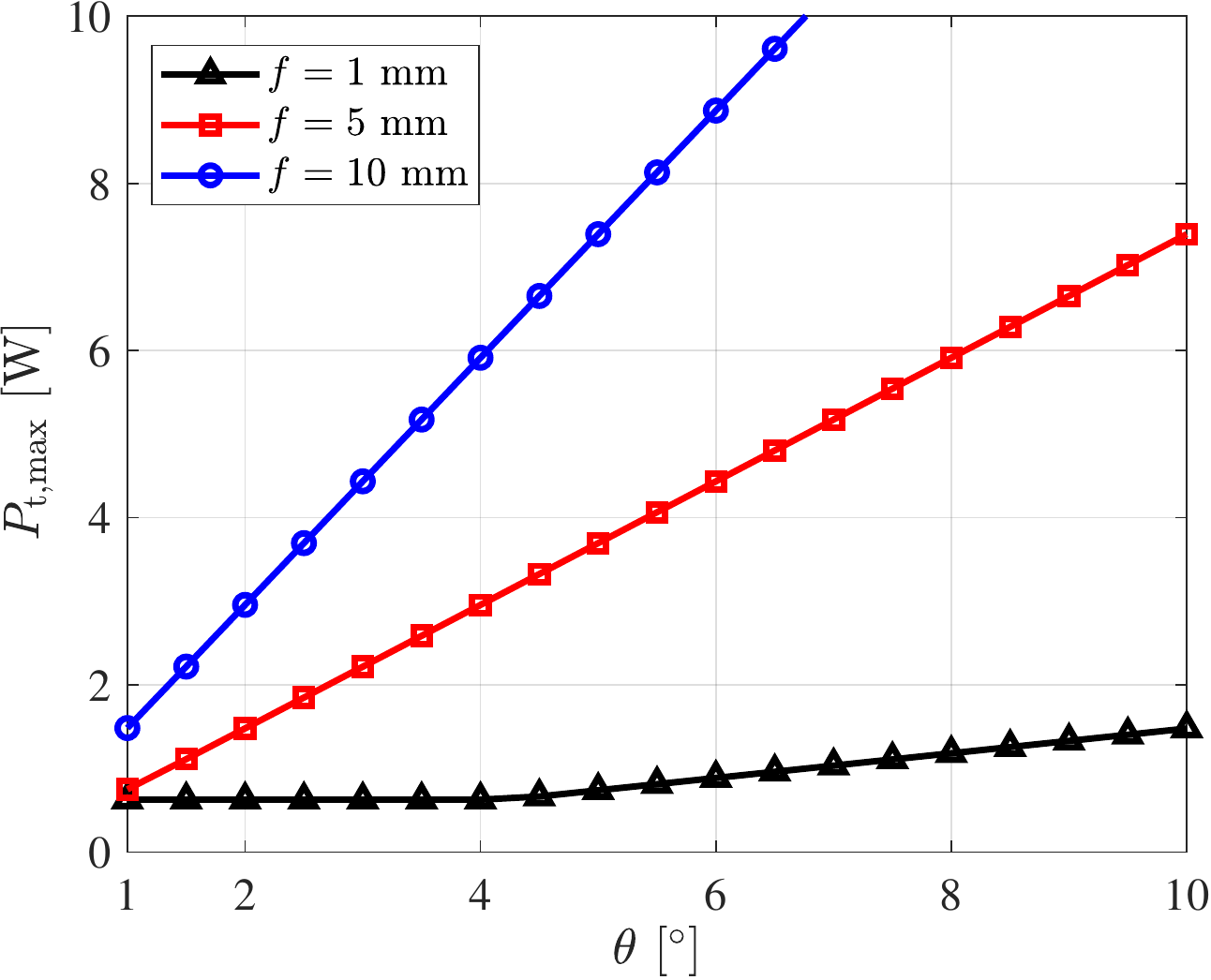}}
	} \quad\ \ 
	\subfloat[$m=5$\label{sub2:Fig-Diffuser-eyesafety}]{%
		\resizebox{.47\linewidth}{!}{\includegraphics{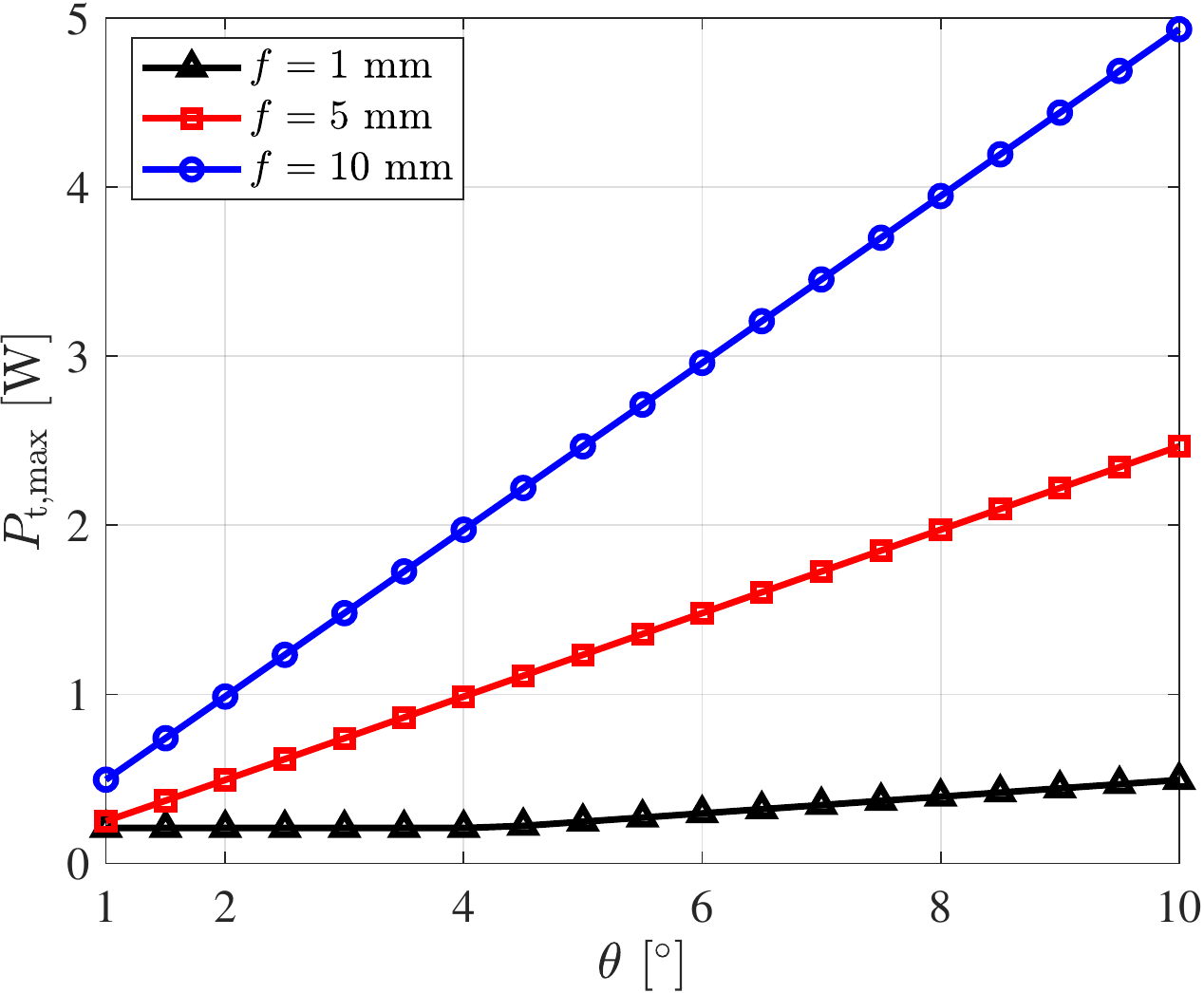}}
	}\\
	\caption{$P_{\rm{t,max}}$ versus $\theta$ for a Lambertian diffuser of order $m$ and with different focal lengths of the lens, $f$.}
	\label{Fig-Diffuser-Lambertian}
\end{figure}

Fig.~\ref{Fig-Diffuser-Lambertian} represents the maximum transmit power, $P_\mathrm{t,max}$, versus the beam divergence angle, $\theta$, with the consideration of eye safety for  Lambertian diffusers of orders $m=1$ and $m=5$. Different lenses with focal lengths of $1$ mm, $5$ mm and $10$ mm are assumed to collimate the beam. The wavelength of the laser source is assumed to be $850$ nm. As it can be inferred from these results, smaller focal lengths impose more constraints on the maximum transmit power. Comparing the sets of results shown in Fig.~\ref{sub1:Fig-Diffuser-eyesafety} and Fig.~\ref{sub2:Fig-Diffuser-eyesafety}, one can observe that as the Lambertian order of the difusser increases, the maximum transmit power increases. This is because  higher Lambertian orders focus the beam more and, as a result, a higher amount of power is collected at the eye.
Another note regarding these results is that, even though the maximum transmit power limit for $f=10$ is theoretically several Watts, this would be limited by the practical maximum transmit power of the laser source reported in the datasheets.

\begin{figure}[t]
		\centering  
	\subfloat[$\Theta_{\rm d}=20^{\circ}$\label{sub1:Fig-Diffuser20-eyesafety}]{%
		\resizebox{.47\linewidth}{!}{\includegraphics{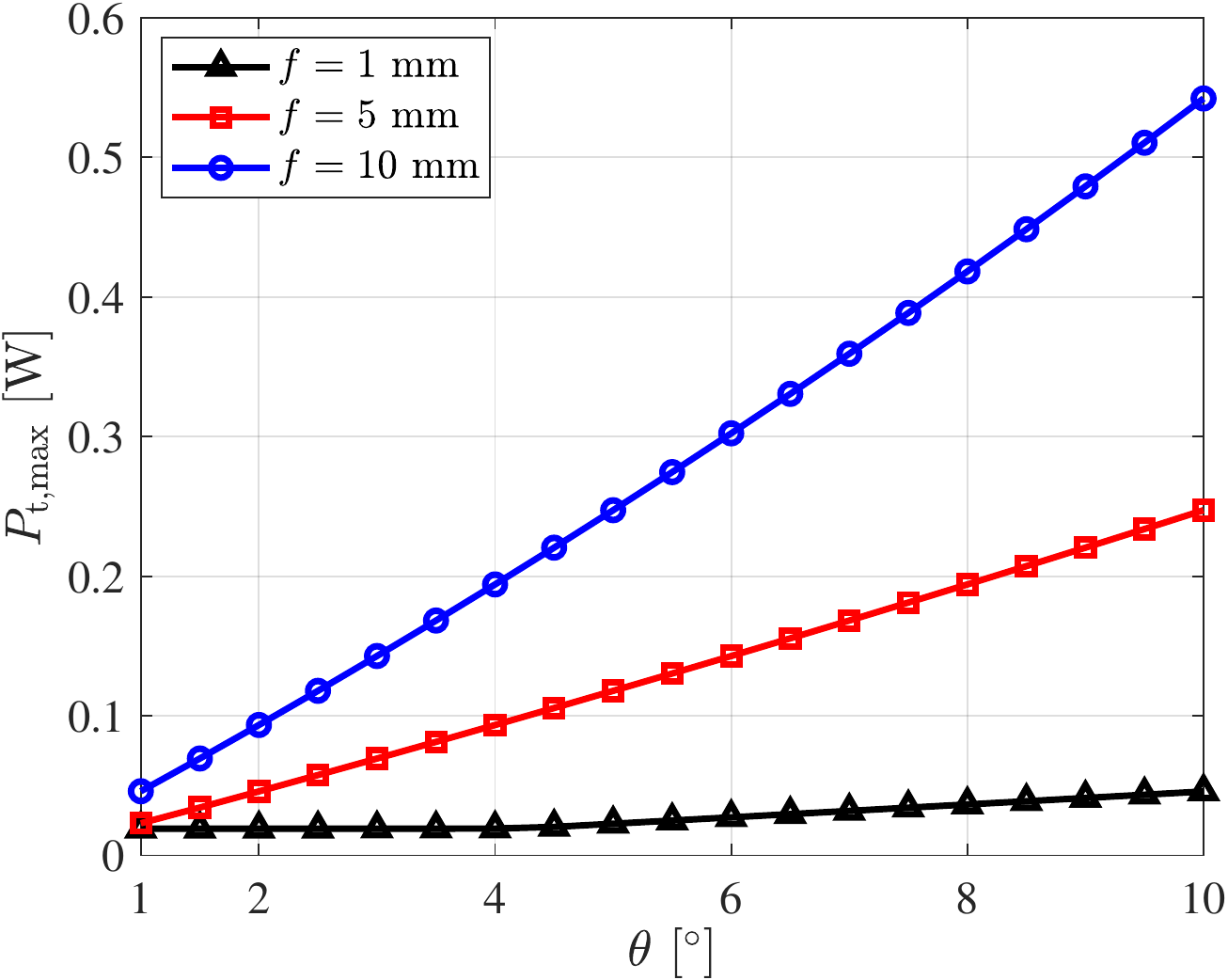}}
	} \quad\ \ 
	\subfloat[$\Theta_{\rm d}=50^{\circ}$\label{sub2:Fig-Diffuser50-eyesafety}]{%
		\resizebox{.47\linewidth}{!}{\includegraphics{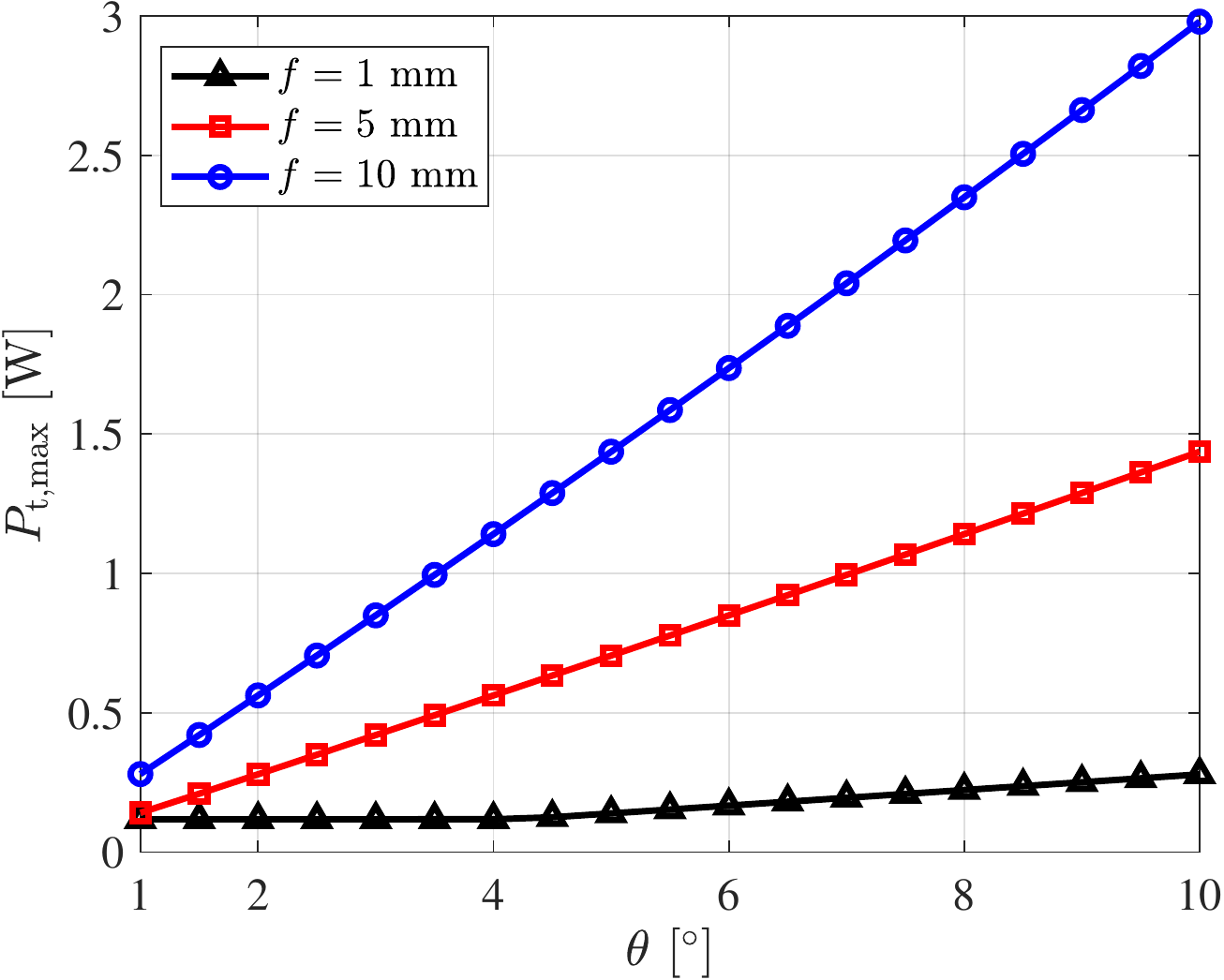}}
	}\\
	\caption{$P_{\rm{t,max}}$ versus $\theta$ for a uniform pattern diffuser with FWHM divergence angle of $\Theta_{\rm d}$ and for different focal lengths of the lens, $f$.}
	\label{Fig-Diffuser-uniform}
\end{figure}

Fig.~\ref{Fig-Diffuser-uniform} represents the maximum transmit power, $P_\mathrm{t,max}$, versus the beam divergence angle, $\theta$, with eye safety considerations for a uniform diffuser with FWHM divergence angles of $20^\circ$ and $50^\circ$. The wavelength of the laser source is assumed to be $850$ nm. The results are obtained based on \eqref{PtmaxUniformPatternEq}. Again, different lenses with focal lengths of $1$ mm, $5$ mm and $10$ mm are assumed to collimate the beam. As it can be observed from these curves, a similar conclusion can be drawn as the one for a Lambertian diffuser, i.e., smaller focal lengths limit the maximum transmit power more. It can be seen that the limitations on the maximum transmit power for the diffuser with the FWHM divergence of $50^\circ$ have been been more relaxed, compared to that of the diffuser with the FWHM divergence of $20^\circ$.


\subsection{Skin Safety}
Injuring the skin is far less likely than injuring the eye unless one is working with very high-powered laser sources and protects only the eye. Skin injuries are typically not possible with low power and medium power lasers. Skin safety analysis depends not only upon the area of the skin absorbing the laser radiation, but also how long the skin is exposed to the laser and also the wavelength of the laser source.
The MPE levels required to cause injury to the skin are quite high, at least several Watts-per-square-centimeter (W/cm$^2$ ) \cite{sliney2013safety}.

\begin{table}[!ht]
\centering
\caption[blah]{MPE calculation for skin safety analysis \cite{StdEye2014}.}
\label{table_skin_safety} 
\begin{tabular}{|c|c|c|c|c|c|c|c|}
\hline    
\multicolumn{2}{|c|}{$\lambda$ [nm]} & \multicolumn{3}{|c|}{$1400\ \leq\lambda< 1500$ } & \multicolumn{3}{c|}{$1500\ \leq\lambda< 10^5$}  \\
\hline
\multicolumn{2}{|c|}{{$t_{\rm{ex}}$ [s]}} & \makecell{$10^{-3}$ to\\ 0.35} & \makecell{0.35 to\\ 10} & \makecell{10 to\\ $3\times10^4$} & \makecell{$10^{-3}$ to\\ 0.35} & \makecell{0.35 to\\ 10} & \makecell{10 to\\ $3\times10^4$} \\
\hline
\makecell{\multirow{3}{*}{MPE [W$/$m$^2$]}} & $A_{\rm s}\leq0.01$
& \multicolumn{2}{|c|}{$5600t_{\rm{ex}}^{-0.75}$} &$1000$ & \multicolumn{2}{|c|}{$\dfrac{10^4}{t_{\rm{ex}}}$} & $1000$ \\
\cline{2-8} & $0.01<A_{\rm s}\leq0.1$
& \multicolumn{2}{|c|}{$\dfrac{56t_{\rm{ex}}^{-0.75}}{A_{\rm s}}$} &$\dfrac{10}{A_{\rm s}}$ & \multicolumn{2}{|c|}{$\dfrac{10^2}{t_{\rm{ex}}A_{\rm s}}$} & $\dfrac{10}{A_{\rm s}}$ \\
\cline{2-8} & $A_{\rm s}>0.1$
& \multicolumn{2}{|c|}{$560t_{\rm{ex}}^{-0.75}$} &100 & \multicolumn{2}{|c|}{$\dfrac{10^3}{t_{\rm{ex}}}$} & $100$ \\
\hline 
\multicolumn{2}{|c|}{\makecell{Limiting aperture\\ diameter (mm) for skin}} & $3.5$ &$3.5$ & $3.5$ &$3.5$ &$3.5$ & $3.5$\\
\hline
\multicolumn{2}{|c|}{\makecell{Limiting aperture\\ diameter (mm) for eye skin}} & 1 &$1.5t_{\rm{ex}}^{3/8}$ & $3.5$ &1 &$1.5t_{\rm{ex}}^{3/8}$ & $3.5$\\
\hline
\end{tabular}
\end{table}

For skin safety, the MHP is where the intensity of the beam is the highest, and in most cases this will be at the position closest to the laser beam waist. 

For most beams in the $700-1400$~nm wavelength range, the transmit power will ideally be limited by retinal power limits. In this case, for the purposes of skin safety, it is only necessary to limit access to the beam closest to the waist, so that the beam irradiance is not high enough to cause damage. Shielding provides sufficient distance to the beam waist and it helps to avoid direct contact with the laser. This allows a further increase in transmit power.

In the 1400 nm -- 10$^5$ nm spectral region, and for exposure durations between 10 and $3 \times 10^4$ seconds, the average irradiance measured through a 3.5 mm diameter aperture should not exceed 1000 W$/$m$^2$, unless the exposed area of skin is greater than 0.1~m$^2$. For values when the exposed area of skin is greater than 0.1~m$^2$, the maximum irradiance is 100~W$/$m$^2$, and for skin areas between 0.01~m$^2$ and 0.1~m$^2$, the maximum permissible exposure is $10/A_\mathrm{s}$ [W$/$m$^2$], where $A_\mathrm{s}$ is the area of exposed skin. In addition, for beams with a 1/e$^2$ diameter of less than 1 mm, the actual radiant exposure (not averaged over the 3.5 mm aperture) should be calculated \cite{StdEye2014}. The MPE values for the spectral region of 1400 nm to 10$^5$ nm for exposure durations higher than $10^{-3}$~s are presented in Table~\ref{table_skin_safety}. It can be seen from this table, the MPE depends on the limiting aperture diameter. For skin safety considerations, the limiting aperture diameter depends on the exposure duration, which are given in the table.
In the following, we will provide the skin safety analysis for single transverse mode and multimode beams.  

\subsubsection{Single Transverse Mode}
In order to calculate the skin-hazard for an ideal Gaussian beam, it is clear that the maximum transmit power should be evaluated at the point at which the accessible beam diameter is smallest, and in the case of an ideal Gaussian laser beam, this will be at the beam waist location. If shielding is used, then the skin hazrdous distance will shift to the location where the shield is placed.
The skin-hazard-limited transmit power for an ideal Gaussian beam with the shielding distance of $d_{\rm sh}$ is given as:
\begin{equation}
\label{SkinSafetyGeneralEq}
P_\mathrm{t,max}=\frac{\pi r_{\rm a}^2\times {\rm MPE} } {1-\exp{\left(-\frac{2r_{\rm a}^2}{W^2\left(d_{\rm sh}\right)}\right)}},
\end{equation}
where $r_{\rm a}$ is the radius of the limiting aperture given in Table~\ref{table_skin_safety}. 
The area of exposed skin at $z=d_{\rm sh}$ is $A_{\rm s}=\pi W^2(d_{\rm sh})$. 
Let's assume $t_{\rm ex}>10$ s, then, $r_{\rm a}=1.75$ mm and the maximum transmit power to ensure skin safety requirements for different conditions of $W(d_{\rm sh})$ are presented in Table~\ref{PtmaxSkinSafety}.

\begin{table}[!b]
\centering
\caption{Maximum permissible transmit power for skin safety.}
\label{PtmaxSkinSafety}
    \begin{tabular}{| l | c | l | }
    \hline
\multicolumn{1}{|c|}{$W(d_{\rm sh})$ [m]} & \multicolumn{1}{|c|}{MPE [W$\cdot$ m$^{-2}$]} & \multicolumn{1}{|c|}{$P_{\rm t,max}$ [W]}\\ \hline
$W(d_{\rm sh})<1.75\times10^{-3}$ & 1000 & 1000$\times{\pi W^2(d_{\rm sh})}$\\
\hline
$1.75\times10^{-3}\leq W(d_{\rm sh})<\sqrt{\dfrac{0.01}{\pi}}$ & 1000 & $\dfrac{1000\pi r_{\rm a}^2}{1-\exp\left(-\dfrac{2r_{\rm a}^2}{W^2(d_{\rm sh})}\right)}$\\ \hline
$\sqrt{\dfrac{0.01}{\pi}}\leq W(d_{\rm sh}) <\sqrt{\dfrac{0.1}{\pi}}$ & $\dfrac{10}{\pi W^2(d_{\rm sh})}$ &$\dfrac{10r_{\rm a}^2}{W^2(d_{\rm sh})\left(1-\exp\left(-\dfrac{2r_{\rm a}^2}{W^2(d_{\rm sh})}\right)\right)}$\\ \hline 
$W(d_{\rm sh})\geq \sqrt{\dfrac{0.1}{\pi}}$ & 100 & $\dfrac{100 \pi r_{\rm a}^2}{1-\exp\left(-\dfrac{2r_{\rm a}^2}{W^2(d_{\rm sh})}\right)}$ \\ \hline
    \end{tabular}
\end{table}


\begin{figure}[t!]
 \centering
\includegraphics[width=0.6\textwidth]{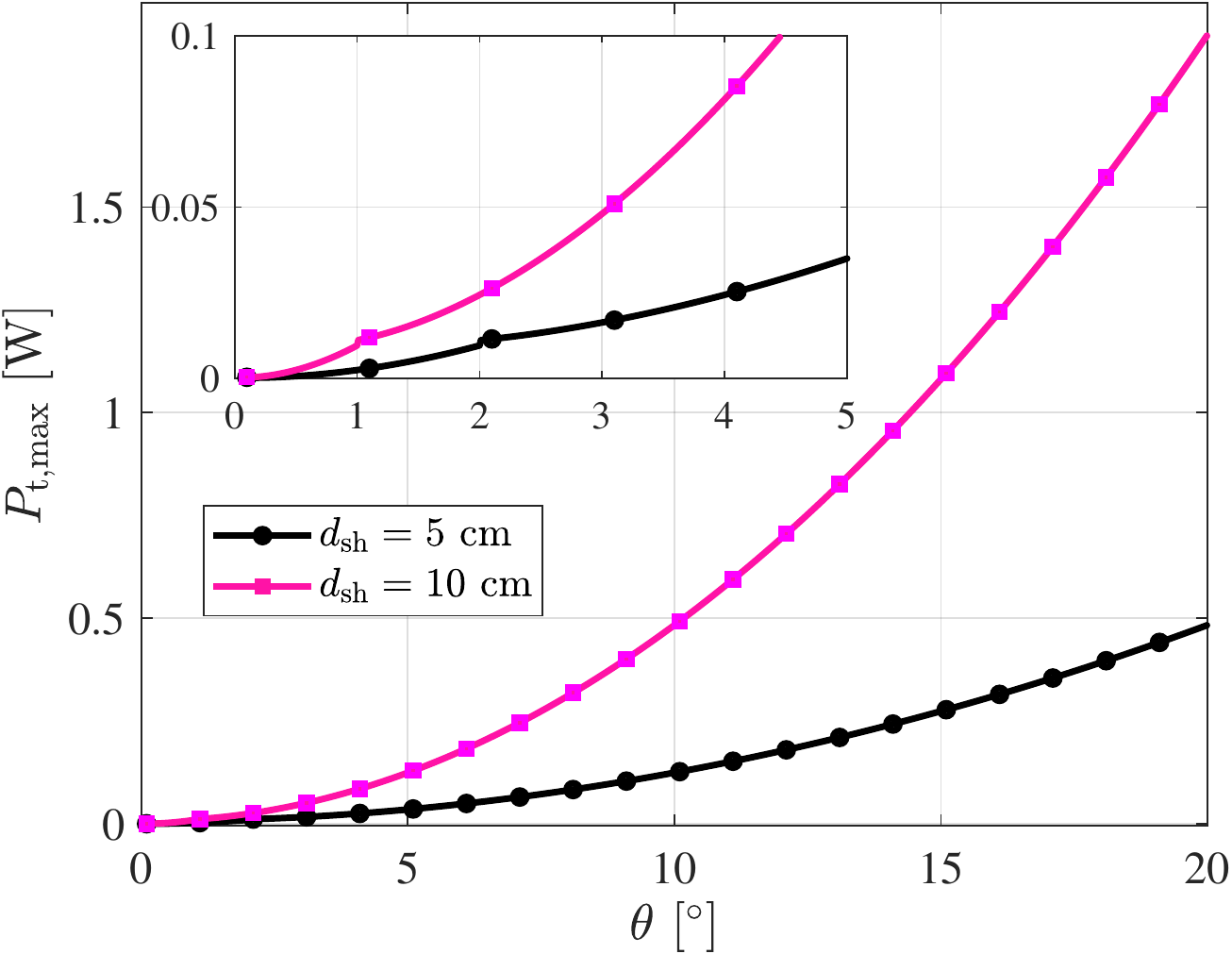} \caption{Skin safety results for $\lambda=1550$ nm.}
\label{Fig-skinsafety}
\end{figure} 

The skin safety results for $\lambda=1550$ nm are illustrated in Fig.~\ref{Fig-skinsafety}. As discussed in Section~\ref{Termin-wavelength}, skin safety becomes notable for higher wavelengths ($>1400$ nm). The wavelength $\lambda=1550$ nm is chosen as an example as there are commercial APDs working at this wavelength. As it can be observed from these results, the maximum transmit power increases as the beam divergence increases. Furthermore, as the shielding distance decreases, the maximum transmit power decreases. For high beam divergence angles, $P_{\rm t,max}$ can be upto few Watts, however, for laser sources with very small beam divergence angles, $P_{\rm t,max}$ is considerably small. Shielding is a practical method that can help to increase $P_{\rm t,max}$ for those lasers.     

\subsubsection{Multimode Laser Sources}
For multimode beams, the maximum transmit power will be dependent upon the maximum power gathered through an aperture of 3.5 mm diameter \cite{StdEye2014}, calculated at the accessible position where the beam has the smallest area. This can be calculated as per the formula presented in \eqref{HermiteGaussianPrEq} and \eqref{Lag:Pr:SafetyAnalysis} for Hermite-Gaussian and Laguerre-Gaussian beams, respectively.

\section{Summary and Concluding Remarks}
\label{Sec: Conclusion}
This tutorial provided technical information necessary for the safety analysis of laser sources under different conditions. Initially, we studied beam propagation for ideal Gaussian and multimode laser beams, including two well-known solutions of the paraxial Helmholtz equation, i.e., Hermite-Gaussian and Laguerre-Gaussian modes. We investigated beam shaping by using optical elements such as lenses and two types of practical diffusers. We also examined the beam propagation for VCSEL arrays, where multimode VCSELs with various mode power coefficients were assumed.

The main objective of this study is to derive the maximum permissible transmit power that fulfills both eye and skin safety requirements for laser products according to the IEC standard regulations. Analytical derivations were developed for various beam propagation models and optics. Specifically, first, we considered the ideal Gaussian beam, for which an algorithm was provided to obtain the maximum transmit power. Then, we looked into higher-mode beams by comparing safety results of the mode decomposition and $M$-squared approaches. It was demonstrated that the $M$-squared method leads to a more conservative maximum transmit power with respect to the precise beam decomposition method in some scenarios, however, for the purpose of power budget improvement, one can use the intricate beam decomposition approach. We also studied the effect of using a lens in front of a laser source for various focal lengths and distances between the lens and laser, showing that the maximum transmit power is minimal when the laser source is placed at the focal point of the lens. 

Thereupon, we analyzed VCSEL arrays by considering  practical aspects such as pitch distance (i.e., the center-to-center distance between adjacent VCSEL elements along rows or columns). The results suggest that by increasing the size of arrays, the transmit power per VCSEL reduces, and this effect is more pronounced for small values of pitch distance. We showed that an increase of $7$ times in the maximum transmit power of a $5\times5$ VCSEL array is possible when the pitch distance increases from $0$ to $250\ \mu$m. For lasers with diffusers, we showed that the transmit power limitation can always be relaxed to some degree, especially when using diffusers with a higher FWHM divergence angle. Finally, we investigated skin safety as another important factor in laser safety analysis, which in particular plays a significant role for higher wavelengths ($\geq1400$ nm). The discussions in the paper were supported by extensive simulation results to characterize the impacts of various parameters of laser sources such as wavelength and beam divergence angle, as well as transmitter optics and different exposure durations. This paper provides guidelines for the design of high-speed laser-based OWC systems in consideration of the safe transmit power for both indoor and outdoor applications.

\section{Acknowledgement}
The authors acknowledge financial support from the EPSRC under program grant EP/S016570/1 `Terabit Bidirectional Multi-User Optical Wireless System (TOWS) for 6G LiFi'. Prof. Harald Haas also acknowledges financial support from the Engineering and Physical Sciences Research Council (EPSRC) under the Established Career Fellowship grant EP/R007101/1 as well as the Wolfson Foundation and the Royal Society. The authors also acknowledge helpful technical discussions provided by Dr. Ravinder Singh.





\bibliography{reference}

\end{document}